\documentclass[10pt, journal]{IEEEtran}
\usepackage[cmex10]{amsmath}
\usepackage{amssymb}                    
\usepackage{array}
\usepackage{mdwmath}
\usepackage{mdwtab}
\usepackage{eqparbox}
\usepackage{url}
\usepackage{cite}                       
\usepackage{float}                      
\usepackage{graphicx}
\usepackage{multicol}
\ifCLASSOPTIONcompsoc
\usepackage[caption=false,font=normalsize,labelfont=sf,textfont=sf]{subfig}
\else
\usepackage[caption=false,font=footnotesize]{subfig}
\fi
\usepackage{multirow}
\usepackage{algorithmicx}
\usepackage{algorithm}
\usepackage{algpseudocode}
\algnewcommand{\Initialization}[1]{
  \State \textbf{Initialization}}
\algnewcommand{\Repeatu}[1]{
  \State \textbf{Repeat}}
\algnewcommand{\until}[1]{
  \State \textbf{until}}
\begin{document}
\bstctlcite{IEEEexample:BSTcontrol}
    \title{Broadened-beam Uniform Rectangular Array Coefficient Design in LEO SatComs Under Quality of Service and Constant Modulus Constraints}
    \author{Weiting Lin,~\IEEEmembership{Student Member,~IEEE,}
            Yuchieh Wu,
            Borching Su,~\IEEEmembership{Member,~IEEE}
}  
\maketitle
\begin{abstract}
Satellite communications (SatComs) are anticipated to deliver global Internet access.
Low Earth orbit (LEO) satellites (SATs) offer the advantage of higher downlink capacity due to their reduced link budget compared to medium Earth orbit (MEO) and geostationary Earth orbit (GEO) SATs.
In this paper, beam broadening methods for uniform rectangular arrays (URAs) in LEO SatComs were studied.
The proposed method is the first of its kind to jointly consider path loss variation from SAT to the user terminal (UT) due to the Earth's curvature to guarantee the quality of service (QoS), 
constant modulus constraints (CMCs) favored for maximizing power amplifier (PA) efficiency, 
and out-of-beam radiation suppression to avoid interference. 
A broadened-beam URA coefficient design problem is formulated and decomposed into two uniform linear array (ULA) design subproblems utilizing Kronecker product beamforming. 
With this decomposition, the number of beamforming coefficients that need to be optimized is significantly reduced compared to the original URA design problem. 
The non-convex ULA subproblems are addressed using the semidefinite relaxation (SDR) technique and a convex iterative algorithm. 
Simulation results reveal the advantages of the proposed method for suppressing the out-of-beam radiation and achieving the design criteria.
In addition, channel capacity evaluations are carried out. 
It demonstrates that the proposed ``broadened-beam" beamformers can offer capacities that are at least four times greater than those of beamformers employing an array steering vector when the beam transition time is considered. 
The proposed method holds potential for LEO SAT broadcasting applications, such as digital video broadcasting (DVB).\end{abstract}
\IEEEpeerreviewmaketitle

\section{Introduction} \label{sec:introduction}
\IEEEPARstart{S}{atellite} communications (SatComs) are expected to provide global Internet access and facilitate seamless connections worldwide \cite{Kodheli2021, LU2020, 3GPPNTN2023_TS18_202312}.
Due to advances in satellite (SAT) payload design and launching processes, the cost of deploying Low Earth orbit (LEO) SATs has become affordable for business service providers \cite{Osoro2021_Starlink_OneWeb_Kuiper_cost_model}. 
LEO SAT constellations, such as Starlink, OneWeb, Kuiper, and Lightspeed, have been deployed in recent years \cite{Osoro2021_Starlink_OneWeb_Kuiper_cost_model}.
LEO SATs have the potential to deliver high-speed broadband services with low latency due to their reduced link budget and shorter propagation delay compared to medium Earth orbit (MEO) and geostationary Earth orbit (GEO) SATs \cite{Kodheli2021}.
To achieve a higher downlink capacity, beamforming is essential to overcome the link budget.
If the transmitted signal phase is adjusted according to the array steering vector for beamforming purposes,
the main lobe beamwidth is inversely proportional to the antenna aperture when the signal wavelength is fixed \cite{Richards2014}.
SATs equipped with a larger array can achieve greater array gain, resulting in higher downlink capacity.
However, the SAT beam coverage areas decrease as the antenna aperture increases and more beams are required to cover specific service areas.     
Although beam hopping technologies can be utilized for beam resource management \cite{Tang2021, Hu2020, Lei2020}, the complexity of the system increases significantly as the number of beams increases or the demand for various services increases \cite{Hu2020, Lei2020}.
In addition, frequent beam switching increases the beam transition time delays resulting from the data queueing time on SATs \cite{Tang2021, Hu2020, Lei2020}. 
In light of this, it is crucial for SATs equipped with a large array size to pursue high downlink capacity while simultaneously serving user terminals (UTs) over wide areas, especially for downlink broadened-beam transmission.

For broadened-beam applications in SatComs,
isoflux radiation patterns have been considered to achieve a uniform received power density within SAT service areas \cite{Vigano2010, Reyna2012, Ibarra2015, Yoshimoto2019, Zeng2021_isoflux, Cai2023}. 
In \cite{Vigano2010}, 
an analytically based synthesis technique for designing sparse arrays with isoflux radiation patterns was presented.
In addition, many isoflux radiation pattern synthesis techniques were based on evolutionary algorithms.
In \cite{Reyna2012, Ibarra2015, Yoshimoto2019, Zeng2021_isoflux, Cai2023}, 
evolutionary algorithms were proposed to minimize fitness functions to approximate a predefined isoflux radiation pattern. 
In \cite{Reyna2012}, a genetic algorithm was applied, and aperiodic planar arrays were designed.
In \cite{Ibarra2015}, an evolutionary multi-objective optimization method was employed, and sparse concentric rings arrays were designed.
In \cite{Yoshimoto2019}, a firefly algorithm was utilized, and non-uniformly spaced linear and planar arrays were implemented.
In \cite{Zeng2021_isoflux}, a genetic algorithm and particle swarm optimization methods were applied, and an elliptical dipole antenna array was designed. 
In \cite{Cai2023}, a differential evolution algorithm was presented, and a uniform linear array (ULA) of a shared subarray architecture was designed.

In SAT applications, beamformer design with constant modulus constraints (CMCs) is desired to enable power amplifiers (PAs) to operate near compression points to maximize efficiency \cite{Li2021, Larsson2018, Tervo2021}.
PA efficiency is essential for reducing power consumption \cite{Li2021}, especially in SAT where only limited power resources are available \cite{Piacibello2022}.
Also, high PA efficiency prevents wasted power that generates heat, potentially leading to circuitry overheating \cite{Larsson2018}.
Nevertheless, \cite{Vigano2010, Reyna2012, Ibarra2015, Yoshimoto2019, Zeng2021_isoflux, Cai2023} did not include CMCs when synthesizing isoflux radiation patterns.  
The beamforming coefficient designed without considering the CMCs in \cite{Reyna2012, Ibarra2015, Yoshimoto2019, Zeng2021_isoflux} may be impractical for SAT applications.
One possible reason for not considering CMCs in \cite{Reyna2012, Ibarra2015, Yoshimoto2019, Zeng2021_isoflux} could be the difficulties in considering equality constraints in evolutionary algorithms \cite{BarkatUllah2012, Mengjun2021, Zhang2021}.

In the literature, beam broadening algorithms considering CMCs or dynamic range ratio (DRR) constraints have been studied in \cite{Fonteneau2021_EuCNC, Xu2019_IEEEAccess, Liang2020_TAP, Zhang2021_AWPL, Lei2023_SP}. 
In \cite{Fonteneau2021_EuCNC},
a closed-form analytic expression for a phase-only broadened beampattern was derived.
In \cite{Xu2019_IEEEAccess}, 
linearly polarized patterns with control over the sidelobe level, cross-polarization level, and DRR constraints were synthesized.
A semidefinite relaxation (SDR) technique and a reweighted minimization method \cite{Fazel_2004} were applied to address semidefinite programming (SDP) with a rank-one constraint.
Also, alternating direction method of multipliers (ADMM)-based methods have been applied in beampattern synthesis  \cite{Liang2020_TAP, Zhang2021_AWPL, Lei2023_SP}. 
In \cite{Liang2020_TAP},
the objective was to minimize the ratio of the peak sidelobe level (PSL) to the main lobe lower bound considering the CMCs.
In \cite{Zhang2021_AWPL},
the objective was to minimize the PSL with a given main lobe variation under the CMCs and DRR constraints.
In \cite{Lei2023_SP}, the minimum array directivity in the main lobe regions was maximized by considering the DRR constraints.
Although beam broadening methods with CMCs or DRR constraints were discussed in  \cite{Vigano2010, Fuchs2014_TAP, Liu2018_TAP, Xu2019_IEEEAccess, Liang2020_TAP, Zhang2021_AWPL, Fonteneau2021_EuCNC, Lei2023_SP},   
none of them were specifically designed for LEO SAT applications. 
They did not consider the variation of path loss from SAT to UT due to the Earth’s curvature, 
and thus may not guarantee quality of service (QoS) (i.e., the received signal-to-noise ratio (SNR)) in SAT service areas. 

In this paper, 
we study beam broadening algorithms for a uniform rectangular array (URA) coefficient design in LEO SatComs.
To guarantee QoS in SAT service areas, 
the path loss variation between SAT and UT due to the Earth's curvature is considered in the beampattern shaping which is inspired by the synthesis of the isoflux radiation pattern in \cite{Reyna2012, Ibarra2015, Yoshimoto2019, Zeng2021_isoflux, Cai2023}.
Other design criteria include achieving CMCs for maximizing the PA efficiency and minimizing the power leakage in out-of-beam areas to prevent interference. 
The URA design problem is formulated and decomposed into ULA design subproblems using Kronecker product beamforming \cite{Wang2021, Frank2022_planner_array_Kronecker_Product_Beamforming, Albagory2022_planner_array, VanTrees2002}, which offers promising properties in terms of computational efficiency
\cite{Wang2021, Frank2022_planner_array_Kronecker_Product_Beamforming, Albagory2022_planner_array}. 
With the decomposition, the number of
beamforming coefficients that need to be optimized in the ULA design subproblems are significantly reduced compared to the original URA design problem.
The non-convex ULA design subproblems are reformulated as semidefinite programming (SDP) with a rank-one constraint and are addressed by the proposed algorithm.
Simulation results reveal the advantages of the proposed method for suppressing the out-of-beam radiation, guaranteeing QoS, and achieving CMCs.
Additionally, channel capacity analyses are conducted to compare the capacity of the proposed ``broadened-beam" beamformers with that of beamformers utilizing an array steering vector when the beam transition time is taken into account.

The remainder of this paper is organized as follows.
In Section \ref{Sec:system_model}, the system model is introduced and the baseband equivalent transceiver model is derived.
In Section \ref{sec:problem_formulation}, the received SNR in SAT beam coverage areas is derived, and the optimization problem of the URA coefficient design in LEO SatComs considering QoS and CMCs is formulated.
In Section \ref{Sec:proposed_algorithm}, the method of decomposing the URA design problem into ULA design subproblems is introduced, and the proposed algorithm is presented.
In Section \ref{sec:numerical_results}, numerical results reveal the advantages of the proposed method to achieve the design criteria and channel capacity evaluation of ``broadened-beam" beamformers and beamformers utilizing an array
steering vector is conducted. 
Finally, conclusions are presented in Section \ref{sec:Conclusion} and suggestions are provided for future research. 

\emph{Notations}:
Boldface upper case letters represent matrices, boldface lower case letters represent column vectors, and italic letters represent scalars, such as $\bf X$, $\bf x$, and $x$. 
The $m$-th entry of $\bf x$ and \mbox{$(m,n)$}-th entry of $\bf X$ are denoted by $x_m$ and $[{\bf X}]_{m,n}$, respectively.
The $M$-dimensional complex vector space is denoted by $\mathbb{C}^{M}$ and the space of all $M\times N$ matrices with complex entries is denoted by \mbox{$\mathbb{C}^{M\times N}$}.
The operators $(\cdot)^*$, $(\cdot)^T$, and $(\cdot)^H$ represent the complex conjugate, transpose, and conjugate transpose, respectively. 
The set of all real numbers is denoted by $\mathbb{R}$.
Zero-based indexing is applied, and $\mathbb{Z}_M$ stands for the set \mbox{$\{0,1,...,M-1\}$} for any positive integer $M$.  The set of all $M\times M$ positive semidefinite matrices is denoted as ${\mathbb H}^M_{+}$. 
The operators \mbox{${\mathbb E}\{\cdot\}$}, \mbox{$\mathfrak{Re}\{\cdot\}$}, \mbox{$|\cdot|$}, and $\otimes$
denote the expectation, real part, modulus of a complex scalar, and Kronecker product, respectively.
The functions $\text{Tr}({\bf X})$, $\rm{vec}({\bf X})$, and $\text{rank}({\bf X})$, are the trace, column-wise vectorization, and rank of matrix ${\bf X}$, respectively.

\section{System Model} \label{Sec:system_model}
\begin{figure}[h!]
    \begin{center}
    \includegraphics[width = 2.5in]{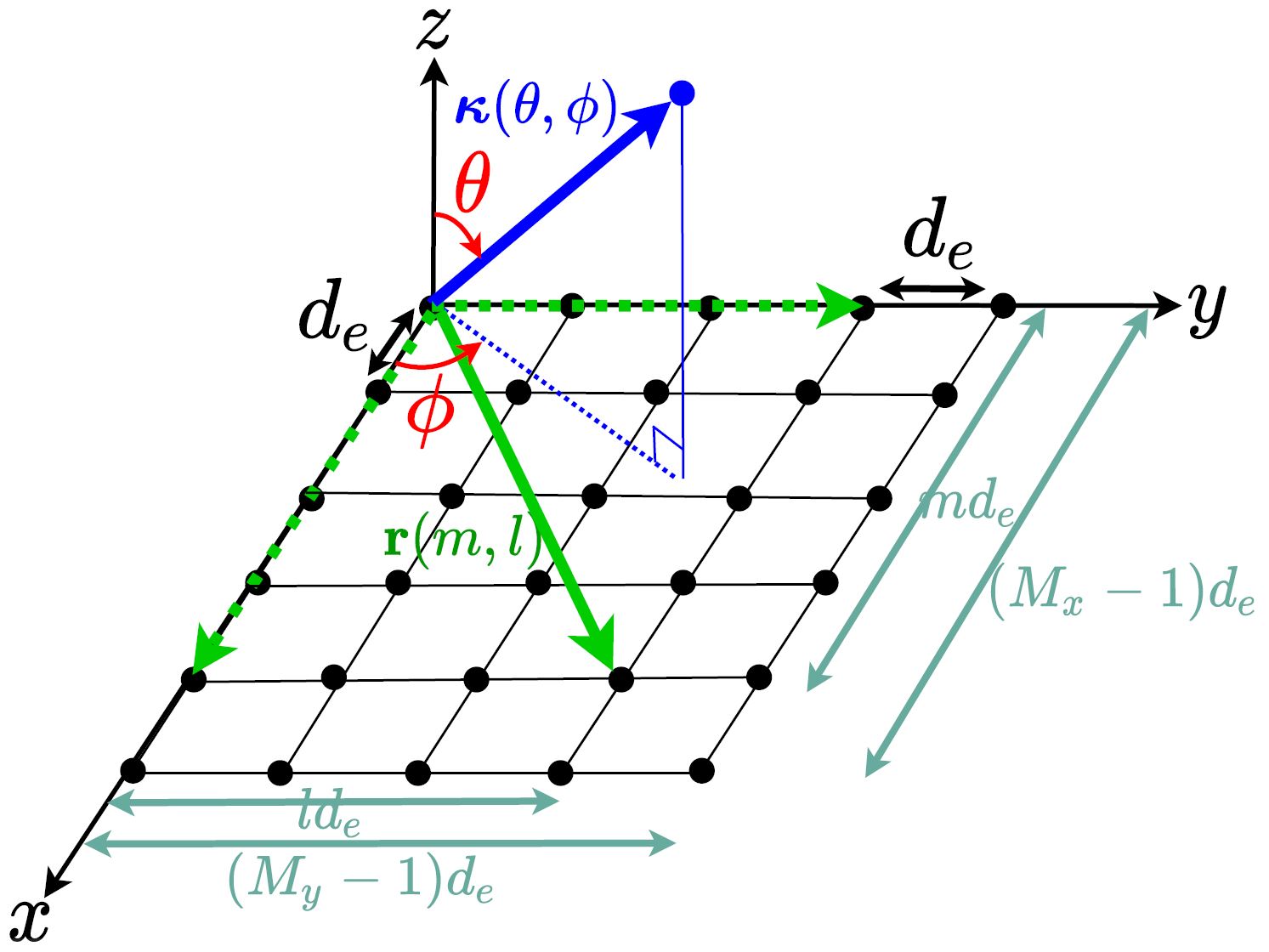}\\
    \caption{Uniform rectangular array.} \label{Fig:URA_array_manifold}
    \end{center}
\end{figure}

\begin{figure*}[h!]
    \begin{center}
    \includegraphics[width = 6.5in]
    {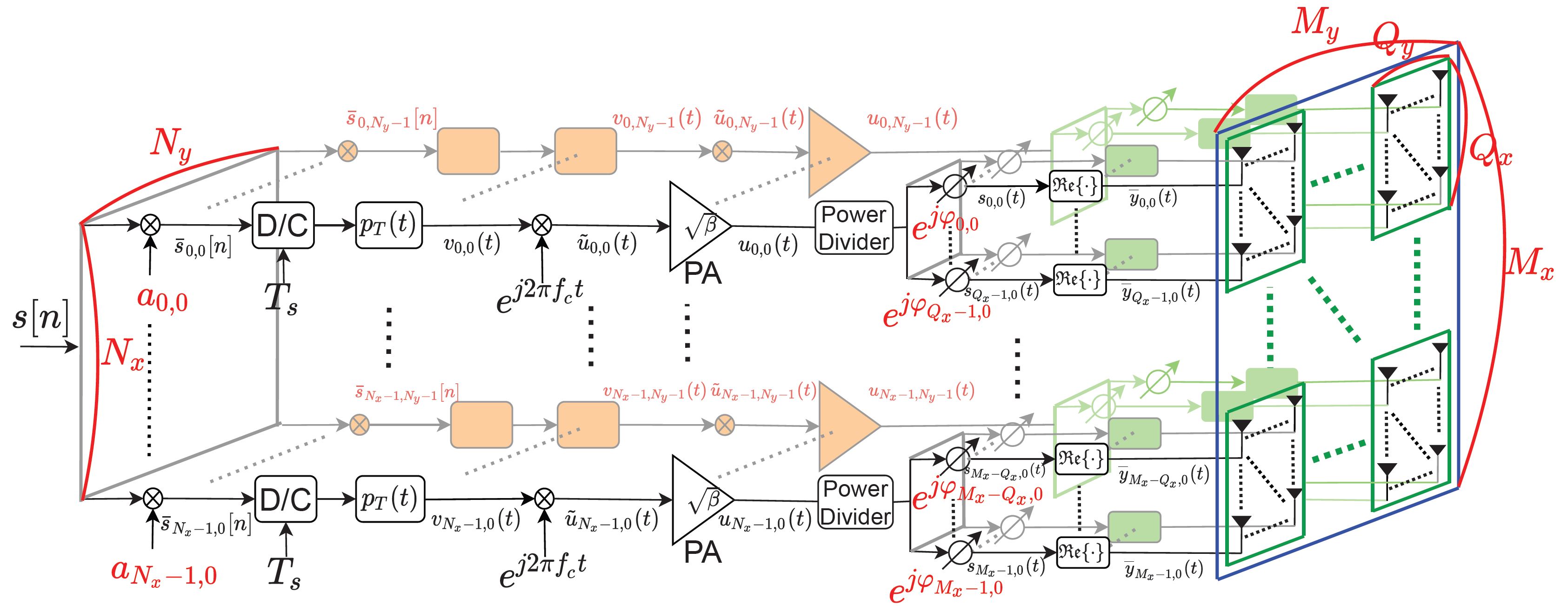}\\
    \caption{SAT transmitter system model.} \label{fig:HB_Tx_circuit_diagram}
    \end{center}
\end{figure*}

Consider a URA consisting of \mbox{$M_x \times M_y$} array elements, as illustrated in Fig. \ref{Fig:URA_array_manifold}. 
The antenna elements are placed at the grid of $x$-$y$ plane with antenna element spacing $d_e$.
The URA element position vector is defined as 
\mbox{${\bf r}(m, l) = [md_e,ld_e,0]^T$}, where \mbox{$m \in \mathbb{Z}_{M_x}$} and \mbox{$l \in \mathbb{Z}_{M_y}$}.
The URA look direction is defined as
\mbox{${\boldsymbol \kappa}(\theta,\phi) = \left[
 \sin (\theta) \cos (\phi),
 \sin (\theta) \sin (\phi),
 \cos (\theta)
\right]^T$},
where \mbox{$\theta \in [0,\frac{\pi}{2}]$} is the SAT elevation angle and \mbox{$\phi \in [0,2\pi]$} is the SAT azimuth angle.
The propagation delay of the antenna element at position \mbox{${\bf r}(m,l)$} is 

\vspace{-2mm}
\begin{footnotesize}
\begin{align} \label{Eq:tau_URA}
    \tau_{m, l}(\theta,\phi)& 
    = \frac{{\bf r}(m, l)^T{\boldsymbol\kappa}(\theta,\phi)}{c}
    =\frac{m d_e \sin(\theta) \cos(\phi) + l d_e\sin(\theta) \sin(\phi)}{c},
\end{align}
\end{footnotesize}
where $c$ is the speed of light.

\subsection {SAT Transmitter and User Terminal Receiver} \label{sec:SAT_Tx}
The SAT transmitter system model is depicted in Fig. \ref{fig:HB_Tx_circuit_diagram}. 
Hybrid beamforming is applied to reduce the number of radio frequency (RF) chains, including upconverters and PAs, to save costs.
Assume that the hybrid beamformer comprises \mbox{$M_x M_y$} antenna elements and \mbox{$N_xN_y$} RF chains.
There are \mbox{$Q_x Q_y$} antenna elements sharing the same RF chain, where
\begin{equation} \label{def:Q_xQ_y}
    Q_x=\frac{M_x}{N_x}, \ Q_y=\frac{M_y}{N_y}.
\end{equation}
Note that $Q_x$ and $Q_y$ are factors of $M_x$ and $M_y$, respectively.
For beamforming purposes, 
amplitude \mbox{$a_{i,j}, \ \forall i \in {\mathbb Z}_{N_x}$}, \mbox{$\forall j \in {\mathbb Z}_{N_y}$}, 
and phase shifters with continuousphase adjustments 
\begin{equation} \label{eq:varphi}
    \varphi_{m, l}\in [-\pi,\pi), \ \forall m \in {\mathbb Z}_{M_x},
    \forall l \in {\mathbb Z}_{M_y}, 
\end{equation}
are applied to the transmitted signal. 
Let $s[n]$ be a constant modulus signal to be transmitted, which is multiplied by the amplitude $a_{i,j}$,
converted to the continuous-time domain through a discrete-time to continuous-time (D/C) converter with sampling period $T_s$, 
shaped using transmitting pulse $p_T(t)$,
up-converted to the carrier frequency $f_c$ and
enlarged by the PA with gain factor $\beta$.
We then have 
\begin{equation} \nonumber
    u_{i, j}(t) = a_{i,j} \sqrt{\beta} e^{j2\pi f_c t} s(t), \ \forall i \in {\mathbb Z}_{N_x}, \forall j \in {\mathbb Z}_{N_y},
\end{equation}
where \mbox{$s(t) =\sum_{n=-\infty}^{\infty}s[n] p_T(t - nT_s)$}.
Pass $u_{i, j}(t)$ through $1$ to $Q_xQ_y$ power dividers and phase shifters, we obtain 
\begin{equation} \nonumber
\begin{aligned}
    s_{m,l}(t) 
    &= \sqrt{\frac{1}{Q_xQ_y}}e^{j\varphi_{m,l}}u_{i, j}(t) \\ 
    &= \sqrt{\frac{1}{Q_xQ_y}} a_{i,j} e^{j\varphi_{m,l}} \sqrt{\beta} e^{j2\pi f_c t} s(t),
\end{aligned}
\end{equation}
where
\mbox{$m \in {\mathbb Z}_{M_x}$}, \mbox{$l \in {\mathbb Z}_{M_y}$}, \mbox{$i = \left\lfloor \frac{m}{Q_x} \right\rfloor$} and \mbox{$j = \left\lfloor \frac{l}{Q_y} \right\rfloor$}.
Note that ``$\sqrt{\frac{1}{Q_xQ_y}}$" arises from the power divider for upholding the conservation of energy such that 
\begin{equation}
    |u_{i, j}(t)|^2 = \sum_{m=i Q_x}^{Q_x(i+1)-1} \ \sum_{l=j Q_y}^{Q_y(j+1)-1} |s_{m,l}(t)|^2.
\end{equation}
Define the URA beamforming coefficient matrix ${\bf W}$ as
\begin{equation}\label{eq:HB_beamforming_coefficients_matrix}
    [{\bf W}]_{m,l} = \sqrt{\frac{1}{Q_xQ_y}} a_{i,j} e^{-j \varphi_{m,l}},
\end{equation}
where \mbox{$m \in {\mathbb Z}_{M_x}$, $l \in {\mathbb Z}_{M_y}$, $i = \left\lfloor \frac{m}{Q_x} \right\rfloor$ and $j = \left\lfloor \frac{l}{Q_y} \right\rfloor$}.
Then, in Fig. \ref{fig:HB_Tx_circuit_diagram}, the transmitted signal of the $(m,l)$-th antenna can be expressed as 
\begin{equation}
\begin{aligned}
    {\overline y}_{m,l}(t) 
    = \mathfrak{Re}\left\{ s_{m,l}(t) \right\} 
    = \mathfrak{Re}\left\{ [{\bf W}]_{m,l}^* \sqrt{\beta} e^{j2\pi f_c t} s(t) \right\}.
\end{aligned}
\end{equation}
Assume $s(t)$ is a narrowband signal.
In addition, the signal transmitted from the $(m,l)$-th antenna experiences a propagation delay $\tau_{m,l}(\theta,\phi)$ defined in (\ref{Eq:tau_URA}).
The signal in the far-field (towards the angle $(\theta, \phi)$) can be written as 
\begin{equation} \label{eq:yt_analog_delay_filter}
\begin{aligned}
    y(t) &= \sum_{m=0}^{M_x-1} \sum_{l=0}^{M_y-1} {\overline y}_{m,l}(t-\tau_{m,l}(\theta,\phi)) \\
         &= \mathfrak{Re}\left\{ \sqrt{\beta}  e^{j2\pi f_c t} \sum_{m=0}^{M_x-1} \sum_{l=0}^{M_y-1} [{\bf W}]_{m, l}^*  e^{-j2\pi f_c \tau_{m,l}(\theta, \phi)} s(t) \right\}.      
\end{aligned}
\end{equation}
From (\ref{eq:yt_analog_delay_filter}), the URA transmit beampattern is defined as \cite{VanTrees2002}
\begin{equation}\label{def:URA_BP}
\begin{aligned}
    \widetilde{\rm B}({\bf W},\theta,\phi) 
    &= \sum_{m=0}^{M_x-1}\sum_{l=0}^{M_y-1}[{\bf W}]_{m,l}^{*}e^{-j2\pi f_c\tau_{m,l}(\theta,\phi)},
\end{aligned}
\end{equation}
where ${\bf W}$ is the URA beamforming coefficient matrix defined in (\ref{eq:HB_beamforming_coefficients_matrix}) and
$\tau_{m,l}(\theta, \phi)$ is the propagation delay defined in (\ref{Eq:tau_URA}). 
Let the distance between the SAT and UT be $d(\theta)$, and the path loss in the SAT downlink be $L(\theta)$, which will be defined later in (\ref{Eq:pathloss_L}).
Assume there is no obstruction between the SAT and the user terminal (UT).
The line-of-sight channel is modeled as
\begin{equation} \label{channel_model}
    h(t)=\frac{1}{\sqrt{L(\theta)}}\delta \left(t-\frac{d(\theta)}{c}\right),
\end{equation}
where $c$ is the speed of light.

\begin{figure}[h!]
    \begin{center}
    \includegraphics[width=3.5in]{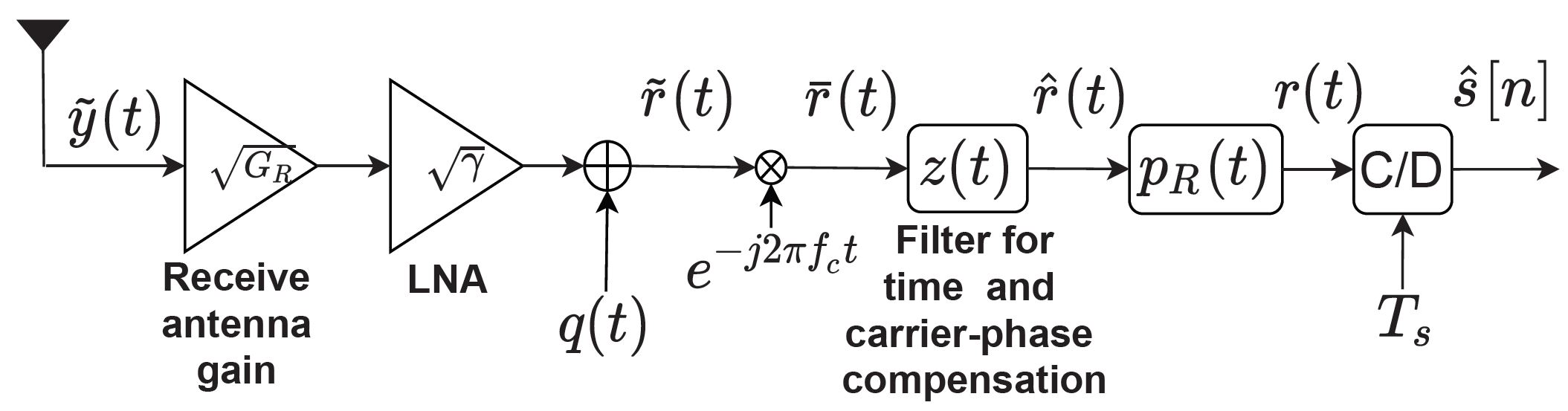}\\
    \caption{User terminal receiver system model.} \label{fig:HB_Rx_circuit_diagram}
    \end{center}
\end{figure}
In Fig. \ref{fig:HB_Rx_circuit_diagram}, UT Rx is shown. 
The received signal is \mbox{$y_r(t) = (y*h)(t)=\frac{1}{\sqrt{L(\theta)}}y\left(t-\frac{d(\theta)}{c}\right)$},
and its pre-envelope is expressed as
\begin{equation}
\begin{aligned}
    {\widetilde y}(t) 
    &= y_r(t) + j {\hat y}_r(t) \\
    &= \sqrt{\frac{\beta}{L(\theta)}} e^{j2\pi f_c (t-\frac{d(\theta)}{c})} \widetilde{\rm B}({\bf W},\theta,\phi)  s\left(t-\frac{d(\theta)}{c}\right),
\end{aligned}
\end{equation}
where ${\hat y}_r(t)$ is the Hilbert transform of $y_r(t)$ \cite{ProakisComm}.
Let $G_R$ be the receive antenna gain and $\gamma$ be the gain factor of the low noise amplifier (LNA).
The noise is modeled as \mbox{$q(t) = \sqrt{\gamma} q_a(t) + q_{\gamma}(t)$} \cite{Taparugssanagorn2014}, 
where $q_a(t)$ is the amount of noise that enters the antenna and $q_{\gamma}(t)$ is the noise introduced by the LNA. 
\mbox{$q_a(t) \sim {\mathcal N}(0, \frac{kT_a}{2})$} is modeled as a zero-mean additive white Gaussian noise (AWGN) with variance $\frac{kT_a}{2}$, where $T_a \ \rm{[K]}$ is the antenna temperature \cite{Dell1963_Book} and \mbox{$k = 1.38 \times 10^{-23}\ [\rm{W \cdot s/K}]$} is the Boltzmann constant.
\mbox{$q_{\gamma}(t) \sim {\mathcal N}(0, \frac{kT_{\gamma}}{2})$}, where $T_{\gamma}\ \rm{[K]}$ is the noise temperature of LNA.
We then have \mbox{${\widetilde r}(t) = \sqrt{\gamma G_R} {\widetilde y}(t) + q(t)$}, where \mbox{$q(t) \sim {\mathcal N}(0, \frac{\gamma kT_a}{2} + \frac{kT_{\gamma}}{2})$}.
Let ${\widetilde r}(t)$ go through the down converter and filter 
\mbox{$z(t) = \delta \left(t + \frac{d(\theta)}{c}\right)e^{j2\pi f_c (d(\theta)/c)}$} 
which aims to compensate for the time misalignment and carrier phase difference due to the signal propagation delay $\frac{d(\theta)}{c}$ introduced in channel $h(t)$ defined in (\ref{channel_model}).
We obtain 
${\hat r}(t) 
= ({\bar r} * z)(t) 
= \sqrt{\frac{\beta \gamma G_R}{L(\theta)}} {\rm {\widetilde B}}({\bf W},\theta, \phi)  s(t) + q'(t)$,
where \mbox{${\bar r}(t) = {\widetilde r}(t)e^{-j2\pi f_c t}$} and \mbox{$q'(t)= (q(t)e^{-j2\pi f_c t}) * z(t)$}. 
If the gain of the first amplifying stage is large, the noise introduced by the subsequent components has a diminishing effect on the SNR according to the Friis formula \cite{Dell1963_Book}.
Because $\gamma$ is typically a large value, the noise introduced by the other components is considered negligible for convenience. 
Applying the receiving filter $p_R(t)$, we obtain \mbox{$r(t)=({\hat r}*p_R)(t)$}.
Suppose the Nyquist pulse-shaping criterion for zero intersymbol interference (ISI) is met (i.e., \mbox{$(p_T*p_R)(nT_s) = \delta[n]$} \cite{ProakisComm}).
The received discrete-time signal is derived as follows:
\begin{equation} \label{eq:s_hat_n}
\begin{aligned} 
    {\hat s}[n] = \sqrt{\frac{\beta \gamma G_R}{L(\theta)}} {\rm {\widetilde B}}({\bf W},\theta, \phi) s[n] + q[n].
\end{aligned}
\end{equation}
The baseband equivalent noise \mbox{$q[n] \sim \mathcal{CN}(0, \gamma k T_{\rm sys} f_{\rm BW})$}
is a zero-mean circularly symmetric Gaussian noise with variance \mbox{$\gamma k T_{\rm sys} f_{\rm BW}$} \cite{Richards2014},   
where
\mbox{$f_{\rm BW} \ [{\rm Hz}]$} is the channel bandwidth, and
the system noise temperature is defined as \mbox{$T_{\rm sys} \ {\rm [K]} = T_a \ {\rm [K]} + T_e \ {\rm [K]}$} \cite{Dell1963_Book},
where $T_a$ is the antenna temperature and $T_e$ is the effective noise temperature. 
Since we only consider the noise introduced by the LNA, \mbox{$T_e = \frac{T_{\gamma}}{\gamma}$} in our case. 
Given the noise factor \mbox{$N_f \ {\rm [dB]}$}, one can calculate \mbox{$T_e = (10^{0.1 N_f \ {\rm [dB]}} - 1)T_0$}, where \mbox{$T_0 = 290 \ \rm{[K]}$} is the standard temperature \cite{Dell1963_Book}.

\subsection{Baseband Equivalent Transceiver Model}
\begin{figure*}
    \centering
    \subfloat[]{\includegraphics[width = 6.5in]
    {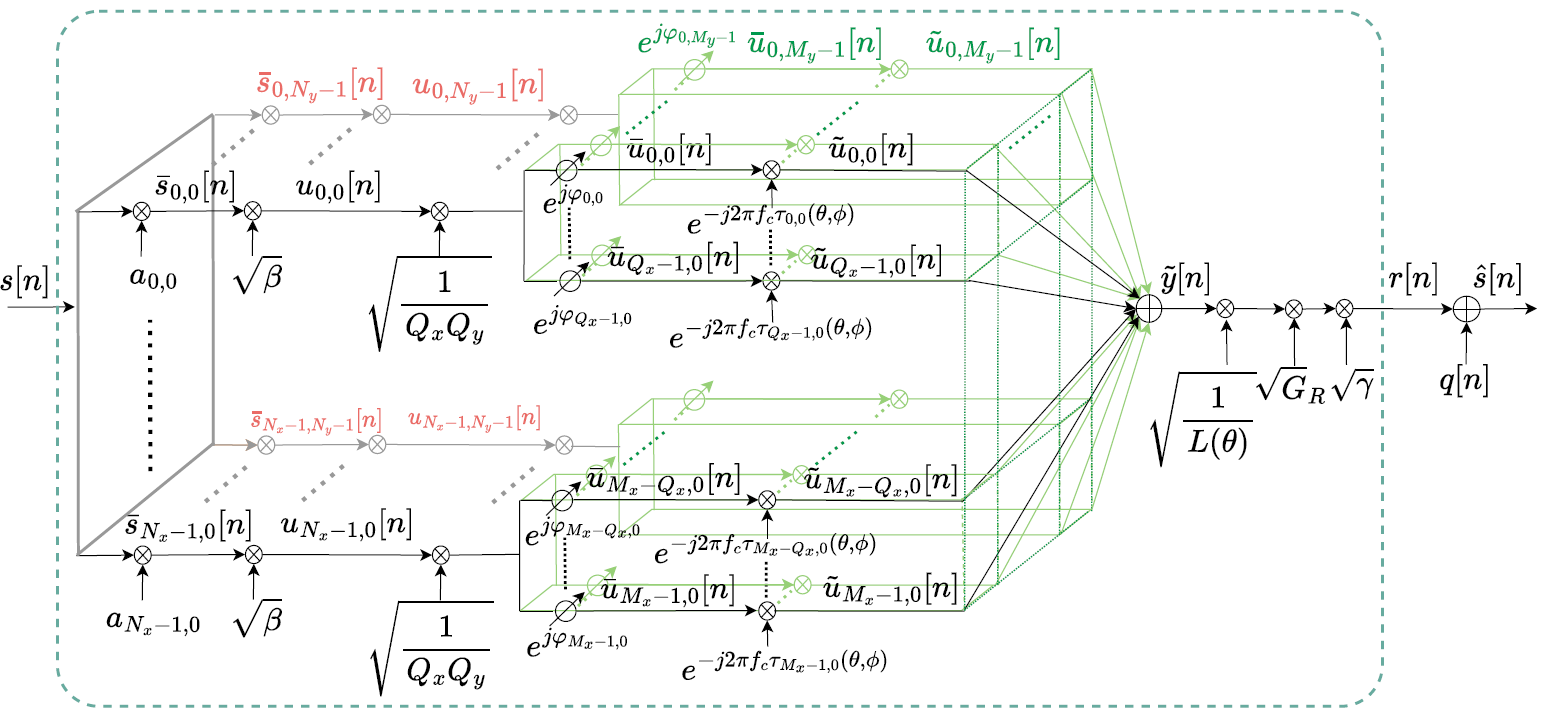} 
    \label{fig:Mathematical model (baseband 1)}} 
    \par 
    \subfloat[]{\includegraphics[width = 4.5in]{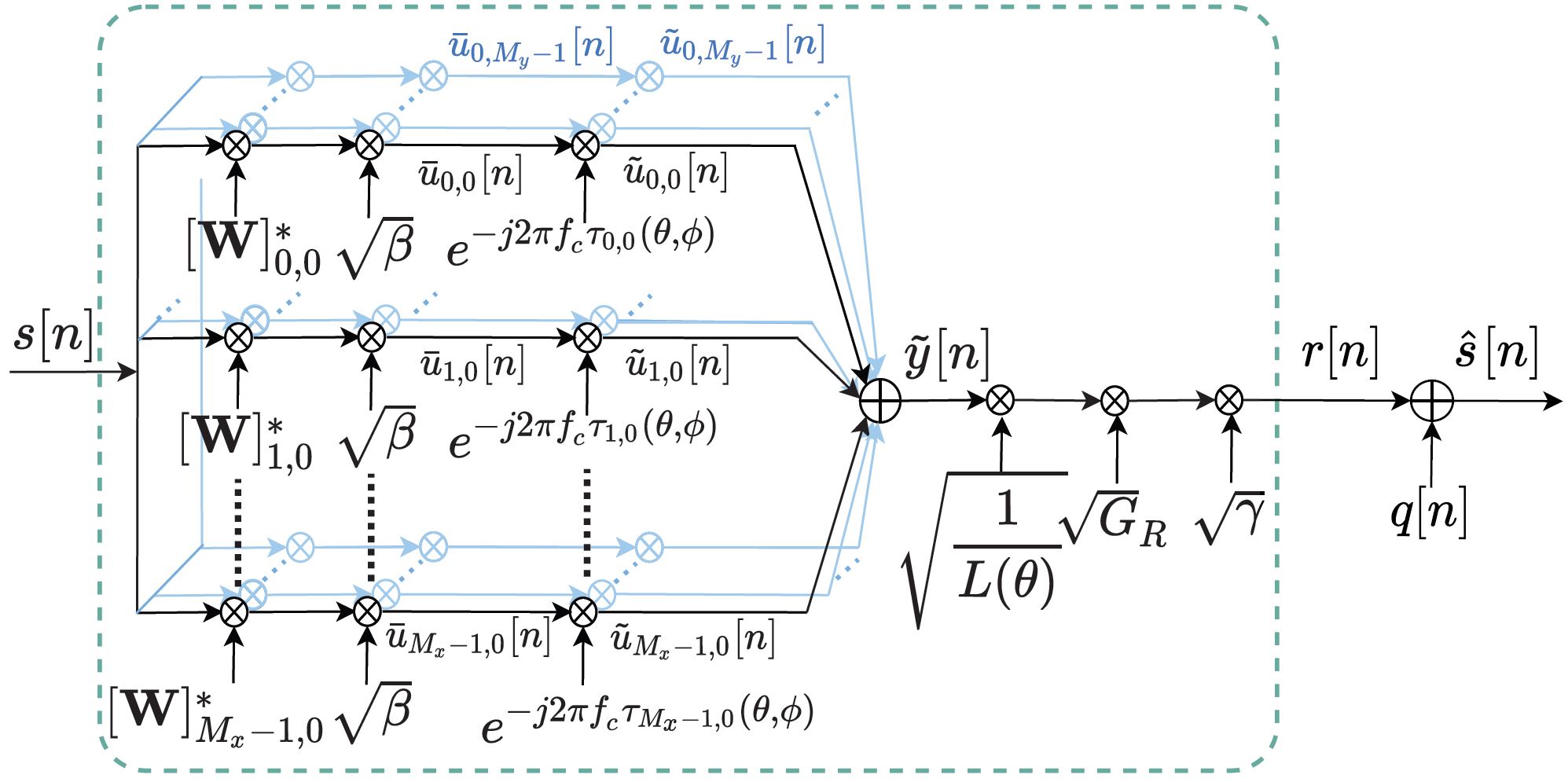}    \label{fig:Mathematical model (baseband 2)}} 
    \caption{Baseband transceiver model with (a) hybrid beamforming structure and (b) fully digital beamforming structure.}
\end{figure*}

According to (\ref{eq:s_hat_n}), the baseband transceiver models with hybrid and fully digital beamforming structures are depicted in Fig. \ref{fig:Mathematical model (baseband 1)} and Fig. \ref{fig:Mathematical model (baseband 2)}, respectively.
From Fig. \ref{fig:Mathematical model (baseband 1)}, it is more explicit for us to calculate the PA average output power, which will be shown in (\ref{def:P_PA}); while, Fig. \ref{fig:Mathematical model (baseband 2)} is a simpler representation.
In fact, Fig. \ref{fig:Mathematical model (baseband 1)} and Fig. \ref{fig:Mathematical model (baseband 2)} are mathematically equivalent. 
In both figures, \mbox{${\widetilde y}[n] = \sqrt{\beta} {\rm {\widetilde B}}({\bf W},\theta, \phi) s[n]$}, where ${\rm {\widetilde B}}({\bf W},\theta, \phi)$ is defined in (\ref{def:URA_BP}).
Additionally, 
${\hat s}[n] = \sqrt{\frac{\gamma G_R}{L(\theta)}}{\widetilde y}[n] + q[n] = \sqrt{\frac{\beta \gamma G_R}{L(\theta)}} {\rm {\widetilde B}}({\bf W},\theta, \phi) s[n] + q[n]$ as derived in (\ref{eq:s_hat_n}).
From Fig. \ref{fig:Mathematical model (baseband 1)}, PA average output power can be computed as, \mbox{$\forall i \in \mathbb{Z}_{N_x}, \forall j \in \mathbb{Z}_{N_y}$},
\begin{equation} \label{def:P_PA}
\begin{aligned} 
    P_{\rm{PA, avg}}^{(i,j)} 
    &=  \frac{ {\mathbb E}\{ |u_{i,j}[n]|^2 \}}{\rm{R}} 
    = \beta P_s a_{i,j}^2 
    = \beta P_s Q_x Q_y |[{\bf W}]_{m,l}|^2,
\end{aligned}
\end{equation}
where \mbox{$u_{i,j}[n] = \sqrt{\beta} a_{i,j} s[n]$} is the PA output signal,
\mbox{${\rm R} = 50 \ [\Omega]$} is the antenna impedance, the source signal power is
\begin{equation} \label{eq:Ps}
    P_s = \frac{{\mathbb E}\{|s[n]|^2\}}{{\rm R}} \ [{\rm W}],
\end{equation}
and the relationship between $a_{i,j}^2$ and $|{\bf W}_{m,l}|^2$ can be seen from (\ref{eq:HB_beamforming_coefficients_matrix}). 
Moreover, the transmit signal power is the summation of the PA average output power in the RF chains:
\begin{equation}  \label{def:P_T}
    P_T = \sum_{i=0}^{N_x-1}\sum_{j=0}^{N_y-1} P_{\rm{PA, avg}}^{(i,j)}. 
\end{equation}

\section{Problem Formulation for Broadened-beam URA Coefficient Design in LEO SatComs}  \label{sec:problem_formulation}
\subsection{LEO SatComs Scenario}
\begin{figure}
    \centering
    \subfloat[]{\includegraphics[width=1.7in]{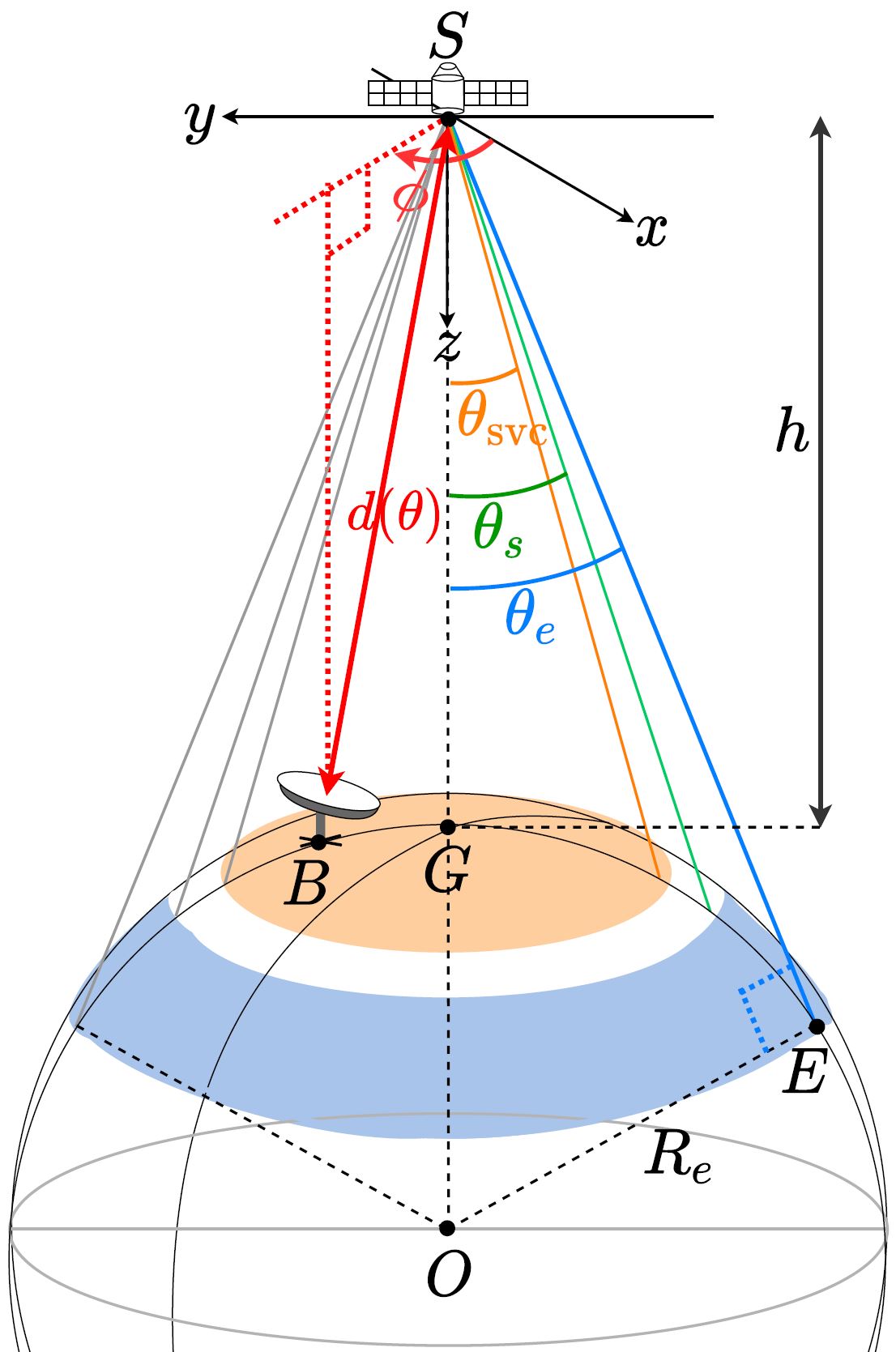} 
    \label{Fig:SATCOM_geometry} }  
    \hspace{-12mm}
    \hfill 
    \subfloat[]{\includegraphics[width=1.6in]{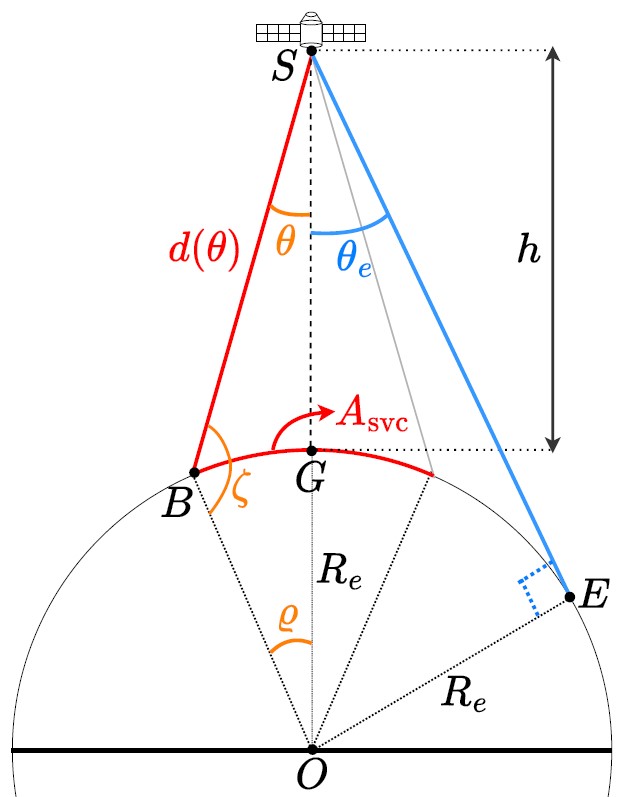}
    \label{Fig:SAT_FoV_svc_area}} 
    \caption{(a) LEO SatComs scenario and (b) side view of (a).}
\end{figure}

A scenario of LEO SatComs is illustrated in Fig. \ref{Fig:SATCOM_geometry}, where the coordinate system $(x, y, z)$ is defined from the SAT perspective, and the $z$ axis represents the direction from the SAT towards the center of the Earth. 
Assume that the SAT is at altitude $h$ with SAT elevation angle, \mbox{$\theta \in [0, \frac{\pi}{2}]$}, and SAT azimuth angle, \mbox{$\phi \in [0, 2\pi]$}.
Suppose that the SAT transmitter depicted in 
Fig. \ref{fig:HB_Tx_circuit_diagram} is used for communication purposes.
Let the SAT service angle be $\theta_{\rm{svc}}$, 
and let the SAT field of view (FoV) angle $\theta_e$ be the angle between the $z$-axis and the line connecting the SAT to the Earth's tangent point ${\rm E}$. 
From $\triangle{\rm{OSE}}$ in Fig. \ref{Fig:SAT_FoV_svc_area}, the SAT FoV angle can be calculated as 
\begin{align} \label{Eq:theta_e}
    \theta_e = \sin^{-1} \left(\frac{R_e}{h + R_e} \right),
\end{align}
where $R_e$ is the Earth's radius.
We regard \mbox{$\theta \in [\theta_e, \frac{\pi}{2}]$} as the ``don't care" angle region since the transmitted signals within the region would not reach the Earth.
The transmit beampattern main lobe angle set is defined as 
\begin{align} \label{Theta_m_bar}
    \Pi_m = \{(\theta,\phi)|\theta \in [0, \theta_{\rm{svc}}],  \phi \in [0,2\pi]\},
\end{align}
and SAT service areas correspond to the orange-colored areas in Fig \ref{Fig:SATCOM_geometry}.
The transmit beampattern sidelobe angle set is defined as
\begin{align} \label{Theta_s_bar}
    \Pi_s = \{(\theta,\phi)|\theta \in [\theta_s,\theta_e],  \phi \in [0,2\pi]\},
\end{align}
and SAT out-of-beam areas correspond to the blue-colored areas in Fig \ref{Fig:SATCOM_geometry}.
Let the distance between SAT and the ground UT be $d(\theta)$.
From $\triangle{\rm{OSB}}$ in Fig. \ref{Fig:SAT_FoV_svc_area}, according to the law of cosines, we have
\begin{equation}
    R^2_E = d(\theta)^2+(h+R_e)^2-2d(\theta)(h+R_e)\cos(\theta).
\end{equation}
Then $d(\theta), \ \forall \theta \in [0,\theta_e]$, is obtained as 
\begin{align} \label{Eq:d}
    d(\theta) = (h+R_e)\cos(\theta)-\sqrt{R^2_E-(h+R_e)^2\sin^2(\theta)}. 
\end{align}
In addition, SAT service areas, $A_{\rm{svc}}$, are areas of the spherical cap in Fig. \ref{Fig:SAT_FoV_svc_area} which can be calculated by \cite{book_spherical_cap}
\begin{align} \label{A_svc}
A_{\rm{svc}} = 2 \pi R_e^2(1-\cos(\varrho)),
\end{align}
where 
\vspace{-3mm}
\begin{align} \label{eq:beta_Spherical_cap}
    \varrho = \pi - \theta - \zeta, \  \theta \in [0,\theta_e],
\end{align}
and \mbox{$\zeta= \pi - \sin^{-1}\left(\frac{R_e+h}{R_e}\sin(\theta)\right) \in [\frac{\pi}{2},\pi]$} is obtained by the law of sines
\mbox{$\frac{R_e}{\sin(\theta)} = \frac{R_e+h}{\sin(\zeta)}$}. 

\subsection{Received SNR Derivation}
In the SAT downlink,
the signal strength is degraded due to transmission loss $L(\theta)$, including
transmitter cable loss $L_{\rm{c,T}}$, 
receiver cable loss $L_{\rm{c,R}}$,
atmospheric path loss $L_{\rm{a}}$,
scintillation loss $L_{\rm{sl}}$,
and free-space propagation loss $L_{\rm fs}(\theta)$.
The transmission loss is considered to be \cite{3GPPNTN2019_TR16}
\begin{equation} \label{Eq:pathloss_L}
    L(\theta) = L_{\rm{c,T}}L_{\rm{c,R}}L_{\rm fs}(\theta)L_{\rm{a}}L_{\rm{sm}}L_{\rm{sl}} = \sigma^2(\theta)L_0,
\end{equation}
where \mbox{$L_{\rm fs}(\theta)=\left(\frac{4 \pi d(\theta)}{\lambda}\right)^2$} with $d(\theta)$ derived in (\ref{Eq:d}),
$\lambda$ is the wavelength of the transmitted signal, 
\begin{align}\label{def:L0} 
L_0 = L_{\rm{c,T}}L_{\rm{c,R}}L_{\rm{a}}L_{\rm{sm}}L_{\rm{sl}}\left(\frac{4 \pi h}{\lambda}\right)^2,
\end{align}

\vspace{-3mm}
\begin{footnotesize}
\begin{align}\label{URA_isoflux_mask} 
    \sigma(\theta) = \frac{d(\theta)}{h} = \frac{(h+R_e)\cos(\theta)-\sqrt{R^2_E-(h+R_e)^2\sin^2(\theta)}}{h}.
\end{align}
\end{footnotesize}

\noindent
Note that \mbox{$\sigma: [0,\theta_e] \rightarrow \mathbb{R}$} with $\theta_e$ defined in (\ref{Eq:theta_e}). 
In Fig. \ref{Fig:URA_isolux_radiation_mask}, $\sigma(\theta)$ is depicted. 

According to the received discrete-time signal ${\hat s}[n]$ derived in (\ref{eq:s_hat_n}),
the received signal power in watts, $\rm{[W]}$, can be calculated as

\vspace{-2mm}
\begin{small}
\begin{equation}
\begin{aligned} \label{eq:P_r}
    P_r({\bf W},\theta,\phi) 
    &= \frac{{\mathbb E}\big\{\big|\sqrt{\frac{\beta \gamma G_R}{L(\theta)}} {\rm {\widetilde B}}({\bf W},\theta, \phi) s[n]\big|^2\big\}}{{\rm R}} \\
    &= \left(\frac{\beta \gamma G_R}{\sigma^2(\theta)L_0}\right) P_s |{\rm {\widetilde B}}({\bf W},\theta, \phi)|^2, 
\end{aligned}
\end{equation}
\end{small}

\vspace{-1mm}
\noindent
where 
\mbox{${\rm R} = 50 \ [\Omega]$} is the antenna impedance, and 
$P_s$ is defined as (\ref{eq:Ps}).
Also, since the baseband noise derived in (\ref{eq:s_hat_n}) is \mbox{$q[n] \sim \mathcal{CN}(0, \gamma k T_{\rm{sys}} f_{\rm{BW}})$},
the noise power in watts, ${\rm{[W]}}$, is evaluated as \cite{Richards2014, Dell1963_Book} 
\begin{equation} \label{eq:P_N}
    P_N =  \gamma k T_{\rm{sys}} f_{\rm{BW}},
\end{equation}
where
\mbox{$k = 1.39 \times 10^{-23}\ [{\rm{W \cdot s/K}}]$} is the Boltzmann constant, 
\mbox{$f_{\rm{BW}} \ [{\rm{Hz}}]$} is the channel bandwidth, 
and \mbox{$T_{\rm{sys}}\ {\rm{[K]}}$} is the system noise temperature.
The received SNR is derived as
\begin{equation}  \label{eq:SNR_r}
\begin{aligned}
    \text{SNR}({\bf W},\theta,\phi) 
    = \frac{P_r({\bf W},\theta,\phi)}{P_N}
    = \frac{\beta G_R P_s|\widetilde{\rm B}({\bf W},\theta,\phi)|^2}{\sigma^2(\theta)L_0 k T_{\rm{sys}} f_{\rm{BW}}}.
\end{aligned}
\end{equation}
In (\ref{eq:SNR_r}), the ratio of $G_R$ to $T_{\rm{sys}}$ is typically referred to as the antenna gain-to-noise-temperature
${\rm{G/T}} \ [{\rm{dB/K}}] = 10 \log_{10}\left(\frac{G_R}{T_{\rm{sys}}}\right)$ \cite{3GPPNTN2019_TR16}.

\subsection{Problem Formulation} \label{subsec:problem_formulation}
\begin{figure}
    \centering
    \hspace{-5mm}
    \subfloat[]{\includegraphics[width = 1.8in]{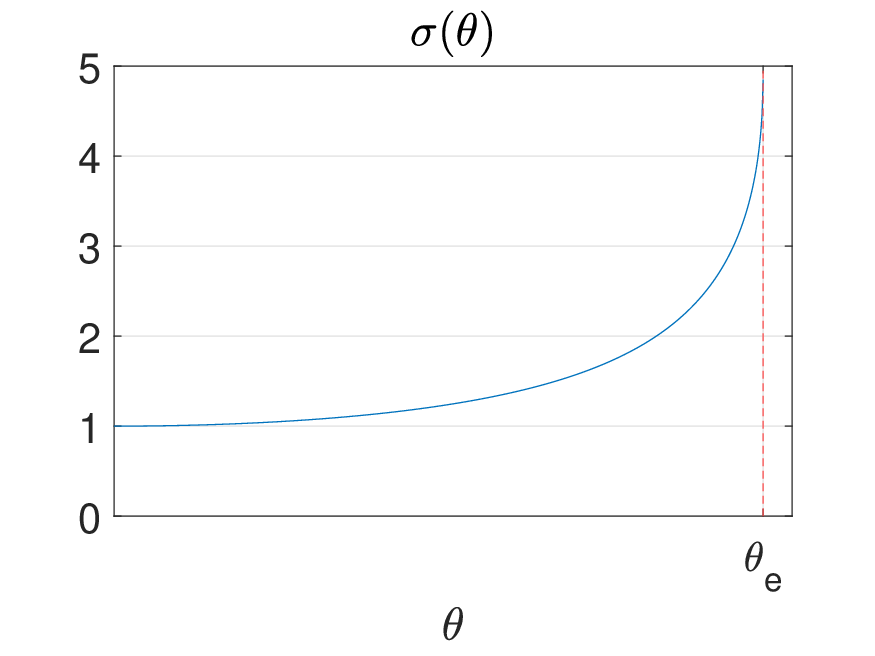} 
    \label{Fig:URA_isolux_radiation_mask}}  
    \hspace{-9mm} 
    \hfill
    \subfloat[]{\includegraphics[width = 1.8in]{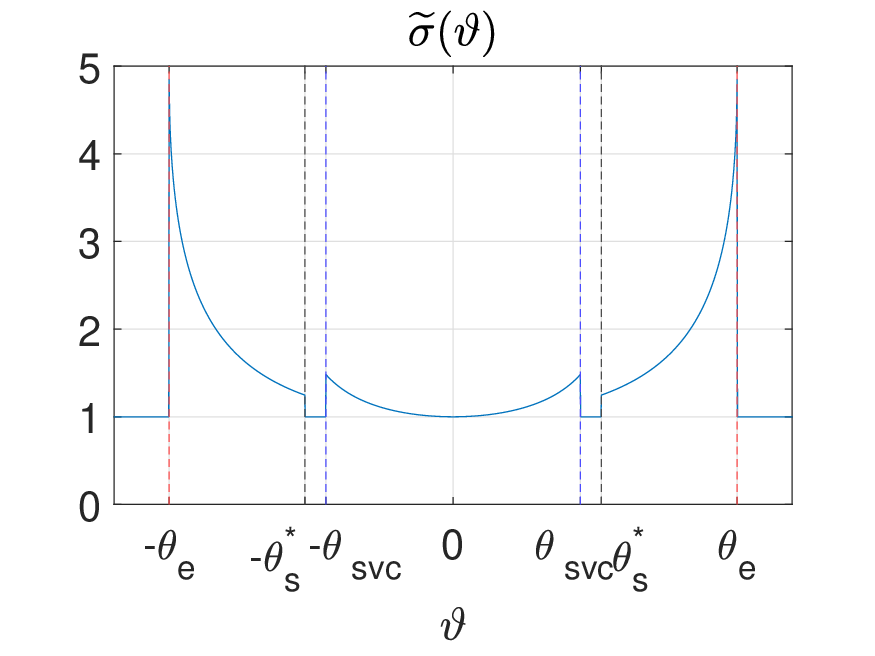}
    \label{Fig:ULA_isolux_radiation_mask}} 
    \caption{(a) $\sigma(\theta)$ defined in (\ref{URA_isoflux_mask}) and (b) ${\widetilde \sigma}(\vartheta)$ defined in (\ref{ULA_isoflux_radiation_mask}).}
\end{figure}

An optimization problem for the broadened-beam URA coefficient design in LEO SatComs under QoS constraints and constant modulus constraints (CMCs) is formulated in (\ref{P1_00}).
The objective function (\ref{P1_00a}) aims to suppress the maximum received signal power in the SAT out-of-beam areas to mitigate interference. 
The QoS in SAT service areas is guaranteed by (\ref{P1_00b}), and CMCs are considered in (\ref{P1_00c}).
\begin{subequations}{\label{P1_00}}
\begin{align} 
    \mathop{\rm{minimize}}\limits_{{\bf W}\in\mathbb{C}^{M_x\times M_y}} \  &\mathop{\sup}\limits_{\forall (\theta,\phi) \in \Pi_s}  P_r({\bf W}, \theta,\phi) {\label{P1_00a}}  \\
    {\rm{subject \ to}} \ & \text{SNR}({\bf W}, \theta,\phi) \geq \text{SNR}_{\min}, \forall (\theta,\phi) \in \Pi_m {\label{P1_00b}} \\
    & \left|[{\bf W}]_{m,l} \right| = 1, \ m\in{\mathbb Z}_{M_x},l \in{\mathbb Z}_{M_y},  {\label{P1_00c}}
\end{align}
\end{subequations}
where $P_r({\bf W}, \theta, \phi)$ is derived in (\ref{eq:P_r}), 
$\text{SNR}({\bf W}, \theta,\phi)$ is derived in (\ref{eq:SNR_r}),
$\text{SNR}_{\min}$ is the specified lower bound for the received SNR in SAT service areas,
$\Pi_m$ is the transmit beampattern main lobe angle set defined in (\ref{Theta_m_bar}) with specified SAT service angle $\theta_{\rm{svc}}$, and
$\Pi_s$ is the transmit beampattern sidelobe angle set defined in (\ref{Theta_s_bar}).
We can reformulate the problem (\ref{P1_00}) as its epigraph representation
\begin{subequations}{\label{P1_0}}
\begin{align}   
    \hspace{-6mm}
    \mathop{\rm{minimize}}\limits_{{\bf W}\in\mathbb{C}^{M_x\times M_y}, t\in{\mathbb R}} \  & t^2  {\label{P1_0a}}  \\
    {\rm{subject \ to}} \ 
    & P_r({\bf W}, \theta,\phi) \leq t^2,
    \forall (\theta,\phi) \in \Pi_s {\label{P1_0b}} \\
    & \text{SNR}({\bf W}, \theta,\phi) \geq \text{SNR}_{\min},
    \forall (\theta,\phi) \in \Pi_m {\label{P1_0c}} \\
    & \left|[{\bf W}]_{m,l} \right| = 1, \ m\in{\mathbb Z}_{M_x},l \in{\mathbb Z}_{M_y}.  {\label{P1_0d}}
\end{align}
\end{subequations}
Also, the equivalent problem of the problem (\ref{P1_0}) is shown as
\begin{subequations} \label{P1_1}
\begin{align} 
    \mathop{\rm{minimize}}\limits_{{\bf W}\in\mathbb{C}^{M_x\times M_y}, t\in{\mathbb R}} &t{\label{P1_1a}} \\
    {\rm{subject \ to }} \quad &|\widetilde{\rm B}({\bf W}, \theta,\phi)|\leq t\sigma(\theta), \forall (\theta,\phi) \in \Pi_s {\label{P1_1b}} \\
    &|\widetilde{\rm B}({\bf W}, \theta,\phi)| \geq \alpha\sigma(\theta), \forall (\theta,\phi) \in \Pi_m {\label{P1_1c}} \\
    & \left|[{\bf W}]_{m,l}\right| = 1, \ m\in{\mathbb Z}_{M_x},l \in{\mathbb Z}_{M_y},
\end{align}
\end{subequations}
where $\sigma(\theta)$ is defined in (\ref{URA_isoflux_mask}),
\begin{align} \label{eq:alpha}
    \alpha = \sqrt{\frac{\text{SNR}_{\min}k T_{\rm{sys}} f_{\rm{BW}}L_0}{\beta G_R P_s}}.  
\end{align}
We can observe that the QoS guarantee constraint (\ref{P1_00b}) results in the beampattern main lobe lower bound constraint (\ref{P1_1c}), and the main lobe lower bound is anticipated to follow the shape of $\sigma(\theta)$,  defined in (\ref{URA_isoflux_mask}), as shown in Fig. \ref{Fig:URA_isolux_radiation_mask}. 
The shape of $\sigma(\theta)$ is related to the distance variation between SAT and UT due to the Earth's curvature which has also been considered in \cite{Vigano2010, Reyna2012, Ibarra2015, Yoshimoto2019, Zeng2021_isoflux, Cai2023} for synthesizing isoflux radiation patterns. 
Moreover, CMCs (\ref{P1_00c}) enable PAs to operate at near compression points to maximize efficiency which has not been considered in  \cite{Vigano2010, Reyna2012, Ibarra2015, Yoshimoto2019, Zeng2021_isoflux, Cai2023}. 
The problem (\ref{P1_1}) is a non-convex problem that is nonsmooth and difficult to solve.
Furthermore, URA with a large number of antenna elements, e.g., \mbox{$32 \times 32$}, is preferable for SAT to enhance the downlink capacity. 
However, intensive computations are required to solve the problem (\ref{P1_1}) when variable ${\bf W}$ has a large size. 
Thus, the nonconvex constraints and large variable sizes make solving the URA design problem (\ref{P1_1}) challenging.

\section{Proposed method} \label{Sec:proposed_algorithm}
A URA with a large number of antenna elements, e.g., \mbox{$32 \times 32$}, is typically preferable for SAT to enhance downlink capacity. 
However, addressing the URA design problem (\ref{P1_1}) is computationally demanding when the variable ${\bf W}$ has a large size.
To enhance computational efficiency,
we propose to decompose the URA design problem (\ref{P1_1}) into ULA design subproblems with the idea of Kronecker product beamforming 
\cite{Wang2021, Frank2022_planner_array_Kronecker_Product_Beamforming, Albagory2022_planner_array, VanTrees2002}. 
The idea of URA design problem decomposition is introduced in Section \ref{sec:idea_kronecker_product} and the resulting ULA design subproblems are presented in Section \ref{sec:URA_to_ULAs}.
The nonconvex ULA design subproblems are reformulated as semidefinite programming (SDP) with a rank-one constraint and
addressed using the semidefinite relaxation (SDR) method accompanied with
a convex iterative algorithm in Section \ref{sec:iterative_algorithm}.
Lastly, the proposed broadened-beam URA coefficient
design method is summarized in Algorithm \ref{algorithm1}.
\subsection{Idea of URA Design Problem Decomposition} \label{sec:idea_kronecker_product}
If we have separable weightings 
\begin{align} \label{URA_weight_matrix}
    {\bf W} = {\bf x}{\bf y}^T \in \mathbb{C}^{M_x \times M_y},
\end{align}
where ${\bf x} \in \mathbb{C}^{M_x}$ and ${\bf y} \in \mathbb{C}^{M_y}$,
or equivalently by the Kronecker product property \mbox{$\rm{vec}({\bf W}) = {\bf x} \otimes {\bf y}$},
the URA beampattern defined in (\ref{def:URA_BP}) can be derived as the product of two ULA beampatterns \cite{VanTrees2002, Frank2022_planner_array_Kronecker_Product_Beamforming}. 
\begin{align} \label{eq:B_tilde}
    \widetilde{\rm B}({\bf W},\theta,\phi)
    &= \sum_{m=0}^{M_x-1}\sum_{l=0}^{M_y-1}[{\bf W}]_{m,l}^{*}e^{-j2\pi f_c\tau_{m,l}(\theta,\phi)} \nonumber \\
    &=\sum_{m=0}^{M_x-1} x_m^{*}e^{-j 2\pi f_c {\overline \tau}_m(\vartheta_x)} \sum_{l=0}^{M_y-1} y_l^{*}e^{-j 2\pi f_c {\overline\tau}_l(\vartheta_y)} \nonumber\\
    &={\rm B}({\bf x},\vartheta_x){\rm B}({\bf y},\vartheta_y),
\end{align}
where 
\begin{equation} \label{eq:vartheta_x}
    \vartheta_x = \sin^{-1}(\sin(\theta)\cos(\phi)), \ \vartheta_y = \sin^{-1}(\sin(\theta)\sin(\phi)),  
\end{equation}
and ${\rm B}({\bf x},\vartheta_x)$ is the ULA beampattern defined as \cite{VanTrees2002}
\begin{equation}{\label{ULA_BP}}
    {\rm B}({\bf x},\vartheta_x) =\sum_{m=0}^{M_x-1}x_m^{*}e^{-j2\pi f_c\overline{\tau}_m(\vartheta_x)} = {\bf x}^H{\bf a}(\vartheta_x),
\end{equation}
where 
\mbox{${\bf x}=[x_0,x_1...,x_{M_x-1}]^T$}, 
\begin{equation}
    {\overline \tau}_{m}(\vartheta_x) =\frac{m d_e \sin (\vartheta_x)}{c}, \ \forall \vartheta_x \in \left[-\frac{\pi}{2},\frac{\pi}{2}\right], 
\end{equation}
\begin{equation}
    {\bf a}(\vartheta_x)=[1, e^{-j2\pi f_c\overline{\tau}_1(\vartheta_x)}, \ ... \ ,e^{-j2\pi f_c\overline{\tau}_{M_x-1}(\vartheta_x)}]^T.
\end{equation}
Similarly, ${\rm B}({\bf y},\vartheta_y)$ is defined as 
\begin{equation}
    {\rm B}({\bf y},\vartheta_y) =\sum_{m=0}^{M_y-1}y_m^{*}e^{-j2\pi f_c\overline{\tau}_m(\vartheta_y)} = {\bf y}^H{\bf a}(\vartheta_y),
\end{equation}
where 
\mbox{${\bf y}=[y_0,y_1,...,y_{M_y-1}]^T$}, 
\begin{equation}
    {\overline \tau}_{m}(\vartheta_y) =\frac{m d_e \sin (\vartheta_y)}{c}, \ \forall \vartheta_y \in \left[-\frac{\pi}{2},\frac{\pi}{2}\right], 
\end{equation}
\begin{equation}
    {\bf a}(\vartheta_y)=[1, e^{-j2\pi f_c\overline{\tau}_1(\vartheta_y)}, \ ... \ ,e^{-j2\pi f_c\overline{\tau}_{M_y-1}(\vartheta_y)}]^T.
\end{equation}
\begin{figure}
    \centering  
    \subfloat[]{\includegraphics[width=2in]{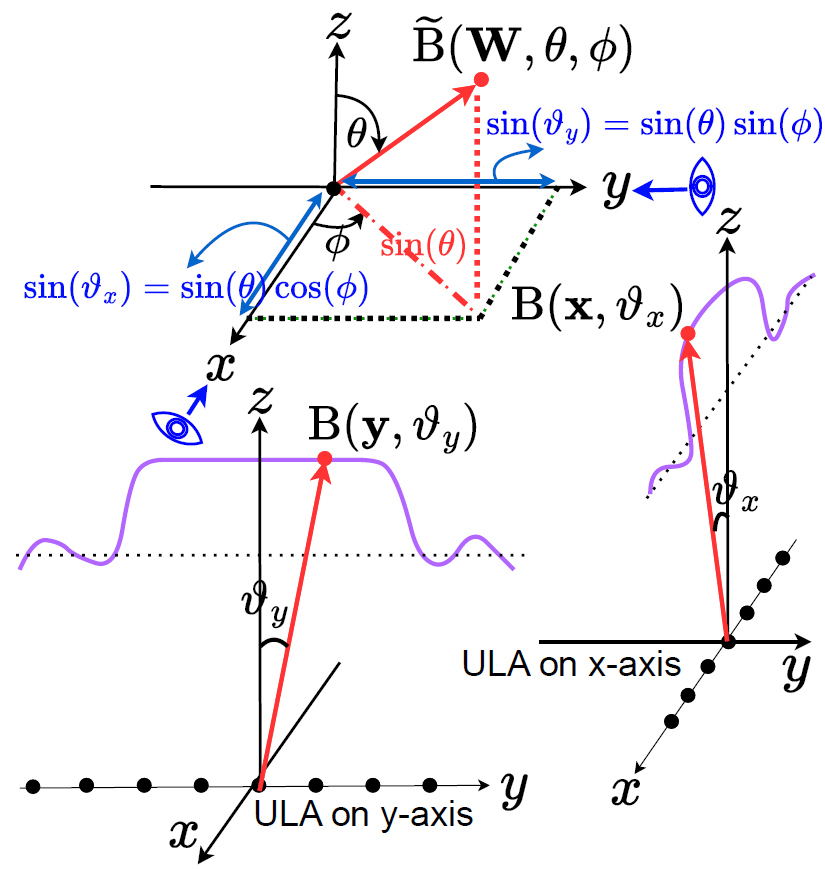} 
    \label{Fig:Visualize_URA_ULAs_array}}  \hspace{-5mm} 
    \par 
    \vspace{-3mm}
    \subfloat[]{\includegraphics[width=2.5in]{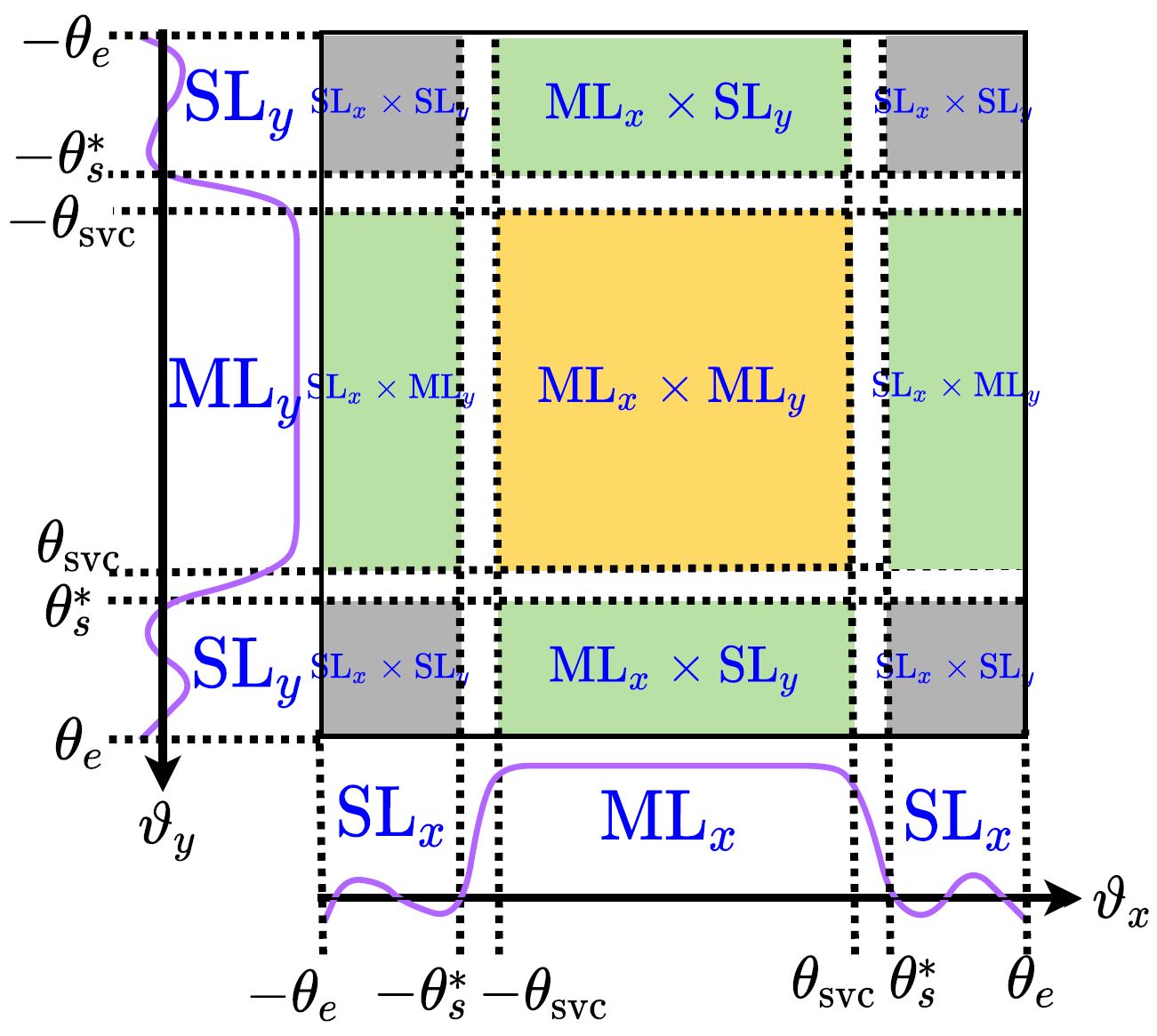}
    \label{Fig:Visualize_URA_ULAs_BP_concept}} 
    \caption{(a) Visualization of the ``composite" URA beampattern based on (\ref{eq:B_tilde}) and (b) Components of the ``composite" URA beampattern ($\rm{ML}_x$/$\rm{ML}_y$ and $\rm{SL}_x$/$\rm{SL}_y$ represent the main lobe/sidelobe region of ULA beampattern with respect to $\vartheta_x$/$\vartheta_y$ axis).}
\end{figure}

\begin{figure} [h]
    \begin{center}
    \includegraphics[width=3in]{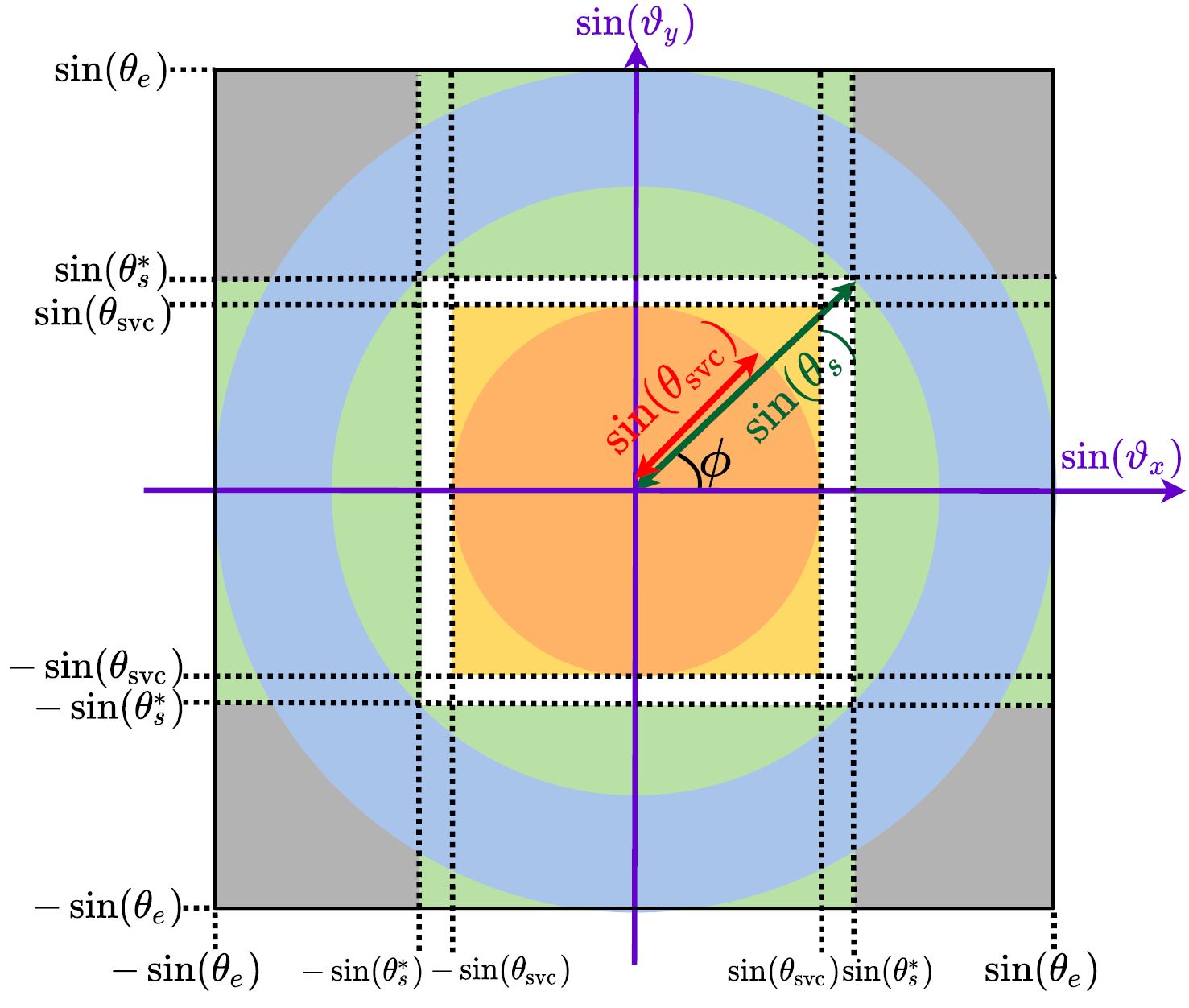}\\
    \vspace{-2mm}
    \caption{Top view of the ``composite" URA beampattern (Square areas in orange and yellow correspond to the main lobe region of the ``composite" URA beampattern; while, green, blue and gray areas correspond to its sidelobe region. Orange circle and blue annular ring areas are the expected URA beampattern main lobe and sidelobe region in the problem (\ref{P1_1}).)} \label{Fig:Visualize_URA_ULAs_BP_topview}
    \end{center}
\end{figure}

Based on (\ref{eq:B_tilde}), 
the URA beampattern, $\widetilde{\rm B}({\bf W},\theta,\phi)$, can be composed of ULA beampatterns, ${\rm B}({\bf x},\vartheta_x)$ and ${\rm B}({\bf y},\vartheta_y)$, which can be interpreted through the beampatterns generated by the ULAs positioned along the x/y-axis as illustrated in Fig. \ref{Fig:Visualize_URA_ULAs_array}.
For convenience, 
we name the URA beampattern generated through (\ref{eq:B_tilde}) as the ``composite" URA beampattern.   
In Fig. \ref{Fig:Visualize_URA_ULAs_BP_concept}, the expected components of the ``composite" URA beampattern are illustrated.
Its main lobe region is highlighted by square areas in yellow which are composed of the ULA beampatterns' main lobe region. 
While, green and gray-colored areas are sidelobe regions of the ``composite" URA beampattern.
Note that the expected main lobe/sidelobe region of the ``composite" URA beampattern are different from that in the URA design problem (\ref{P1_1}).
In the problem (\ref{P1_1}),
with the main lobe angle set $\Pi_m$ defined in (\ref{Theta_m_bar}), the expected URA beampattern main lobe region corresponds to the orange-colored circle in Fig. \ref{Fig:Visualize_URA_ULAs_BP_topview}.
While, with the sidelobe angle set $\Pi_s$ defined in (\ref{Theta_s_bar}), the expected URA beampattern sidelobe region corresponds to the blue-colored annular ring in Fig. \ref{Fig:Visualize_URA_ULAs_BP_topview}.
To decompose the URA design problem (\ref{P1_1}) into ULR design subproblems and ensure the main lobe/sidelobe region of the ``composite" URA beampattern contains the expected URA beampattern main lobe/sidelobe region in the problem (\ref{P1_1}),  
we now define the ULA main lobe/sidelobe angle sets that will be applied to the ULA design subproblems (\ref{P1_2}).
The ULA main lobe angle set is defined as
\begin{align} \label{def:Theta_m}
    \Theta_m = [-\theta_{\rm{svc}}, \theta_{\rm{svc}}].
\end{align}
The ULA sidelobe angle set is defined as
\begin{align} \label{def:Theta_s}
    \Theta_s = [-\theta_e,-\theta_s^*]\cup[\theta_s^*,\theta_e],
\end{align}
where 
\begin{align} \label{def:Theta_s_star}
    \theta_s^* = \sin^{-1}\left(\frac{\sin(\theta_s)}{\sqrt{2}}\right).
\end{align}
Also, we restrict \mbox{$\theta_{\rm{svc}} \leq \sin^{-1} \left(\frac{\sin(\theta_s)}{\sqrt{2}}\right)$} to prevent the ``composite" URA beampattern main lobe region (i.e., square areas in orange and yellow in Fig. \ref{Fig:Visualize_URA_ULAs_BP_topview}) from overlapping the sidelobe region expected in the problem (\ref{P1_1}) (i.e., blue-colored annular ring in Fig. \ref{Fig:Visualize_URA_ULAs_BP_topview}). 

Furthermore, $\sigma(\theta)$ defined in (\ref{URA_isoflux_mask}) needs to be modified for formulating the problem (\ref{P1_2}).
Firstly, inequalities in (\ref{sidelobe_requirments}) are a sufficient condition for (\ref{P1_1b}).
\begin{subequations} \label{sidelobe_requirments}
\begin{align} 
   &|{\rm B}({\bf x},\vartheta_x)| \leq \sqrt{t{\widetilde \sigma}(\vartheta_x)},  \ \forall \vartheta_x \in \Theta_s, \\
   &|{\rm B}({\bf y},\vartheta_y)| \leq \sqrt{t{\widetilde \sigma}(\vartheta_y)}, \ \forall \vartheta_y \in \Theta_s, \label{sidelobe_requirments_2}\\
   &\sqrt{\widetilde{\sigma}(\vartheta_x)\widetilde{\sigma}(\vartheta_y)} \leq \sigma(\theta), \ \forall \vartheta_x, \vartheta_y \in \Theta_s, \forall \theta \in [\theta_{\rm s}, \theta_e], \label{sidelobe_requirments_3}
\end{align}
\end{subequations}
such that \mbox{$|\widetilde{\rm B}({\bf W}, \theta,\phi)| = |{\rm B}({\bf x}, \vartheta_x)||{\rm B}({\bf y}, \vartheta_y)| \leq t {\sigma}(\theta)$}.
To satisfy (\ref{sidelobe_requirments_3}), we can select
\mbox{$\widetilde{\sigma}(\vartheta_x)=\sigma(\vartheta_x)$},
\mbox{$\widetilde{\sigma}(\vartheta_y)=\sigma(\vartheta_y)$}, \mbox{$\forall \vartheta_x, \vartheta_y \in \Theta_s$}.
To show that the selection satisfies (\ref{sidelobe_requirments_3}),
we have \mbox{$\forall \vartheta_x \in \Theta_s$}: 
\begin{equation}
\begin{aligned}
    \widetilde{\sigma}(\vartheta_x)
    = {\sigma}(\vartheta_x)
    &\overset{\text{(a)}} = \sigma(\sin^{-1}(\sin(\theta)\cos(\phi))) \\
    &\overset{\text{(b)}} \leq\sigma(\sin^{-1}(\sin(\theta)))=\sigma(\theta),
\end{aligned} \nonumber
\end{equation} 
where (a) holds by (\ref{eq:vartheta_x}) and (b) holds because both \mbox{$\sigma(\theta)$} and \mbox{$\sin^{-1}(\theta)$} are non-decreasing functions \mbox{$\forall \theta \in (0,\theta_e)$}.
Similarly, it can be shown that \mbox{$\widetilde{\sigma}(\vartheta_y)= {\sigma}(\vartheta_y)\leq \sigma(\theta)$}.
Secondly, inequalities in (\ref{main_lobe_requirments_a}) are a sufficient condition for (\ref{P1_1c}).
\begin{subequations} \label{main_lobe_requirments_a}
\begin{align}
   &|{\rm B}({\bf x},\vartheta_x)| \geq \sqrt{\alpha{\widetilde \sigma}(\vartheta_x)}, \ \forall \vartheta_x \in  \Theta_m, \\ 
   &|{\rm B}({\bf y},\vartheta_y)| \geq \sqrt{\alpha{\widetilde \sigma}(\vartheta_y)}, \ \forall \vartheta_y \in  \Theta_m,  \label{main lobe_requirments_a2}\\
   &\sqrt{\widetilde{\sigma}(\vartheta_x)\widetilde{\sigma}(\vartheta_y)}\geq\sigma(\theta), \forall \vartheta_x, \vartheta_y \in \Theta_m , \forall \theta \in [0, \theta_{\rm{svc}}], \label{main_lobe_requirments_a3}
\end{align}
\end{subequations}
such that \mbox{$|\widetilde{\rm B}({\bf W}, \theta,\phi)| =|{\rm B}({\bf x}, \vartheta_x)||{\rm B}({\bf y}, \vartheta_y)| \geq \alpha\sigma(\theta)$}.
To satisfy (\ref{main_lobe_requirments_a3}), we can select
\begin{equation}
\widetilde{\sigma}(\vartheta_x) = \widetilde{\sigma}(\vartheta_y) = \sigma(\theta), \forall \vartheta_x, \vartheta_y \in \Theta_m , \forall \theta \in [0, \theta_{\rm{svc}}].
\end{equation}
By the relationship $\sin(\theta) = \sqrt{2} \sin(\vartheta_x) = \sqrt{2} \sin(\vartheta_y)$,
we can choose
\begin{equation}
\widetilde{\sigma}(\vartheta_x) = {\sigma}(\theta) |_{\theta = \sin^{-1}(\sqrt{2}\sin(\vartheta_x))}, \forall \vartheta_x \in  \Theta_m,
\end{equation}
\begin{equation}
\widetilde{\sigma}(\vartheta_y) = {\sigma}(\theta) |_{\theta = \sin^{-1}(\sqrt{2}\sin(\vartheta_y))}, \forall \vartheta_y \in  \Theta_m.
\end{equation}
As a result, ${\widetilde \sigma}: [-\theta_e, \theta_e] \rightarrow {\mathbb R}$, plotted in Fig. \ref{Fig:ULA_isolux_radiation_mask}, is defined as

\vspace{-3mm}
\begin{footnotesize}
\begin{align} \label{ULA_isoflux_radiation_mask}
\widetilde{\sigma}(\vartheta) = \left\{
\begin{aligned}
&{\sigma}(\theta) |_{\theta = \sin^{-1}(\sqrt{2}\sin(\vartheta))}, && \vartheta \in  \Theta_m = [-\theta_{\rm{svc}}, \theta_{\rm{svc}}], \theta_{\rm{svc}} \leq \frac{\pi}{4}\\
&{\sigma}(\vartheta), && \vartheta \in  \Theta_s = [-\theta_e,-\theta_s^*]\cup[\theta_s^*,\theta_e]\\
& 1, && \mbox{else}.
\end{aligned}
\right.
\end{align}
\end{footnotesize}

\subsection{Decomposition of the URA Design Problem (\ref{P1_1}) into ULA Design Subproblems} \label{sec:URA_to_ULAs}
According to (\ref{eq:B_tilde}), (\ref{sidelobe_requirments}) and (\ref{main_lobe_requirments_a}),
the URA design problem in (\ref{P1_1}) is transformed to 

\vspace{-5mm}
\begin{small}
\begin{subequations}{\label{P1_2}} 
\begin{align}
    \mathop{\rm{minimize}}\limits_{{\bf x}\in\mathbb{C}^{M_x}, {\bf y}\in\mathbb{C}^{M_y}, t\in {\mathbb R}}& \ t  {\label{P1_2a}}\\
    {\rm{subject \ to }}& \ |{\rm B}({\bf x},\vartheta_x)||{\rm B}({\bf y},\vartheta_y)|\leq t\sqrt{{\widetilde\sigma}(\vartheta_x){\widetilde\sigma}(\vartheta_y)}, \nonumber \\
    & \qquad\qquad\qquad\qquad\qquad\quad \forall \vartheta_x,\vartheta_y \in\Theta_s {\label{P1_2b}} \\
    &|{\rm B}({\bf x},\vartheta_x)||{\rm B}({\bf y},\vartheta_y)|\geq\alpha\sqrt{\widetilde\sigma(\vartheta_x)\widetilde\sigma(\vartheta_y)}, \nonumber \\
    &\qquad\qquad\qquad\qquad\qquad\quad \forall \vartheta_x,\vartheta_y \in\Theta_m {\label{P1_2c}} \\
    &|x_m||y_l|= 1, \quad m \in {\mathbb Z}_{M_x},l\in{\mathbb Z}_{M_y},  {\label{P1_2d}}
\end{align}
\end{subequations} 
\end{small}

\noindent
where $\widetilde\sigma(\vartheta)$ is defined in (\ref{ULA_isoflux_radiation_mask}). 
We then decompose the problem (\ref{P1_2}) into two identical ULA design subproblems (\ref{P1_3}) and (\ref{P1_3Y}) through Kronecker product beamforming \cite{Wang2021, Frank2022_planner_array_Kronecker_Product_Beamforming, Albagory2022_planner_array, VanTrees2002},
\begin{subequations}{\label{P1_3}} 
\begin{align}
    \mathop{\rm{minimize}}\limits_{{\bf x}\in\mathbb{C}^{M_x}, t\in {\mathbb R}} \ &t{\label{P1_2a}}\\
    {\rm{subject \ to }} \ &|{\rm B}({\bf x},\vartheta_x)|\leq\sqrt{t{\widetilde\sigma}(\vartheta_x)}, \ \forall\vartheta_x\in\Theta_s {\label{P1_2b}} \\
    &|{\rm B}({\bf x},\vartheta_x)|\geq{\sqrt{\alpha\widetilde\sigma(\vartheta_x)}}, \ \forall\vartheta_x\in\Theta_m {\label{P1_2c}} \\
    &|x_m|=1, \ m\in{\mathbb Z}_{M_x}.  {\label{P1_2d}}
\end{align}
\end{subequations}
\vspace{-5mm}
\begin{subequations}{\label{P1_3Y}} 
\begin{align}
    \mathop{\rm{minimize}}\limits_{{\bf y}\in\mathbb{C}^{M_y}, t\in {\mathbb R}} \ &t{\label{P1_2Ya}}\\
    {\rm{subject \ to }} \ &|{\rm B}({\bf y},\vartheta_y)|\leq\sqrt{t{\widetilde\sigma}(\vartheta_y)}, \ \forall\vartheta_y\in\Theta_s {\label{P1_2Yb}} \\
    &|{\rm B}({\bf y},\vartheta_y)|\geq{\sqrt{\alpha\widetilde\sigma(\vartheta_y)}}, \ \forall\vartheta_y\in\Theta_m {\label{P1_2Yc}} \\
    &|y_m|=1, \ m\in{\mathbb Z}_{M_y}.  {\label{P1_2Yd}}
\end{align}
\end{subequations}
By utilizing decomposition based on Kronecker product beamforming, the beamforming coefficients that need to be optimized in the ULA design subproblems (\ref{P1_3}) and (\ref{P1_3Y}) are ${\bf x}$ of size $M_x$ and ${\bf y}$ of size $M_y$, respectively. This results in a significant reduction in the size of the optimized variables compared with the original URA design problem (\ref{P1_1}), where the URA beamforming coefficient matrix ${\bf W}$ has a size of $M_x \times M_y$.
Because the two subproblems (\ref{P1_3}) and (\ref{P1_3Y}) are identical,
we will only show the method to solve the subproblem (\ref{P1_3}).
The quadratic form of the problem (\ref{P1_3}) is
\begin{subequations}{\label{P1_5}}
\begin{align}
    \mathop{\rm{minimize}}\limits_{{\bf x}\in\mathbb{C}^{M_x},t\in {\mathbb R}} \ &t {\label{P1_5a}} \\
    {\rm{subject \ to }} \ &{\bf x}^H{\bf A}(\vartheta_x){\bf x}\leq{t{\widetilde\sigma}(\vartheta_x)}, \ \forall\vartheta_x\in\Theta_s{\label{P1_5b}} \\
    &{\bf x}^H{\bf A}(\vartheta_x){\bf x}\geq \alpha\widetilde\sigma(\vartheta_x), \ \forall\vartheta_x\in\Theta_m {\label{P1_5c}} \\
    &{\bf x}^H{\bf E}_m{\bf x}= 1, \ m\in{\mathbb Z}_{M_x}{\label{P1_5d}},
\end{align}
\end{subequations}
where 
\mbox{${\bf A}(\vartheta_x) = {\bf a}(\vartheta_x){\bf a}^H(\vartheta_x)$},  
\mbox{${\bf E}_m = {\bf e}_m{\bf e}_m^H$},
and ${\bf e}_m$ is the $m$-th $M_x$-dimensional standard vector
\begin{equation}
\begin{aligned}
  {\bf e}_m(i) = \left\{
  \begin{aligned}
    1,           && &\mbox{$i=m$}\\
    0,           && &\mbox{else.}\\
  \end{aligned}
  \right.
\end{aligned}        
\end{equation}
Uniformly sampling the main lobe angle set $\Theta_m$ by $\delta$ with $N_{\rm{svc}}$ points, we have \mbox{$\vartheta_i = -\theta_{\rm{svc}}+\delta \cdot i$}
for \mbox{$i \in {\mathbb Z}_{N_{\rm{svc}}}$} with \mbox{$N_{\rm{svc}} = \frac{2 \theta_{\rm{svc}}}{\delta}+1$}.
Also, uniformly sample the sidelobe angle set $\Theta_s$ by $\delta$ with $N_s$ points, we then have 

\vspace{-2mm}
\begin{small}
\begin{equation}
\begin{aligned}
  \vartheta_k = \left\{
  \begin{aligned}
    &-\theta_e + \delta \cdot k, \ \hspace{19.5mm} \mbox{$\forall k \in \{0,1,...,\frac{\theta_e-\theta_s^*}{\delta}\}$}\\
    &\theta_s^* + \delta \cdot \big(k-\frac{\theta_e-\theta_s^*}{\delta}-1\big),             \ \mbox{$\forall k \in \{\frac{\theta_e-\theta_s^*}{\delta}+1,...,N_s-1\}$,}\\
  \end{aligned}
  \right.
\end{aligned}        
\end{equation}
\end{small}

\noindent
where \mbox{$N_s = 2(\frac{\theta_e-\theta_s^*}{\delta}+1)$}.
Define ${\bf X} = {\bf x}{\bf x}^H$, then the problem we intend to solve becomes 
\begin{subequations}\label{P1_6}
\begin{align}
    \mathop{\rm{minimize}}\limits_{{\bf X}\in{\mathbb H}_{+}^{M_x}, t \in {\mathbb R}} \ &t{\label{P1_6a}} \\
    {\rm{subject \ to }} \ &\text{Tr}({\bf X}{\bf A}(\vartheta_k))\leq{t{\widetilde\sigma}(\vartheta_k)},\quad\forall k \in{\mathbb Z}_{N_s}{\label{P1_6b}} \\
    &\text{Tr}({\bf X}{\bf A}(\vartheta_i))\geq{\alpha \widetilde \sigma(\vartheta_i)},\quad\forall i \in{\mathbb Z}_{N_{\rm{svc}}}{\label{P1_6c}} \\
    &\text{Tr}({\bf X}{\bf E}_m)= 1, \quad m\in{\mathbb Z}_{M_x}{\label{P1_6e}}\\
    &\text{rank}({\bf X})=1.{\label{P1_6f}}
\end{align}
\end{subequations}
Note that the problem (\ref{P1_6}) is a semidefinite programming (SDP) with a rank-one constraint (\ref{P1_6f}), which is a non-convex optimization problem. 

\subsection{Proposed Convex Iterative Algorithm} \label{sec:iterative_algorithm}
A convex iterative algorithm is proposed to address the non-convex SDP (\ref{P1_6}) with the rank-one constraint (\ref{P1_6f}).
The semidefinite relaxation (SDR) technique is applied by dropping (\ref{P1_6f}).
Then, the rank-one constraint is gradually approached by suppressing the sum of the second largest eigenvalue to the smallest eigenvalue in each iteration until the ratio of the second largest eigenvalue to the largest eigenvalue is smaller than a specified threshold value $\varepsilon_{\mathrm{rank}}$. 
In the \mbox{($\psi-1$)}-th iteration, the eigendecomposition of ${\bf X}^{(\psi-1)}$ is performed: 
\begin{align} \label{eq:egen_decomp}
    {\bf X}^{(\psi-1)}={\widetilde{\bf U}}^{(\psi-1)}{\bf D}^{(\psi-1)}({\widetilde{\bf U}}^{(\psi-1)})^H,
\end{align}
\begin{align}
{\widetilde{\bf U}}^{(\psi-1)}=[{\bf u}_0^{(\psi-1)},...,{\bf u}_{M_x-1}^{(\psi-1)}],
\end{align}
\begin{align}
{\bf D}^{(\psi-1)} = \rm{diag}(\Lambda_0^{(\psi-1)},...,\Lambda_{M_x-1}^{(\psi-1)}),
\end{align}
where ${\bf u}_{m}^{(\psi-1)}$ are the eigenvectors of ${\widetilde{\bf U}}^{(\psi-1)}$ corresponding to eigenvalues $\Lambda_m^{(\psi-1)}$ in a non-increasing order.
As the eigenvector corresponding to the largest eigenvalue (i.e., ${\bf u}_0^{(\psi-1)}$) is regarded as the favorable direction for achieving \text{rank}({\bf X}) = 1,
the unfavorable direction in vector space ${\mathbb H}_{+}^{M_x}$ is constructed as
\begin{align} \label{eq:V_phi}
{\bf V}^{(\psi-1)} = {\bf U}^{(\psi-1)}({\bf U}^{(\psi-1)})^H,
\end{align}
where
${\bf U}^{(\psi-1)}=[{\bf u}_1^{(\psi-1)},...,{\bf u}_{M_x-1}^{(\psi-1)}]$. 
To suppress unfavorable directions and achieve a rank-one matrix,
a penalty function is introduced \cite{Dattorro2005}
\begin{align} \label{penalty_function}
    \text{Tr}({\bf X}^{(\psi)}{\bf V}^{(\psi-1)}).
\end{align}
In the ($\psi$)-th iteration, 
minimizing (\ref{penalty_function}) implies suppressing the sum of the second largest eigenvalue to the smallest eigenvalue.
The penalty function (\ref{penalty_function}) is applied to (\ref{P1_7a}) with penalty parameter $\rho$.
In the ($\psi$)-th iteration, the convex problem in (\ref{P1_7}) is solved.
\begin{subequations}\label{P1_7}
\begin{align}
    \mathop{\rm{minimize}}\limits_{{\bf X}^{(\psi)}\in{\mathbb H}_{+}^{M_x}, t\in {\mathbb R}} &t + \rho\text{Tr}({\bf X}^{(\psi)}{\bf V}^{(\psi-1)}){\label{P1_7a}} \\
    {\rm{subject \ to }} \ &\text{Tr}({\bf X}^{(\psi)}{\bf A}(\vartheta_k))\leq t{\widetilde\sigma}(\vartheta_k),\ \forall k\in{\mathbb Z}_{N_s}{\label{P1_7b}}\\
    &\text{Tr}({\bf X}^{(\psi)}{\bf A}(\vartheta_i))\geq\alpha \widetilde\sigma(\vartheta_i),\ \forall i\in{\mathbb Z}_{N_{\rm{svc}}} {\label{P1_7c}}\\
    & \text{Tr}({\bf X}^{(\psi)}{\bf E}_m) = 1, \  m\in{\mathbb Z}_{M_x}.{\label{P1_7d}}
\end{align}
\end{subequations}   
Similarly, to tackle the non-convex problem (\ref{P1_3Y}), the problem (\ref{P1_7Y}) is solved in each iteration.
\begin{subequations}\label{P1_7Y}
\begin{align}
    \mathop{\rm{minimize}}\limits_{{\bf Y}^{(\psi)}\in{\mathbb H}_{+}^{M_y}, t\in {\mathbb R}} &t + \rho\text{Tr}({\bf Y}^{(\psi)}{\bf V}^{(\psi-1)}){\label{P1_7Ya}} \\
    {\rm{subject \ to }} \ &\text{Tr}({\bf Y}^{(\psi)}{\bf A}(\vartheta_k))\leq t{\widetilde\sigma}(\vartheta_k),\ \forall k\in{\mathbb Z}_{N_s}{\label{P1_7Yb}}\\
    &\text{Tr}({\bf Y}^{(\psi)}{\bf A}(\vartheta_i))\geq\alpha \widetilde\sigma(\vartheta_i),\ \forall i\in{\mathbb Z}_{N_{\rm{svc}}} {\label{P1_7Yc}}\\
    & \text{Tr}({\bf Y}^{(\psi)}{\bf E}_m) = 1, \  m\in{\mathbb Z}_{M_y}.{\label{P1_7Yd}}
\end{align}
\end{subequations}   

The penalty parameter $\rho$ in (\ref{P1_7a}) and (\ref{P1_7Ya}) balances the trade-off between suppressing PSL and achieving a rank-one matrix.
A smaller value of $\rho$ places greater emphasis on PSL suppression, whereas a larger value of $\rho$ focuses more on tackling the rank-one constraint. 
We increase $\rho$ if the ratio of the largest eigenvalue to the second largest eigenvalue in the consecutive iteration is smaller than $\kappa$.
Given an initial penalty parameter value $\rho^{(0)}$, the update strategy of $\rho^{(\psi)}$ is 
\begin{align} \label{penalty_parameter_update}
   \rho^{(\psi)}=\begin{cases} 
                    \rho^{(\psi-1)} \cdot (1+p), & {\text{if }\frac{\Lambda_0^{(\psi)}}{\Lambda_1^{(\psi)}}-\frac{\Lambda_0^{(\psi-1)}}{\Lambda_1^{(\psi-1)}} \leq \kappa}\\
                     \rho^{(\psi-1)}, & {\text{else}},
                 \end{cases}
\end{align}    
where $\kappa$ and $p$ are the preset positive numbers.
The pseudocode of the proposed broadened-beam URA coefficient design method is shown in Algorithm \ref{algorithm1}. 
Note that the convergence properties of Algorithm \ref{algorithm1} can be influenced by the settings of the penalty parameter update strategy (\ref{penalty_parameter_update}) and the initial point selection (${\bf x}_{\rm{init}}$, ${\bf y}_{\rm{init}}$), which will be mentioned in Section \ref{sec:Simulation_results_and_discussions}.

\begin{algorithm}[h] 
\begin{footnotesize}
\captionsetup{font=footnotesize}
\caption{Proposed Broadened-beam URA Coefficient Design Method} \label{algorithm1} 
\hspace*{\algorithmicindent} \textbf{Input: } 
$M_x$, $M_y$, $\theta_{\rm{svc}}$, $\theta_s$, $\theta_e$, $N_{\rm{svc}}$, $N_s$,  
${\bf x}_{\rm{init}}$, ${\bf y}_{\rm{init}}$, $\text{SNR}_{\min}$, $f_{\rm{BW}}$, $\beta$, \\

\vspace{-2mm}
\hspace{13mm} 
$P_s$, $L_0$, ${\rm G}/{\rm T}$, $\rho^{(0)}$, $p$, $\kappa$, $\varepsilon_{\mathrm{rank}}$ \\ 
\hspace*{\algorithmicindent} \textbf{Output: }${\bf W}_{\rm{opt}}$
\begin{algorithmic}[1]
    \State Calculate $\theta_s^*$ by (\ref{def:Theta_s_star}).
    \State Calculate $\alpha$ by (\ref{eq:alpha}).
    \If{{\textbf{initial point ${\bf x}_{\rm{init}}$ is available}}}
        \State 
        \mbox{${\bf X}^{(0)} = {\bf x}_{\rm{init}}{\bf x}_{\rm{init}}^H$}.
        \State 
        Perform the eigendecomposition of \mbox{${\bf X}^{(0)}$} according to (\ref{eq:egen_decomp}).
        \State
        Set ${\bf V}^{(0)}$ according to (\ref{eq:V_phi}).
    \Else    
        \State ${\bf V}^{(0)} = {\bf 0}$.
    \EndIf
    \State Let $\psi \leftarrow 1$.           
    \Repeat
        \State 
        Solve the problem (\ref{P1_7}) and obtain ${\bf X}^{(\psi)}$. 
        \State 
        Perform the eigendecomposition \mbox{${\bf X}^{(\psi)} =\widetilde{\bf U}^{(\psi)}{\bf D}^{(\psi)}\widetilde{\bf U}^{(\psi)H}$} \\
        \hspace{3.5mm}
        according to (\ref{eq:egen_decomp}).
        \State
        Let \mbox{$\Lambda_0^{(\psi)} = [{\bf D}^{(\psi)}]_{0,0}$} and \mbox{$\Lambda_1^{(\psi)} = [{\bf D}^{(\psi)}]_{1,1}$}.
        \State
        Update ${\bf V}^{(\psi)}$ according to (\ref{eq:V_phi}).
        \State Update penalty parameter $\rho^{(\psi)}$ based on (\ref{penalty_parameter_update}).
        \State \mbox{${\psi}\leftarrow{\psi}+1$}.
    \Until 
    \mbox{$\Lambda_1^{(\psi)}/\Lambda_0^{(\psi)} \leq \varepsilon_{\mathrm{rank}}$} 
        \State        
        Obtain \mbox{${\bf x}_{\rm{opt}} = \sqrt{\Lambda_0^{(\psi)}}{\bf u}_0^{(\psi)}$}.
    \State 
    Similarly, solve the problem (\ref{P1_7Y}) to obtain ${\bf y}_{\rm{opt}}$ following the same procedure as was used for ${\bf x}_{\rm{opt}}$.
    \State
    Obtain the URA beamforming coefficient by (\ref{URA_weight_matrix}): \mbox{${\bf W}_{\rm{opt}} = {\bf x}_{\rm{opt}}{\bf y}_{\rm{opt}}^T$}.
\end{algorithmic}
\end{footnotesize}
\end{algorithm} 

We would like to mention that beampattern synthesis problems in the form of SDP with a rank-one constraint were also considered in \cite{Fuchs2014_TAP, Liu2018_TAP, Xu2019_IEEEAccess}, where the SDR method was also applied by dropping the rank-one constraint.
However, the method for obtaining a low-rank matrix in \cite{Fuchs2014_TAP, Liu2018_TAP, Xu2019_IEEEAccess} differs from ours.
In \cite{Fuchs2014_TAP, Liu2018_TAP, Xu2019_IEEEAccess}, the reweighted minimization method \cite{Fazel_2004} is applied which iteratively minimizes the trace of the matrix (as ${\bf X}$ in the problem (\ref{P1_7})), equivalent to minimizing the sum of the eigenvalues of ${\bf X}$ in each iteration.
In our method, we incorporate a penalty term for minimizing the sum of the second largest to the smallest eigenvalues \cite{Dattorro2005}, along with PSL suppression in the objective function.
Additionally, in \cite{Fuchs2014_TAP, Liu2018_TAP, Xu2019_IEEEAccess}, the sidelobe level is constrained by a predefined value, whereas in our problem, the PSL suppression is considered within the objective function.
Lastly, we highlight that the DRR constraints or CMCs were not considered in \cite{Fuchs2014_TAP, Liu2018_TAP}, and only the DRR constraints were addressed in \cite{Xu2019_IEEEAccess}. 
The proposed algorithm successfully achieves CMCs, as will be demonstrated in Section \ref{sec:Simulation_results_and_discussions}.

\section{Numerical evaluation} \label{sec:numerical_results}
\subsection{Definition of Evaluation Metrics} \label{subec:def_eval_metrics}
The beampatterns obtained through Algorithm \ref{algorithm1} are expected to feature curved shapes in both sidelobe upper bound and main lobe lower bound due to ${\widetilde\sigma}(\vartheta)$ applied in the constraint (\ref{P1_7b})/(\ref{P1_7Yb})) and (\ref{P1_7c})/(\ref{P1_7Yc})). 
However, such curved shapes make it difficult for us to evaluate the peak sidelobe level (PSL).
Consequently, when evaluating PSL, we divide the ULA beampatterns by $\sqrt{{\widetilde \sigma}(\vartheta)}$ to have a constant level of sidelobe upper bound and main lobe lower bound. 
The normalized peak sidelobe level (NPSL) of the ULA beampattern is defined in decibels (dB) as 

\begin{small}
\begin{equation} \label{def:NPSL_ULA}
   \rm{NPSL_{ULA}} = 20 \log_{10}\left(\frac{\mathop{\max}_{\vartheta \in {\Theta_s}} \left\{ |{\rm B}({\bf x},\vartheta) |/ \sqrt{{\widetilde \sigma}(\vartheta)} \right\} }{\mathop{\min}_{\vartheta \in \Theta_{\rm m}} \left\{ | {\rm B}({\bf x},\vartheta) |/ \sqrt{{\widetilde \sigma}(\vartheta)} \right\}}\right),
\end{equation}
\end{small}

\noindent
where ${\rm B}({\bf x},\vartheta)$ is defined in (\ref{ULA_BP}), 
$\Theta_m$ is defined in (\ref{def:Theta_m}), 
$\Theta_s$ is defined in (\ref{def:Theta_s}), and 
${\widetilde \sigma}(\vartheta)$ is defined in (\ref{ULA_isoflux_radiation_mask}).
Similarly, the NPSL of the URA beampattern is defined in dB as 

\begin{small}
\begin{equation}
    \rm{NPSL_{URA}} =  20 \log_{10} \left(\frac{\mathop{\max}_{(\theta,\phi) \in \Pi_s} \left\{ |\widetilde{\rm B}({\bf W},\theta,\phi)| / \sigma(\theta) \right\} }{\mathop{\min}_{(\theta,\phi) \in \Pi_m} \left\{ |\widetilde{\rm B}({\bf W},\theta,\phi)|/\sigma(\theta) \right\}}\right),
\end{equation}
\end{small}

\noindent
where $ \widetilde{\rm B}({\bf W},\theta,\phi)$ is defined in (\ref{def:URA_BP}), 
$\Pi_m$ is defined in (\ref{Theta_m_bar}), 
$\Pi_s$ is defined in (\ref{Theta_s_bar}), and
$\sigma(\theta)$ is defined in (\ref{URA_isoflux_mask}).
To evaluate the QoS, the minimum received SNR in the SAT service areas is defined in dB as 
\begin{equation}
    \text{SNR}_{\rm{svc}} = 10 \log_{10}\left( \mathop{\min}_{(\theta,\phi) \in  \Pi_m} \text{SNR}({\bf W}, \theta, \phi) \right),
\end{equation}
where $\text{SNR}({\bf W}, \theta, \phi)$ is defined in (\ref{eq:SNR_r}).
To evaluate SAT out-of-beam power leakage, the peak received signal power at the SAT out-of-beam areas is defined in dBm as 
\begin{equation} \label{def:P_r_oob}
    P_{r, \rm{oob}} = 10 \log_{10}\left( \mathop{\max}_{(\theta,\phi) \in  \Pi_s} P_r({\bf W}, \theta, \phi) \right) + 30,
\end{equation}
where $P_r({\bf W}, \theta, \phi)$ is defined as (\ref{eq:P_r}).
Finally, to demonstrate the achievability of CMCs, we define
\begin{equation}
   \eta_{\rm{CMC}} = \frac{\max\{|[{\bf W}]_{m,n}|\}}{\min\{|[{\bf W}]_{m,n}|\}}, \ \forall m \in {\mathbb Z}_{M_x}, n \in {\mathbb Z}_{M_y}. 
\end{equation}
\subsection{Parameter Settings} \label{sec:parameter_settings}
LEO SAT system parameters for Ku-band user downlink is shown in Table \ref{table:LEO_systems_param}.
We select URA size as \mbox{$M_x \times M_y = 32 \times 32$}, subarray size as \mbox{$Q_x \times Q_y = 8 \times 8$}, 
the total number of RF chains and PAs is \mbox{$N_x \times N_y = 4 \times 4$}, 
and the antenna element spacing is $d_e=\frac{c}{2f_c}=0.0125\ [{\rm m}]$.
In addition, we assume that the array unit cell radiation patterns are omni-directional.
Also, we suppose that the same PA is utilized across the array. 
We set PA gain factor \mbox{$\beta = 1000$}, i.e., \mbox{$30 \ [\rm{dB}]$}, 
the PA maximum output power be \mbox{$2 \ [{\rm W}]$}, and
the PA backoff value be \mbox{$5 \ [\rm{dB}]$} \cite{3GPPNTN2019_TR16}.
Then the PA average power can be calculated as \mbox{$P_{\rm{PA, avg}} = 2 \cdot 10^{-5/10} = 0.63 \ [{\rm W}]$}, and the transmit signal power can be obtained by (\ref{def:P_T}) as 
\mbox{$P_T = P_{\rm{PA, avg}} \cdot (N_x \times N_y) = 10.08 \ [{\rm W}]$}.
 According to (\ref{def:P_PA}) and the CMCs (i.e., \mbox{$\left|[{\bf W}]_{m,l} \right| = 1$}), 
the source signal power is set as
\mbox{$P_s = \frac{P_{\rm{PA, avg}}}{\beta Q_x Q_y} = 9.84 \cdot 10^{-6} \ [{\rm W}]$}. 
In addition, the losses in (\ref{Eq:pathloss_L}) are set as 
\mbox{$L_{\rm{c,T}}=L_{\rm{c,R}}=1 \ [{\rm{dB}}]$}, 
\mbox{$L_{\rm{a}} = 0.5 \ [{\rm{dB}}]$},
\mbox{$L_{\rm{sm}} = 0 \ [{\rm{dB}}]$} and
\mbox{$L_{\rm{sl}} = 0.3 \ [{\rm{dB}}]$} \cite{3GPPNTN2019_TR16}.
Then we obtain \mbox{$L_0=171.63 \ [{\rm{dB}}]$} by (\ref{def:L0}).
\begin{table}[h!]
    \centering
    \begin{scriptsize}
    \caption{\ LEO SAT system parameters for Ku-band user downlink} \label{table:LEO_systems_param}
    \vspace{-2mm}
    \begin{tabular}{ |c|c|c|c|} 
    \hline
    \textbf{Parameter} & \textbf{Symbol} & \textbf{Value} & \textbf{Units} \\
    \hline \hline
    SAT altitude  & $h$ & $550$  & km   \\ 
    \hline
    Earth radius  & $R_e$ & $6370$ & km  \\
    \hline
    Carrier frequency  & $f_c$ & $12$  & GHz \\ 
    \hline
    Channel bandwidth  & $f_{\rm{BW}}$ & $500$  & MHz \\
    \hline
    Path loss parameter (\ref{def:L0})  & $L_0$ & $171.63$ & dB \\ 
    \hline 
    Receive antenna gain  & $G_R$  & $39.7$ \cite{3GPPNTN2019_TR16} & dBi  \\ 
    \hline  
    Noise factor  & $N_{f}$ & $1.2$ \cite{3GPPNTN2019_TR16} & dB \\ 
    \hline
    Antenna temperature  & $T_{a}$ & $150$ \cite{3GPPNTN2019_TR16}  & K \\
    \hline  
    Standard temperature  & $T_0$ & $290$ & K \\ 
    \hline 
    Boltzmann constant & $k$ & $-228.6$ & dBW/K/Hz \\ 
    \hline 
    Antenna-gain-to-noise-temperature  & ${\rm G/T}$  & $16$ \cite{3GPPNTN2019_TR16} & dB/K  \\ 
    \hline
    \end{tabular}
    \end{scriptsize}
\end{table} 

The SAT service beamwidths $\Theta_{\rm{BW}}$ are specified as $10^{\circ}$, $30^{\circ}$, and $60^{\circ}$. 
In Table \ref{tb:Angular_parameters}, $\Theta_m$ and $\Theta_s$ are listed with varying choices of $\theta_{\rm{svc}}$ and $\theta_s^*$.
Additionally, the SAT FoV angle $\theta_e \approx 67^{\circ}$ can be obtained by (\ref{Eq:theta_e}).
In all the cases, we let the transmit signal power be the same. 
To provide a larger capacity in SAT service areas, $\text{SNR}_{\min}$ in the constraint (\ref{P1_00b}) is eager to be set as large as possible. 
However, according to Parseval's theorem, the total power in the spatial domain should satisfy the total power limitation \cite{VanTrees2002}.  
We can infer that a lower main lobe level can be attained as the beamwidth becomes larger, and the problems (\ref{P1_7}) and (\ref{P1_7Y}) would be infeasible if $\text{SNR}_{\min}$ is set too large. 
Given this, $\text{SNR}_{\min}$ values of different beamwidths are set in Table \ref{tb:Angular_parameters}. 
Suppose that each UT can select proper modulation and coding schemes (MODCODs) based on the received SNR for decoding. 
We refer to the bit error rate (BER) performance of standard MODCODs used in DVB-S2 provided in \cite{Rahman2021_DVBS_MCS}, and suggest applying MODCODs of 8PSK 8/9, QPSK 4/5 and QPSK 1/4 when the received SNRs are guaranteed to be larger than $11 \ \rm{[dB]}$, $5 \ \rm{[dB]}$ and $-2 \ \rm{[dB]}$ in our application.
According to the above settings, $\alpha$ values listed in Table \ref{tb:Angular_parameters} are calculated by (\ref{eq:alpha}). 
The remaining input parameters in Algorithm \ref{algorithm1} are set as follows: 
\mbox{$N_{\rm{svc}} = \frac{2 \theta_{\rm{svc}}}{0.1}+1$}, \mbox{$N_s = 2(\frac{\theta_e-\theta_s^*}{0.1}+1)$}, \mbox{$\rho^{(0)}=0.1$}, \mbox{$p = 0.1$}, \mbox{$\kappa = 5$}, \mbox{$\varepsilon_{\mathrm{rank}} = 10^{-5}$}.
Settings of the penalty parameter update strategy in (\ref{penalty_parameter_update}) and the initial point (i.e., ${\bf x}_{\rm{init}}$ and ${\bf y}_{\rm{init}}$) selection are discussed in more detail in Section \ref{sec:Simulation_results_and_discussions}.
The ${\rm{CVX}}$ toolbox \cite{cvx2014} is applied to solve the problems (\ref{P1_7}) and (\ref{P1_7Y}), 
and simulations are carried out on a personal computer with 3.60 GHz AMD Ryzen 7 3700X, 32 GB.
\begin{table*}[h!] 
    \centering
    \begin{footnotesize}
    \caption{ULA main lobe/sidelobe angle set, $\text{SNR}_{\rm{min}}$, and $\alpha$ settings} \label{tb:Angular_parameters}
    \vspace{-2mm}
    \begin{tabular}{||c c c c c||} 
    \hline 
    \textbf{Beamwidth $\Theta_{\rm{BW}}$} &   & $10^{\circ}$ & $30^{\circ}$ & $60^{\circ}$ \\ 
    \hline \hline
    ULA main lobe angle set $\Theta_m$ & $[-\theta_{\rm{svc}}, \theta_{\rm{svc}}]$ & $[-5^{\circ},5^{\circ}]$  & $[-15^{\circ}, 15^{\circ}]$ & $[-30^{\circ},30^{\circ}]$ \\
    ULA sidelobe angle set $\Theta_s$ & $[-\theta_e,-\theta_s^*]\cup[\theta_s^*,\theta_e]$ & $[-67^{\circ}, -10^{\circ}]\cup[10^{\circ}, 67^{\circ}]$  & $[-67^{\circ}, -20^{\circ}]\cup[20^{\circ}, 67^{\circ}]$ & $[-67^{\circ}, -35^{\circ}]\cup[35^{\circ}, 67^{\circ}]$ \\
    $\text{SNR}_{\rm{min}}$ & (\ref{P1_00b})  & $11 \ [{\rm{dB}}$] & $5 \ [{\rm{dB}}$] & $-2 \ [{\rm{dB}}$] \\ 
    $\alpha$ & (\ref{eq:alpha}) & $179.35 \ (22.54 \ [{\rm{dB}}]$) & $89.89 \ (19.54 \ [{\rm{dB}}]$) & $40.15 \ (16.04 \ [{\rm{dB}}]$) \\
    \hline
    \end{tabular}
    \end{footnotesize}
\end{table*}

\subsection{Numerical Results and Performance Evaluation} \label{sec:Simulation_results_and_discussions}
Simulations are carried out to demonstrate the effectiveness of Algorithm \ref{algorithm1} in achieving out-of-beam radiation suppression, QoS guarantee, and CMCs when the SAT service beamwidths are \mbox{$\Theta_{\rm{BW}}=10^{\circ}$, $30^{\circ}$, and $60^{\circ}$}.
In our trials, we found that the convergence properties of Algorithm \ref{algorithm1} can be influenced by the penalty parameter update strategy in (\ref{penalty_parameter_update}) and initial point selection.
We at first set a fixed penalty parameter \mbox{$\rho^{(\psi)}=0.1 \ \forall \psi$} instead of using the penalty parameter update strategy in (\ref{penalty_parameter_update}). 
However, Algorithm \ref{algorithm1} does not converge when \mbox{$\Theta_{\rm{BW}}=10^{\circ}$, $30^{\circ}$, and $60^{\circ}$}, and \mbox{$\Lambda_1^{(\psi)}/\Lambda_0^{(\psi)}$} remains at $0.24$, $0.13$, and $0.06$ after $200$ iterations, respectively.
We suppose that the reason for not achieving rank-one convergence could be insufficient suppression of unfavorable directions (i.e., $\text{Tr}({\bf X}^{(\psi)}{\bf V}^{(\psi-1)})$ in (\ref
{P1_7a})).
From this inference, we found that it is helpful for Algorithm \ref{algorithm1} to converge when (\ref{penalty_parameter_update}) is applied.
If the ratio of the largest eigenvalue to the second largest eigenvalue in the consecutive iterations is smaller than $\kappa$, the penalty parameter is increased to suppress unfavorable directions.
Conversely, if the algorithm finds the correct direction for rank-one convergence, the penalty parameter remains.
Throughout our trials, different $\rho^{(0)}$, $p$ and $\kappa$ settings in (\ref{penalty_parameter_update}) may lead to different convergence results.
We set \mbox{$\rho^{(0)}=0.1$, $p = 0.1$, and $\kappa = 5$} to obtain good results in terms of obtaining a low ${\rm{NPSL}}_{\rm{ULA}}$.
Moreover, if we select a zero vector as the initial point of Algorithm \ref{algorithm1}  (i.e., \mbox{${\bf x}_{\rm{init}}={\bf y}_{\rm{init}}={\bf 0}$)} and apply (\ref{penalty_parameter_update}),  
Algorithm \ref{algorithm1} converges when $\Theta_{\rm{BW}}=10^{\circ}$ (at the $33$rd iteration) and $30^{\circ}$ (at the $33$rd iteration), but does not converge when $\Theta_{\rm{BW}}=60^{\circ}$ 
(\mbox{$\Lambda_1^{(\psi)}/\Lambda_0^{(\psi)}$} is stuck at $0.07$ after $200$ iterations).
While, we found that if \cite{Fonteneau2021_EuCNC} is selected as the initial point, Algorithm \ref{algorithm1} converges when \mbox{$\Theta_{\rm{BW}}=60^{\circ}$} (at the $67$th iteration), as do \mbox{$\Theta_{\rm{BW}}=10^{\circ}$} (at the $33$rd iteration) and $30^{\circ}$ (at the $48$th iteration).
We suppose \cite{Fonteneau2021_EuCNC} a good initial point, particularly for broadened beam cases, due to its constant modulus property and the convenience of obtaining it through a closed-form solution.

\subsubsection{Numerical Results} \label{sec:Simulation_results}
\begin{figure*}
    \centering
    \subfloat[]{\includegraphics[width=.3\linewidth]{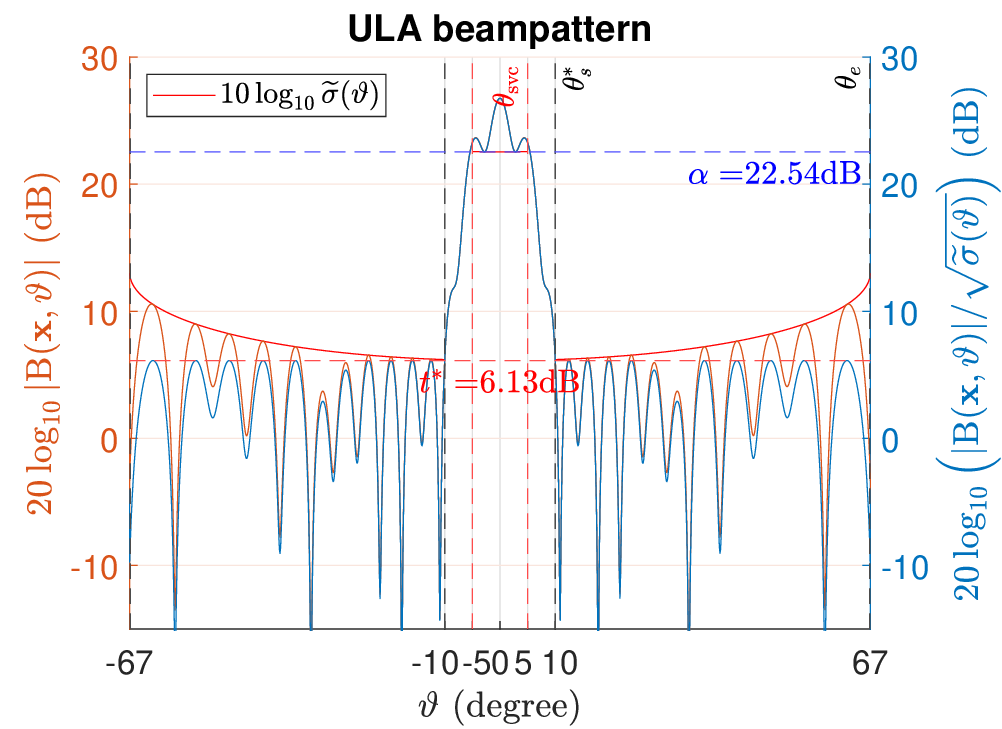} 
    \label{fig:M32_BW10_CMC_figa}}  \hfill
    \subfloat[]{\includegraphics[width=.3\linewidth]{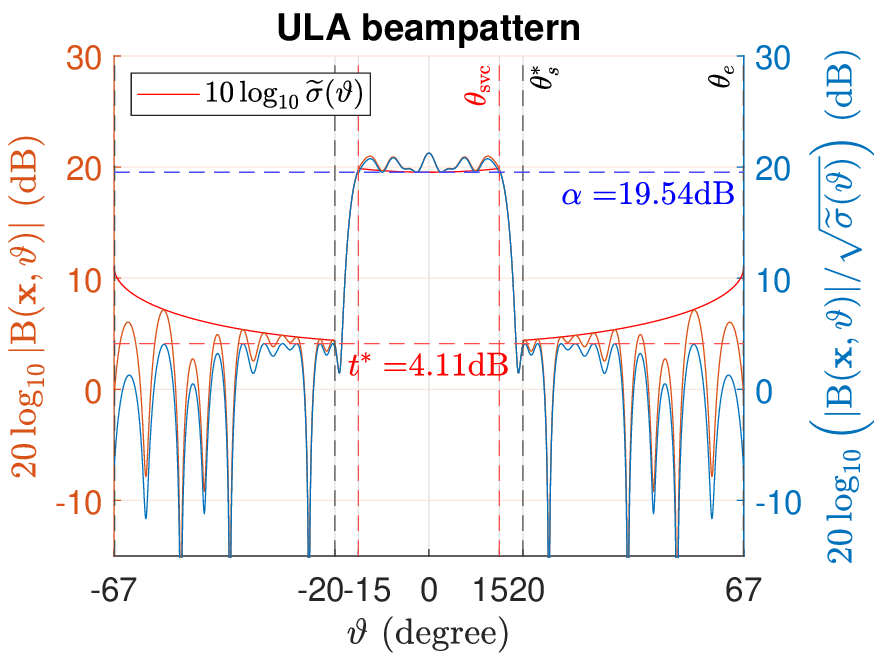}
    \label{fig:M32_BW30_CMC_figa}} \hfill
    \subfloat[]{\includegraphics[width=.3\linewidth]{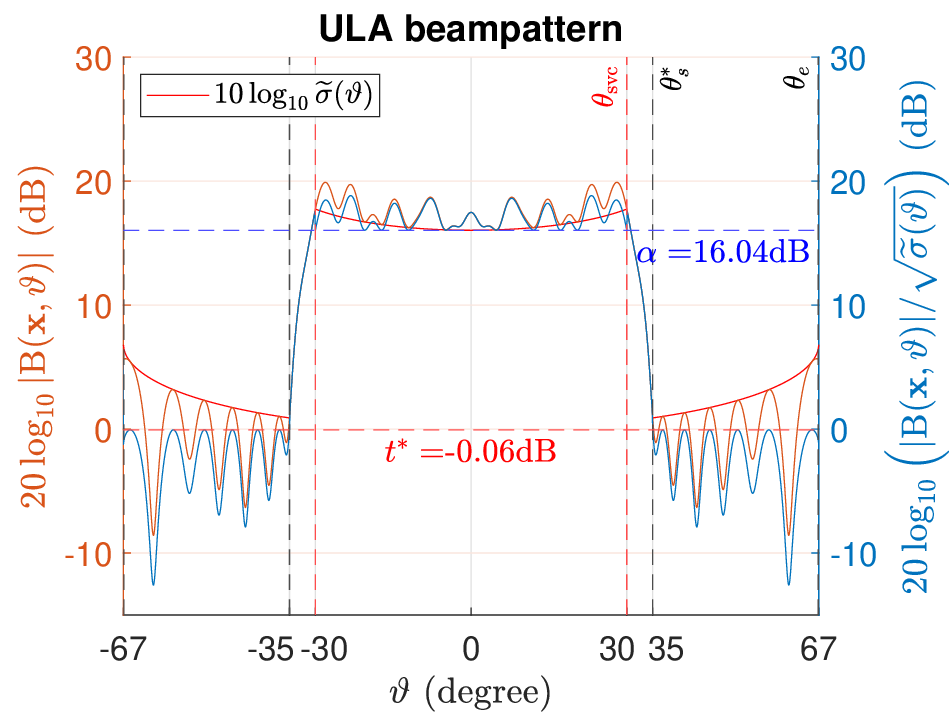}
    \label{fig:M32_BW60_CMC_figa}} 
    \par \vspace{-0.3cm}
    \subfloat[]{\includegraphics[width=.3\linewidth]{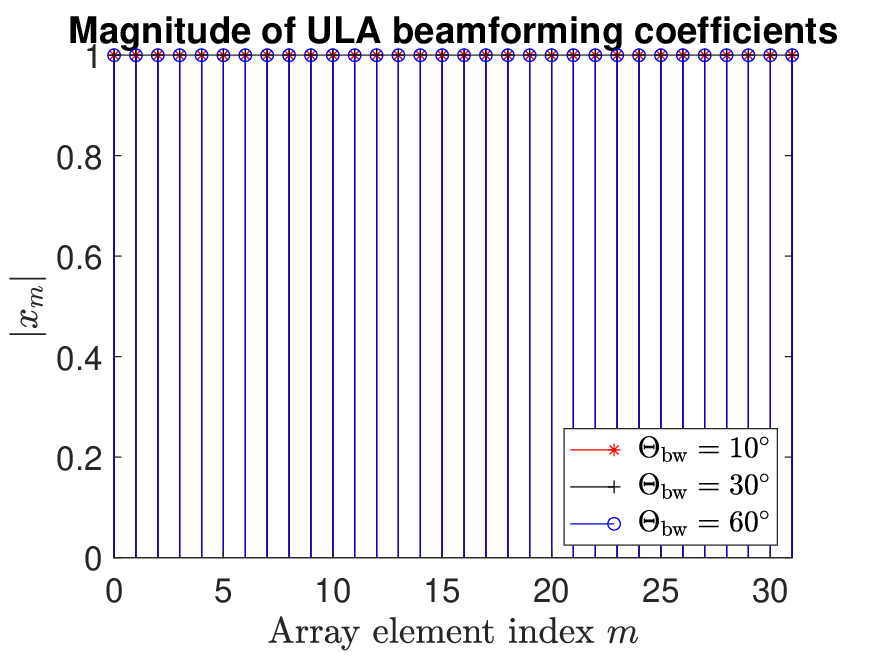}  
    \label{fig:M32_CMC}} 
    \subfloat[]{\includegraphics[width=.3\linewidth]{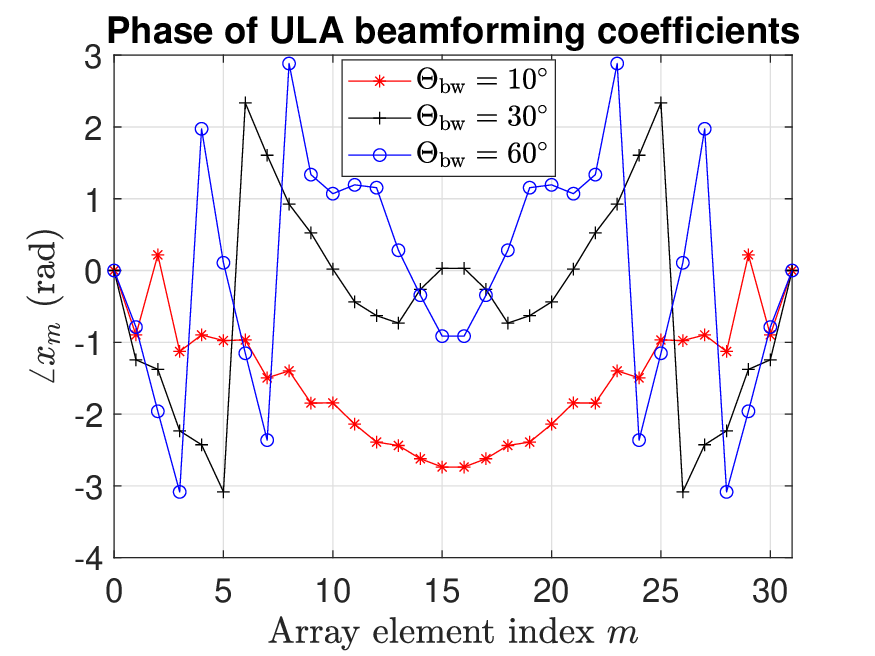}
    \label{fig:M32_phase}} 
    \caption{ ULA beampattern $|{\rm B}({\bf x}, \vartheta)|$ and pattern of $\frac{|{\rm B}({\bf x}, \vartheta)|}{\sqrt{{\widetilde \sigma}(\vartheta)}}$ obtained by Algorithm \ref{algorithm1} when beamwidths are (a) $10^{\circ}$, (b) $30^{\circ}$, and (c) $60^{\circ}$. (d) Magnitude of ULA beamforming coefficients $|x_m|$, and (e) Phase of ULA beamforming coefficients $\angle x_m$.} \label{fig:M32_ULA_BW10_30_60} 
\end{figure*}

\begin{figure}
    \centering  
    \includegraphics[width=3.5in]{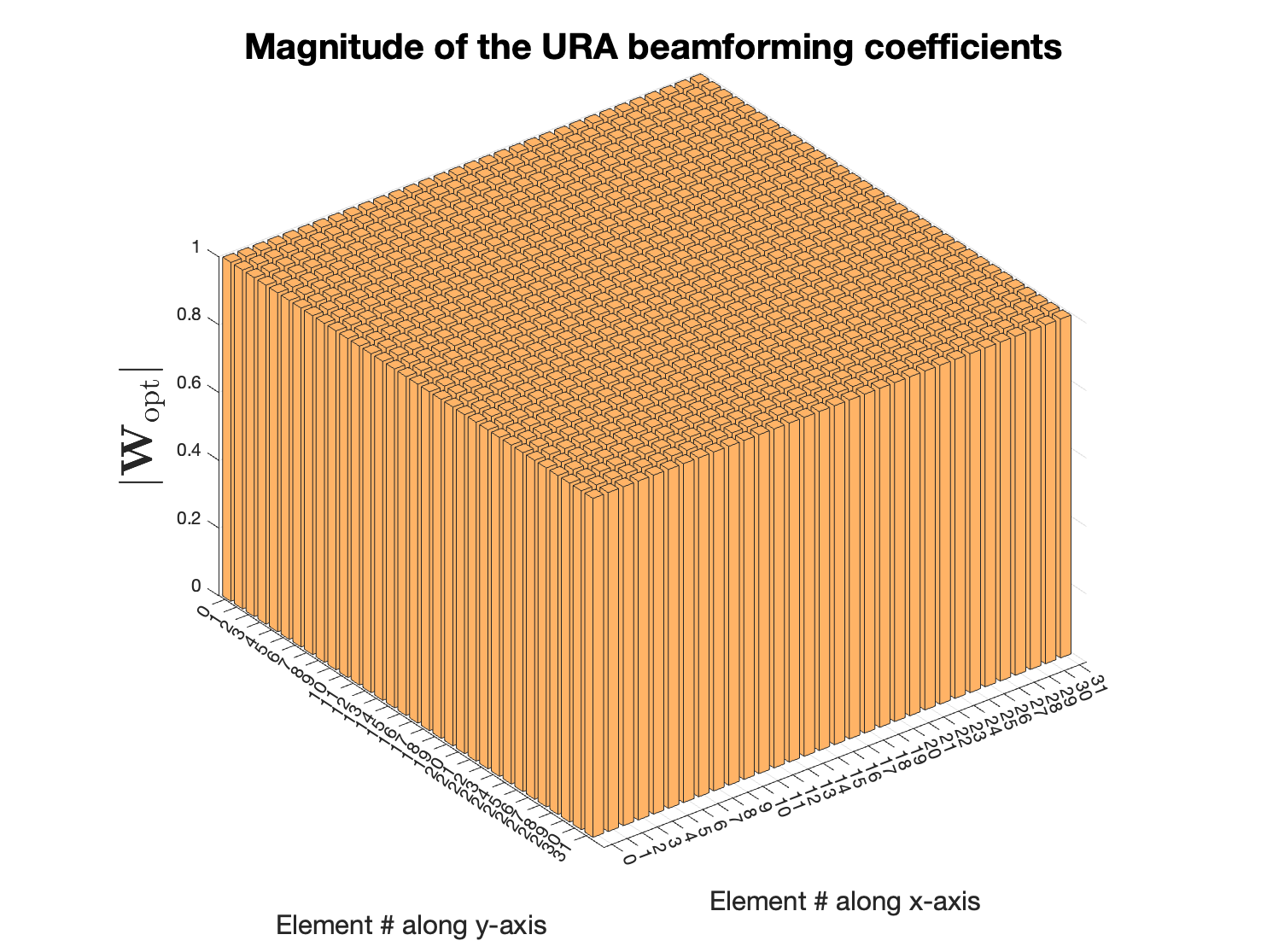} 
    \caption{Magnitude of the URA beamforming coefficients.}
    \label{fig:BF_weight}
\end{figure}

\begin{figure}
    \centering
    \subfloat[]{\includegraphics[width=1.7in]{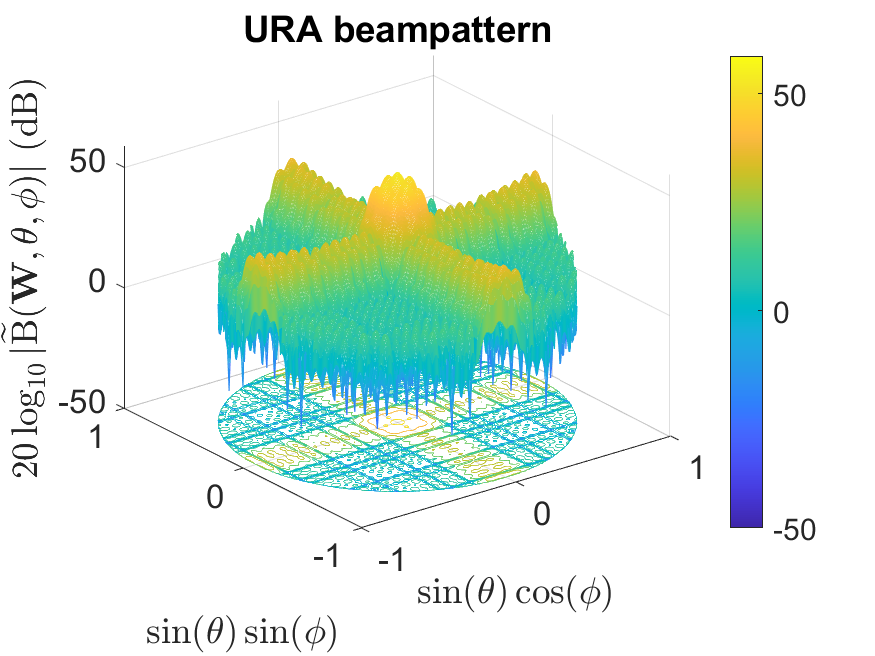} 
    \label{fig:M32_BW10_CMC_figc}} 
    \hspace{-0.5cm}  
    \subfloat[]{\includegraphics[width=1.7in]{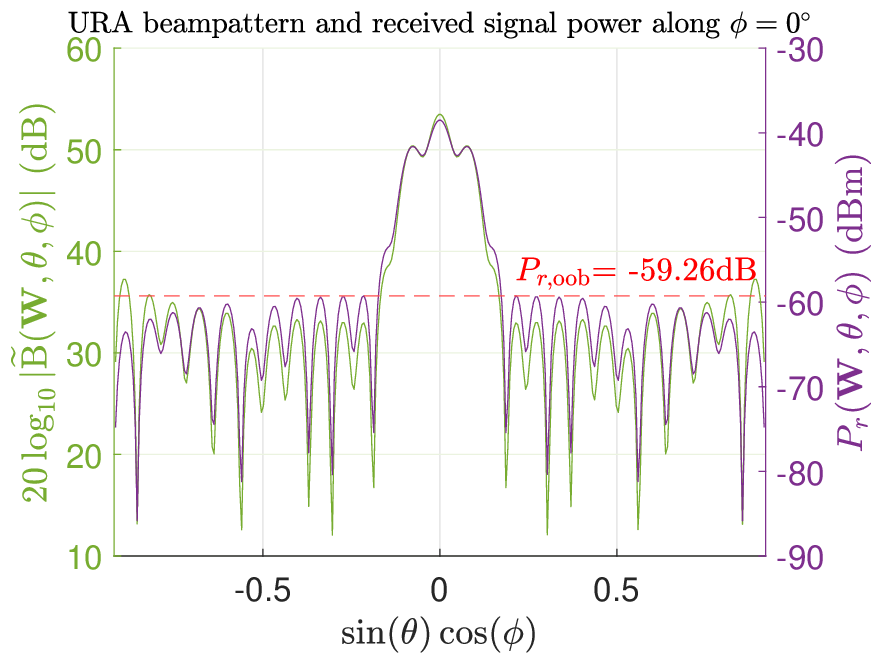}
    \label{fig:M32_BW10_CMC_figd}} 
    \par \vspace{-0.3cm}  
    \subfloat[]{\includegraphics[width=1.7in]{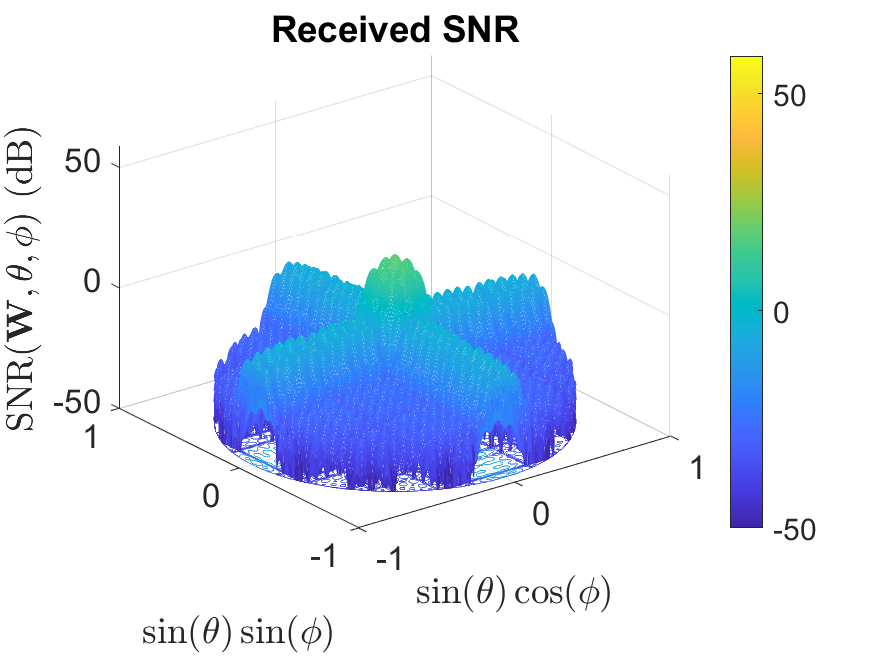}
    \label{fig:M32_BW10_CMC_fige}} 
    \hspace{-0.5cm} 
    \subfloat[]{\includegraphics[width=1.7in]{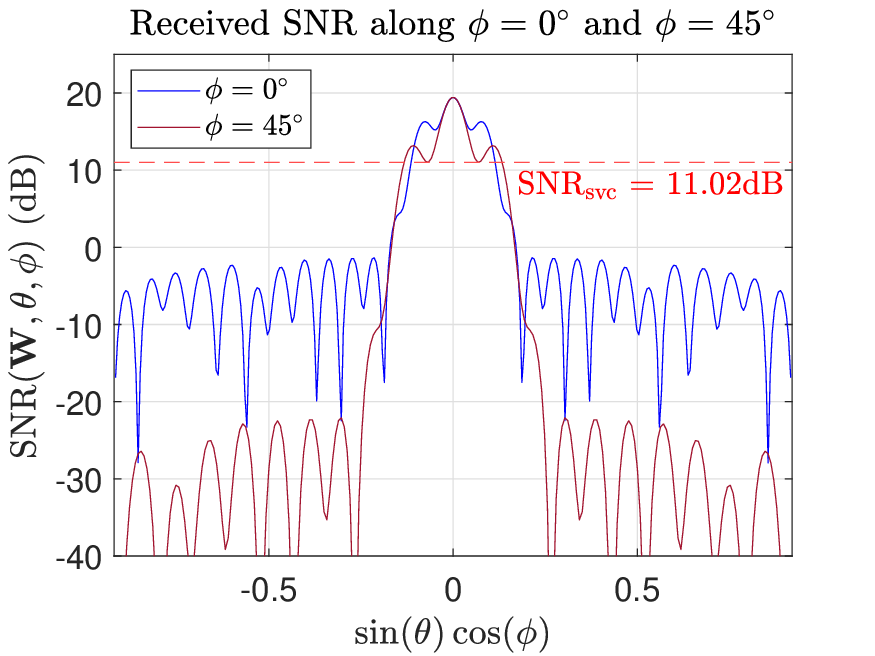}  
    \label{fig:M32_BW10_CMC_figf}} 
    \caption{ Proposed beamformer with $10^{\circ}$ beamwidth (a) URA beampattern, (b) URA beampattern and received signal power along \mbox{$\phi = 0^{\circ}$}, (c) Received SNR, and (d) Received SNR along \mbox{$\phi=0^{\circ}/45^{\circ}$}. }  \label{fig:M32_BW10_CMC}   
\end{figure}

\begin{figure}
    \centering  
    \subfloat[]{\includegraphics[width=1.7in]{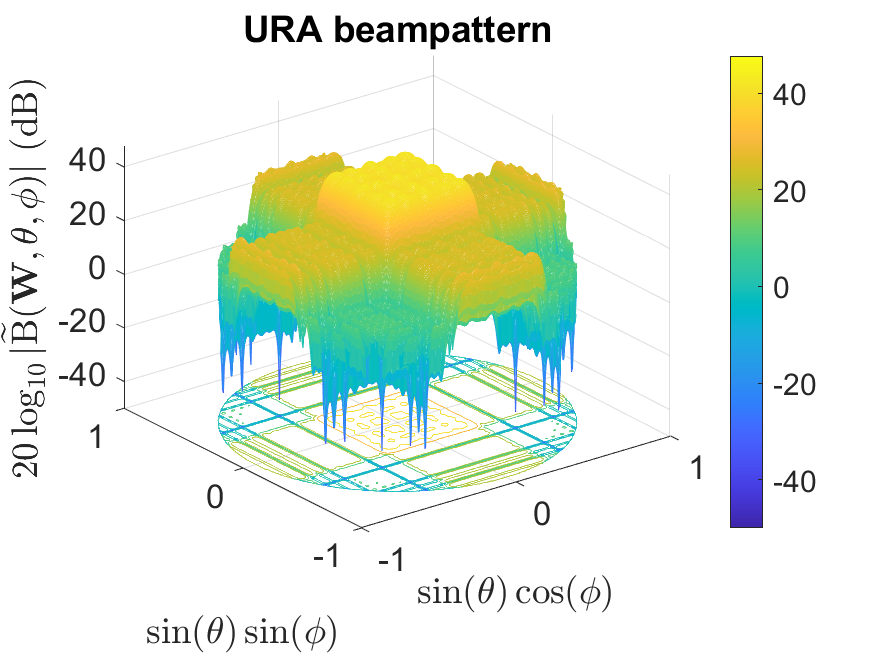} 
    \label{fig:M32_BW30_CMC_figc}} \hspace{-0.5cm} \hfill  
    \subfloat[]{\includegraphics[width=1.7in]{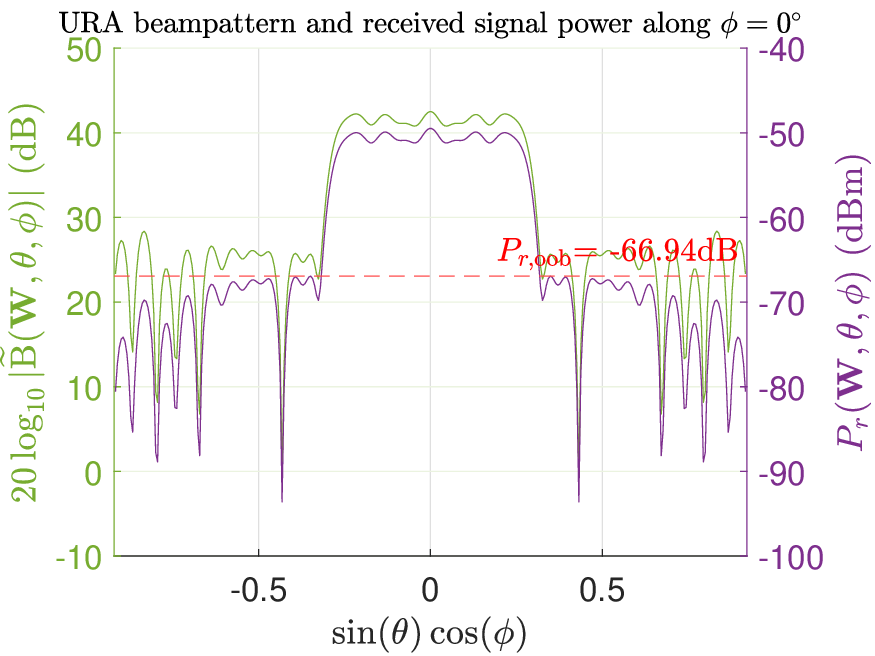}
    \label{fig:M32_BW30_CMC_figd}} \par
    \vspace{-0.3cm}
    \subfloat[]{\includegraphics[width=1.7in]{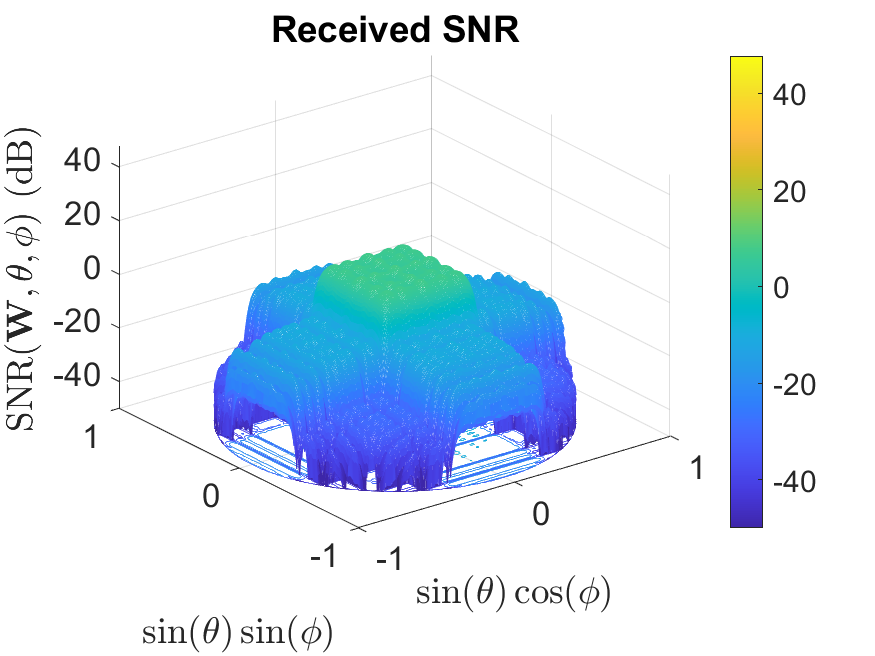}
    \label{fig:M32_BW30_CMC_fige}} \hspace{-0.5cm} \hfill
    \subfloat[]{\includegraphics[width=1.7in]{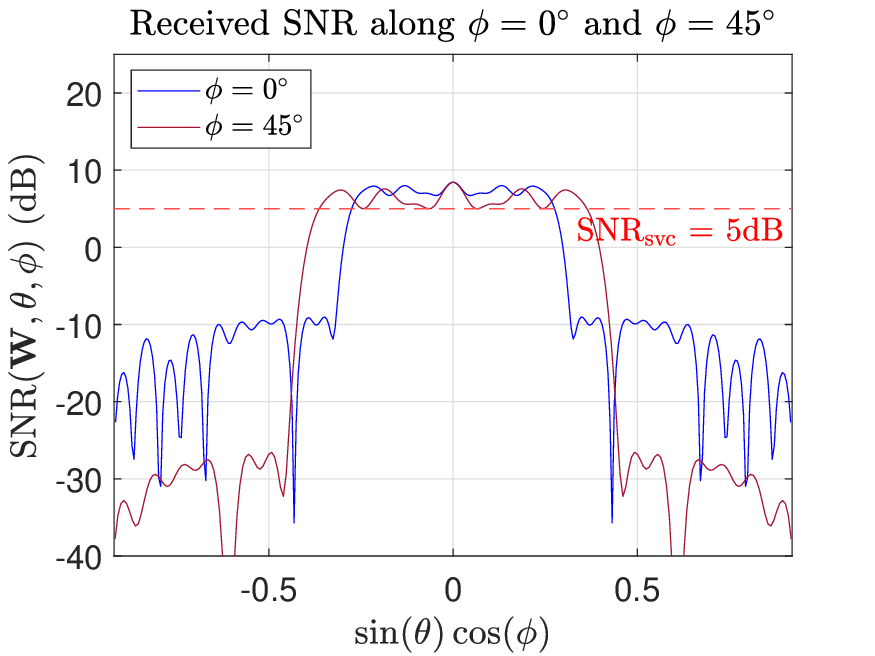} 
    \label{fig:M32_BW30_CMC_figf}} 
    \caption{ Proposed beamformer with $30^{\circ}$ beamwidth (a) URA beampattern, (b) URA beampattern and received signal power along \mbox{$\phi = 0^{\circ}$}, (c) Received SNR, and (d) Received SNR along \mbox{$\phi=0^{\circ}/45^{\circ}$}. } 
    \label{fig:M32_BW30_CMC}
\end{figure}

\begin{figure}
    \centering
    \subfloat[]{\includegraphics[width=1.7in]{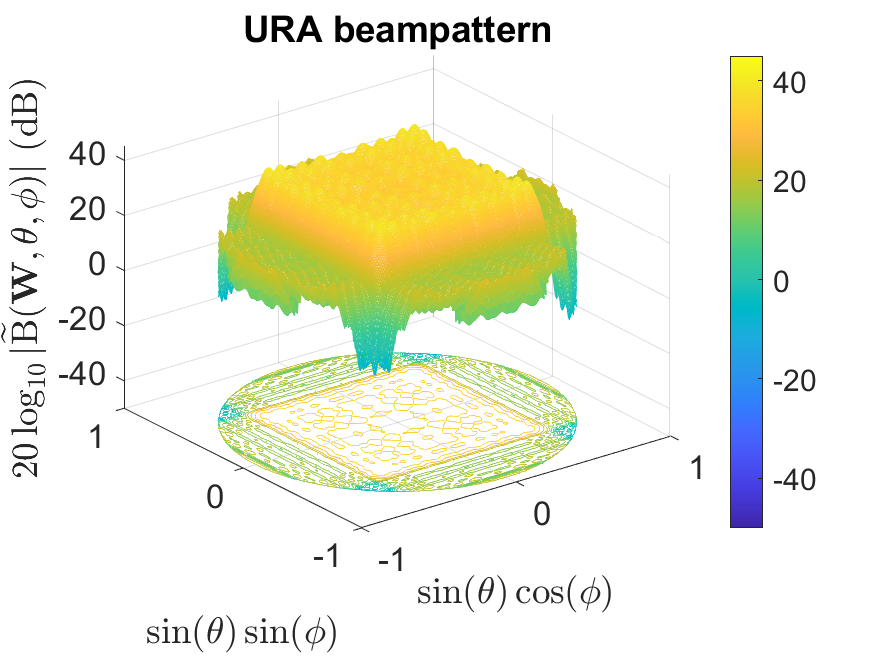} 
    \label{fig:M32_BW60_CMC_figc}} \hspace{-0.5cm} \hfill 
    \subfloat[]{\includegraphics[width=1.7in]{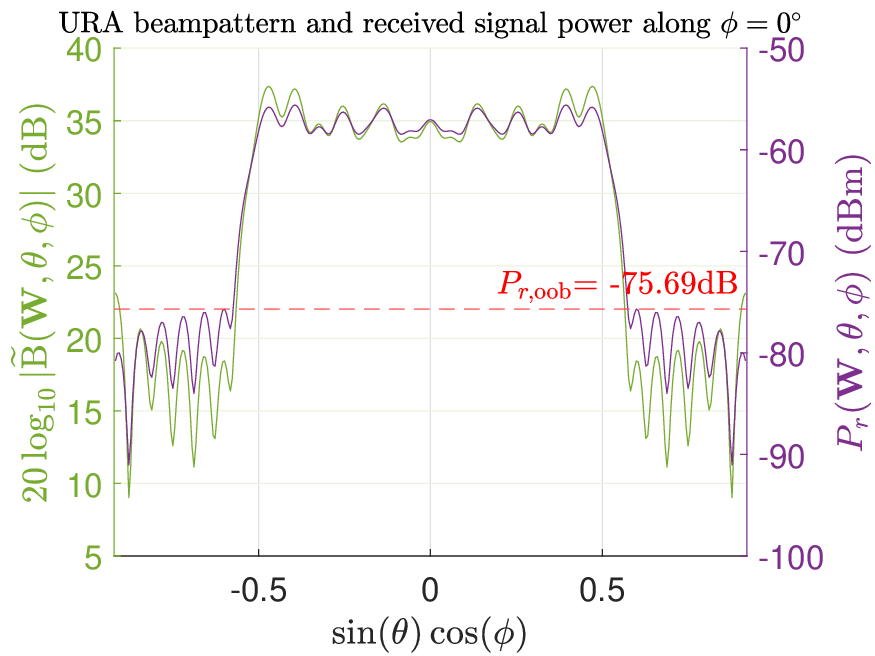}      
    \label{fig:M32_BW60_CMC_figd}} \par
    \vspace{-0.3cm}
    \subfloat[]{\includegraphics[width=1.7in]{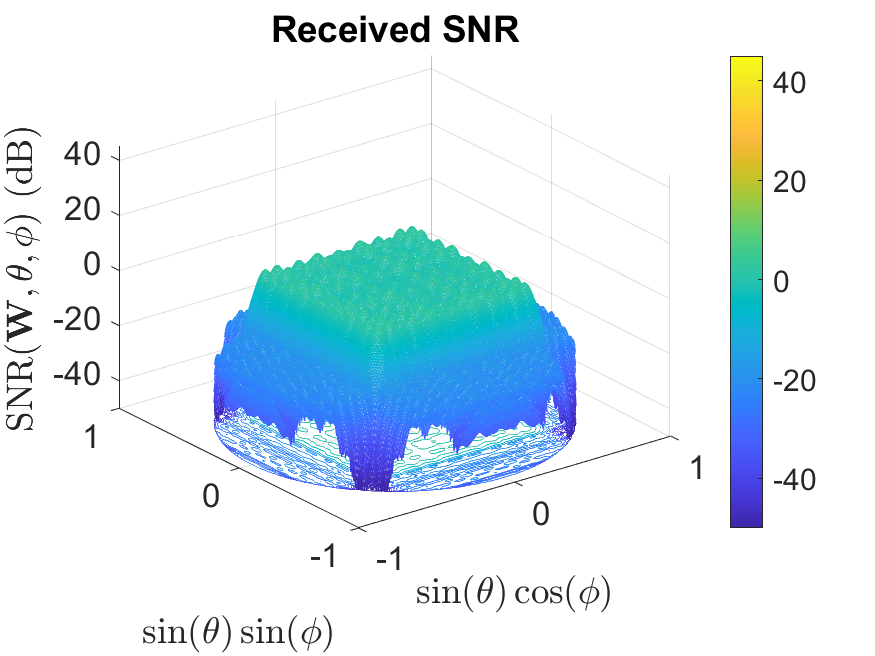}
    \label{fig:M32_BW60_CMC_fige}} \hspace{-0.5cm} \hfill 
    \subfloat[]{\includegraphics[width=1.7in]{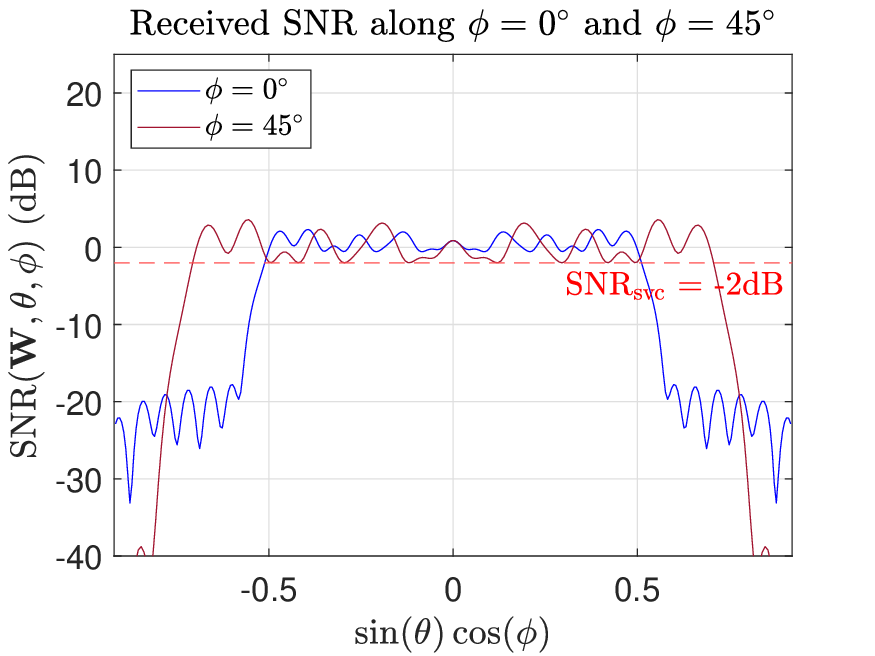}
    \label{fig:M32_BW60_CMC_figf}} 
    \caption{ Proposed beamformer with $60^{\circ}$ beamwidth (a) URA beampattern, (b) URA beampattern and received signal power along \mbox{$\phi = 0^{\circ}$}, (c) Received SNR, and (d) Received SNR along \mbox{$\phi=0^{\circ}/45^{\circ}$}. } 
    \label{fig:M32_BW60_CMC}
\end{figure}

In Fig. \ref{fig:M32_BW10_CMC_figa}, \ref{fig:M32_BW30_CMC_figa} and \ref{fig:M32_BW60_CMC_figa}, ULA beampatterns $|{\rm B}({\bf x}, \vartheta)|$ and patterns of $\frac{|{\rm B}({\bf x}, \vartheta)|}{\sqrt{{\widetilde \sigma}(\vartheta)}}$ are plotted in orange and blue line, respectively.
The ULA beampatterns $|{\rm B}({\bf x}, \vartheta)|$ feature the main lobe lower bound and sidelobe upper bound follow the shape of ${\widetilde \sigma}(\vartheta)$, defined in (\ref{ULA_isoflux_radiation_mask}), which is highlighted in red curves.
While, pattern of $\frac{|{\rm B}({\bf x}, \vartheta)|}{\sqrt{{\widetilde \sigma}(\vartheta)}}$ have a constant level of main lobe lower bound $\alpha$ and sidelobe upper bound $t^*$.
The $\rm{NPSL_{ULA}}$ defined in (\ref{def:NPSL_ULA}) can then be calculated by $20 \log_{10}\left(\frac{t^*}{\alpha}\right)$ listed in Table \ref{tb:Numerical results}.
In each case, the ULA beamforming coefficients meet the CMCs as shown in Fig. \ref{fig:M32_CMC} and the phase of the beamforming coefficients are depicted in Fig. \ref{fig:M32_phase}. 
Moreover, as illustrated in Fig. {\ref{fig:BF_weight}}, the magnitude of the URA beamforming coefficients composed by the ULA beamforming coefficients in Fig. {\ref{fig:M32_CMC}} by ({\ref{URA_weight_matrix}}) are constant at 1 which demonstrates that the CMCs were satisfied.
The URA beampatterns $|{\widetilde {\rm B}}({\bf W},\theta,\phi)|$ are generated by the ULA beampatterns using (\ref{eq:B_tilde}).
In Fig. \ref{fig:M32_BW10_CMC_figc}, \ref{fig:M32_BW30_CMC_figc} and \ref{fig:M32_BW60_CMC_figc}, the URA beampatterns are depicted in \mbox{$\sin(\theta)\cos(\phi)$-$\sin(\theta)\sin(\phi)$} plane.
Through our method, the URA beampattern main lobe region is square, and the PSL of the URA beampattern lies in its cross-region (i.e., when \mbox{$\phi=0^{\circ}/90^{\circ}$}) which verifies the inference in Fig. \ref{Fig:Visualize_URA_ULAs_BP_concept}.
To evaluate $P_{r, \rm{oob}}$ defined in (\ref{def:P_r_oob}), the received signal power $P_r({\bf W},\theta,\phi)$ at $\phi = 0^{\circ}$ for each case is plotted in Fig. \ref{fig:M32_BW10_CMC_figd}, \ref{fig:M32_BW30_CMC_figd} and \ref{fig:M32_BW60_CMC_figd}.
Low $P_{r, \rm{oob}}$ can be achieved to prevent interference with UTs in SAT out-of-beam areas.
Moreover, the received SNR of each case is calculated as (\ref{eq:SNR_r}), and depicted in Fig. \ref{fig:M32_BW10_CMC_fige}, \ref{fig:M32_BW30_CMC_fige} and \ref{fig:M32_BW60_CMC_fige}. 
The received SNR along \mbox{$\phi = 0^{\circ}/45^{\circ}$} are plotted in Fig. \ref{fig:M32_BW10_CMC_figf}, \ref{fig:M32_BW30_CMC_figf} and \ref{fig:M32_BW60_CMC_figf}.
We can observe that the SNR lower bound in SAT service areas is greater than or equal to $\text{SNR}_{\rm{min}}$ set in Table \ref{tb:Angular_parameters} in all cases.
The QoS is guaranteed because we consider the path loss variation from SAT to UT owing to the Earth's curvature when synthesizing beampatterns inspired by the isoflux radiation patterns in \cite{Reyna2012, Ibarra2015, Yoshimoto2019, Zeng2021_isoflux, Cai2023}.
In Table \ref{tb:Numerical results}, 
the performance evaluationsd on the metrics defined in Section \ref{subec:def_eval_metrics} are presented.

\begin{table}[H]
    \centering
    \begin{small}
    \caption{Performance evaluation of the proposed algorithm with the metrics defined in Section \ref{subec:def_eval_metrics}} \label{tb:Numerical results}  \vspace{-2mm} 
    \begin{tabular}{ |c|c|c|c|} 
    \hline
    \bf{Beamwidth $\Theta_{\rm{BW}}$} & {$10^{\circ}$} & {$30^{\circ}$} & $60^{\circ}$  \\ 
    \hline \hline
      $\rm{NPSL_{ULA}}$ $\rm{[dB]}$  & $-16.41$ & $-15.43$ &  $-16.1$  \\ 
    \hline
      $\rm{NPSL_{URA}}$ $\rm{[dB]}$ & $-7.39$ & $-14.59$ & $-16.39$ \\ 
    \hline
      $\text{SNR}_{\rm{svc}}$ ${\rm{[dB]}}$ & $11.02$ & $5$ & $-2$ \\ 
    \hline
      $P_{r, \rm{oob}}$ ${\rm{[dBm]}}$ & $-59.26$ & $-66.94$ & $-75.69$ \\ 
    \hline
      $\eta_{\rm{CMC}}$  & $1$ & $1$ & $1$ \\ 
    \hline
    \end{tabular}
    \end{small}
\end{table} 

\subsubsection{Comparison with Other Works}\label{sec:Comparison_with_Other_Works}
\begin{figure}
    \centering 
    \subfloat[]{\includegraphics[width=2.2in]{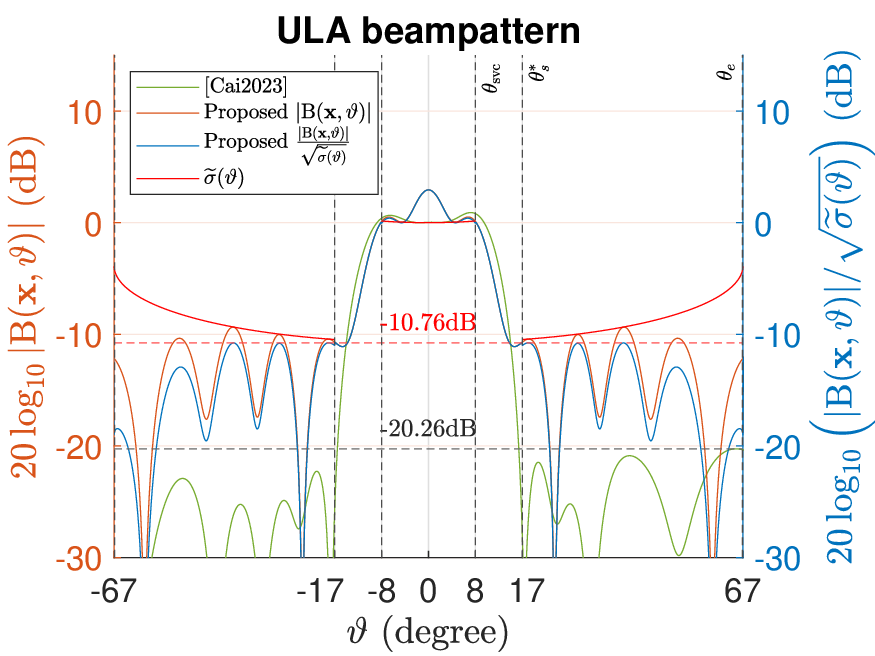} 
    \label{Fig:comparision_Cai2023_BP}}  
    \par \vspace{-0.3cm}
    \subfloat[]{\includegraphics[width=1.7in]{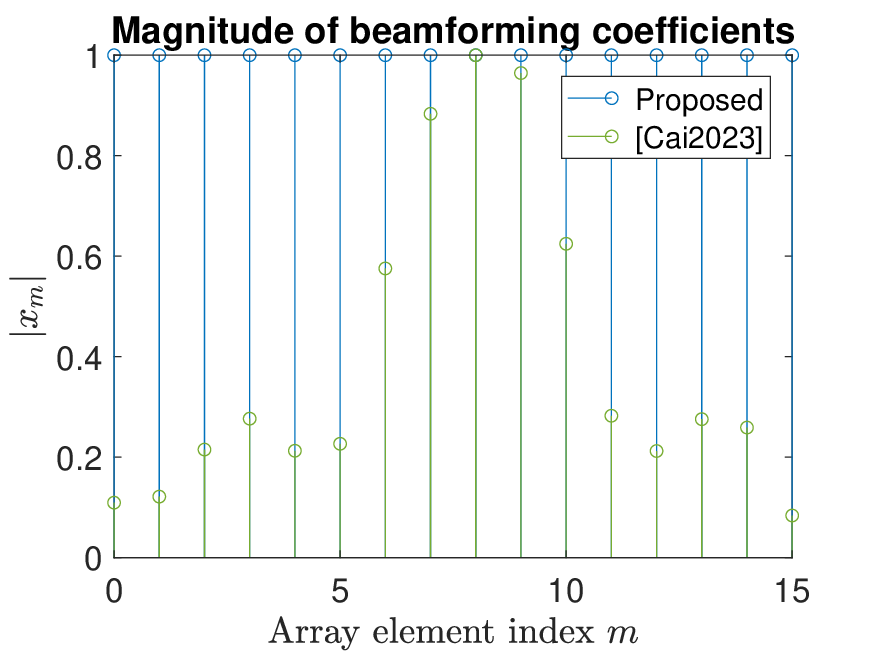}      
    \label{Fig:comparision_Cai2023_weight}} 
    \hspace{-0.5in} 
    \hfill 
    \subfloat[]{\includegraphics[width=1.7in]{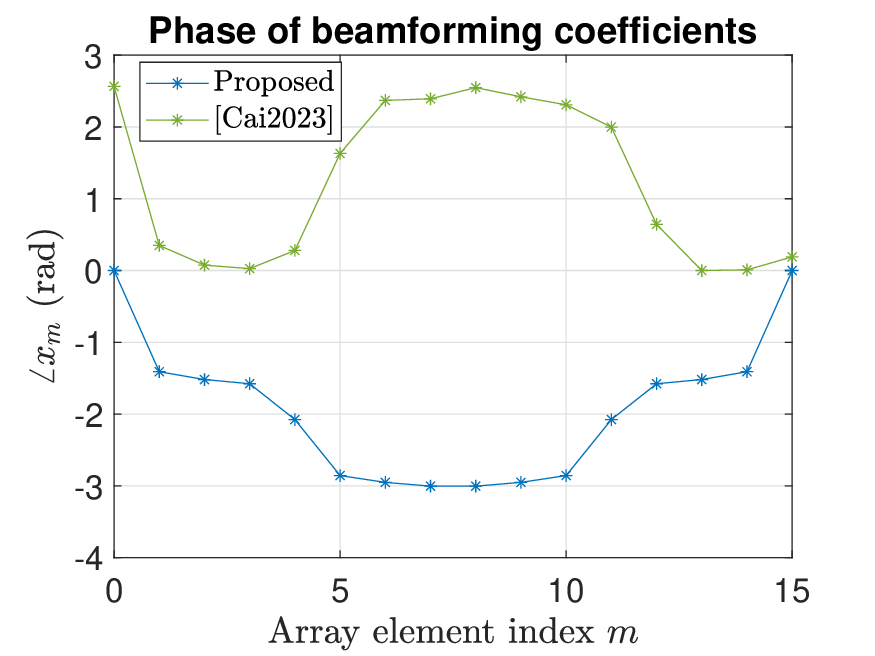}      
    \label{Fig:comparision_Cai2023_phase}}
    \caption{ Comparison of the proposed algorithm with [Cai2023] \cite{Cai2023} which considered the synthesis of isoflux radiation pattern but without CMCs (a) ULA beampattern, (b) Magnitude of beamforming coefficients, and (c) Phase of beamforming coefficients. } 
    \label{Fig:comparision_Cai2023}
\end{figure}

First, the proposed algorithm is compared with \cite{Reyna2012, Ibarra2015, Yoshimoto2019, Cai2023} which considers the synthesis of the isoflux radiation pattern without considering CMCs. 
The numerical results of each paper are listed in Table \ref{table:Comparision_isoflux}.
In \cite{Reyna2012, Ibarra2015, Yoshimoto2019, Cai2023}, evolutionary algorithms are applied and \mbox{$\eta_{\rm{CMC}} \neq 1$} as they did not consider CMCs. 
In Fig. \ref{Fig:comparision_Cai2023}, the results of \cite{Cai2023} are compared with those of the proposed algorithm. 
A ULA with \mbox{$M=10$} is considered.
Parameters of the the proposed algorithm are set as \mbox{$(\theta_{\rm{svc}}, \theta_s^*, \theta_e) = (8^{\circ}, 17^{\circ}, 67^{\circ})$}, \mbox{$\text{SNR}_{\rm{min}} = 0  \ [{\rm{dB}}]$} and other parameters remain the same in Section \ref{sec:parameter_settings}. 
In Fig. \ref{Fig:comparision_Cai2023_BP}, the ${\rm{NPSL_{ULA}}}$ obtained through \cite{Cai2023} and the proposed algorithm are $-20.26 \ [\rm{dB}]$ and $-10.76 \ [\rm{dB}]$, respectively.
As shown in Fig. \ref{Fig:comparision_Cai2023_weight}, the proposed algorithm achieves CMCs with $\eta_{\rm{CMC}} = 1$. 
However, \cite{Cai2023} did not achieve CMCs and its $\eta_{\rm{CMC}} = 11.95$.
As CMCs greatly reduce the degrees of freedom (DoFs), 
it is foreseeable that a lower $\rm{NPSL_{ULA}}$ can be achieved in \cite{Cai2023}.  
Although \cite{Cai2023} can attain lower ${\rm{NPSL_{ULA}}}$, its $\eta_{\rm{CMC}}$ is as high as $11.95$ which is not favorable for maximizing PAs efficiency and may not be practical for LEO SAT applications. 
One possible reason for not considering CMCs in \cite{Reyna2012, Ibarra2015, Yoshimoto2019, Cai2023} could be the difficulty of considering equality constraints in evolutionary algorithms \cite{BarkatUllah2012, Mengjun2021, Zhang2021}.

\begin{table*}[h!]
    \centering
    \begin{normalsize}
    \caption{Comparison of the synthesis of isoflux radiation pattern in \cite{Reyna2012, Ibarra2015, Yoshimoto2019, Cai2023} and the proposed algorithm} \label{table:Comparision_isoflux}   
    \vspace{-2mm}
    \begin{tabular}{ |c|c|c|c|c|} 
    \hline
    \textbf{Paper} & \textbf{Array type / $\#$ of array elements} & \textbf{Beamwidth $\Theta_{\rm{BW}}$} & $\eta_{\rm{CMC}}$ & ${{\rm{NPSL}}} \ [{\rm{dB}}]$ \\  
    \hline \hline
    [Cai2023] \cite{Cai2023}                & ULA/$16$ & $16^{\circ}$  & $11.95$ & $-20.26$ \\ 
    \hline
    [Yoshimoto2019] \cite{Yoshimoto2019}    & Non-uniformly spaced linear array/$12$ & $90^{\circ}$  & $7.91$ &  No sidelobe  \\ 
    \hline
    [Ibarra2015] \cite{Ibarra2015}          & Sparse concentric rings array/$6+8$ & $102.66^{\circ}$  & $66.31$ &  No sidelobe \\ 
    \hline
    [Reyna2012] \cite{Reyna2012}            & URA/$10 \times 10$ & $18^{\circ}$  & $549.5$ &  $-16.7$ \\ 
    \hline
    [Reyna2012] \cite{Reyna2012}            & Aperiodic planar array/$64$ & $18^{\circ}$  & $7.12$ &  $-17$ \\ 
    \hline
    Proposed algorithm                      & ULA/$16$ & $16^{\circ}$  & $1$ &  $-10.76$ \\ 
    \hline
    Proposed algorithm                      & URA/$32 \times 32$ & $10^{\circ}/30^{\circ}/60^{\circ}$  & $1/1/1$ &  $-7.39/-14.59/-16.39$ \\ 
    \hline
    \end{tabular}
    \end{normalsize}
\end{table*} 
\begin{figure}
    \centering
    \subfloat[]{\includegraphics[width=2.3in]{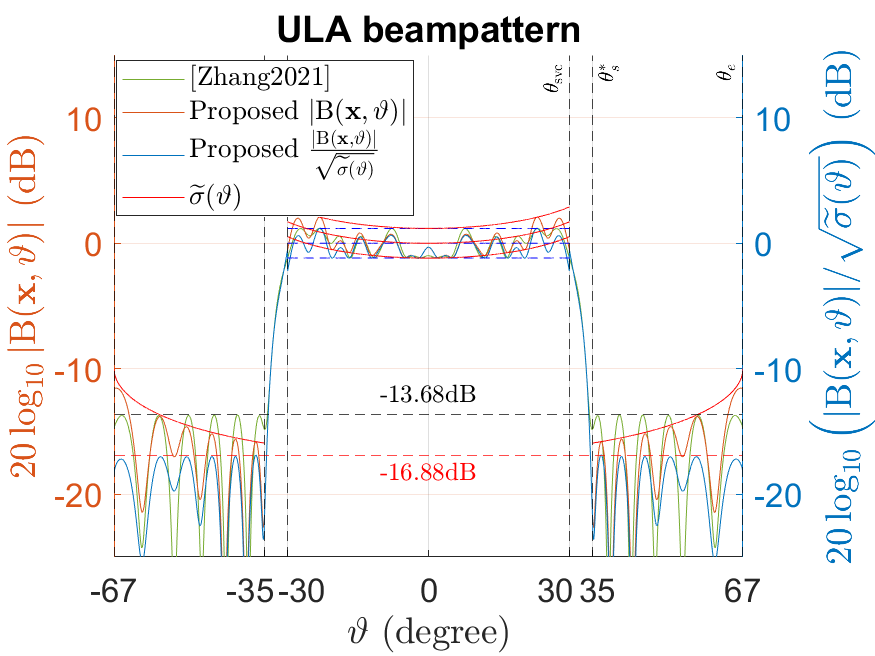} 
    \label{Fig:comparision_[Zeng2021]_BP}}  
    \par \vspace{-0.3cm}
    \subfloat[]{\includegraphics[width=1.7in]{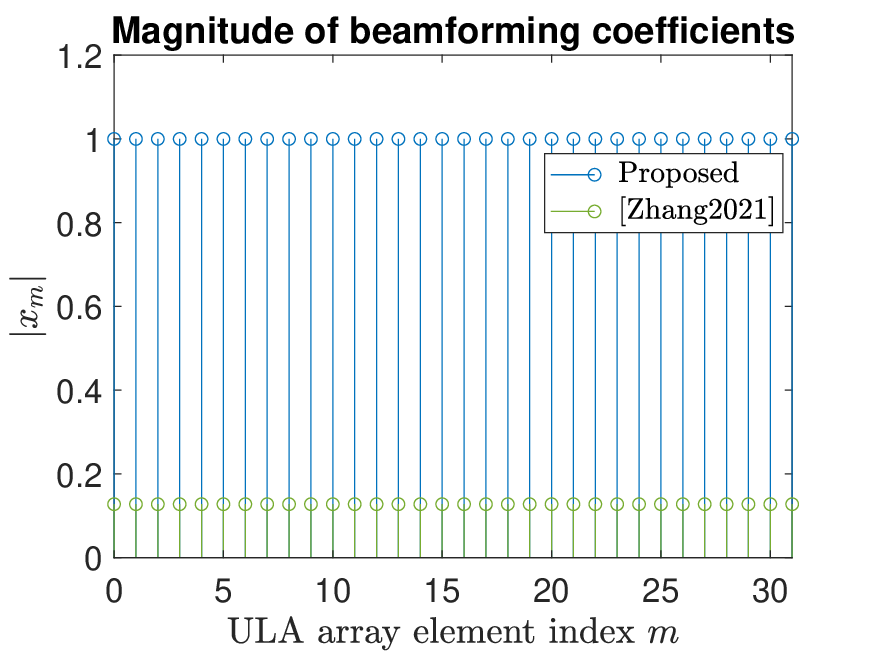}  
    \label{Fig:comparision_[Zeng2021]_weight}}  
    \hspace{-0.5in} \hfill 
    \subfloat[]{\includegraphics[width=1.7in]{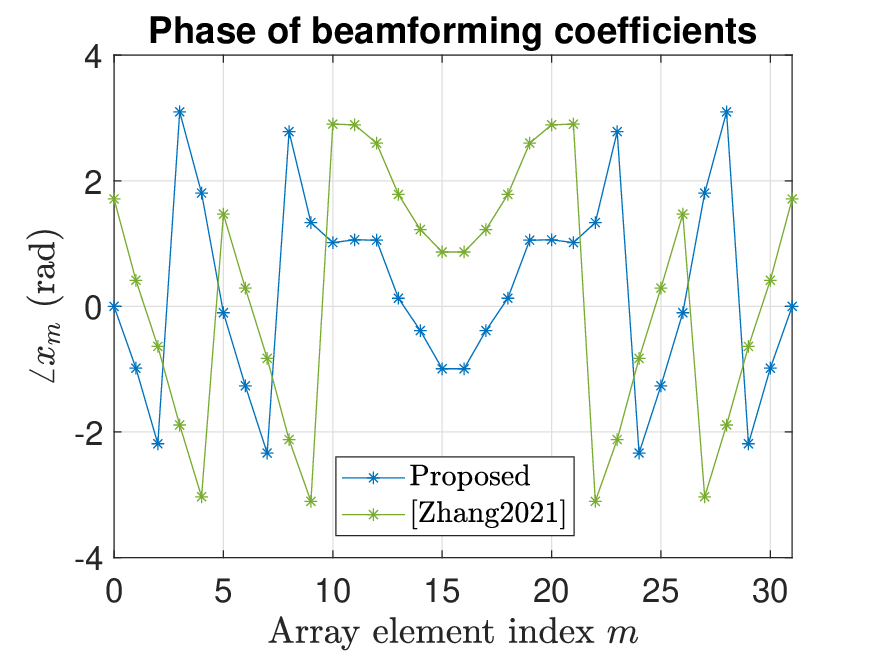}      
    \label{Fig:comparision_[Zeng2021]_phase}} 
    \caption{ Comparison of the proposed algorithm with [Zhang2021] \cite{Zhang2021_AWPL} which considered CMCs but without the synthesis of isoflux radiation pattern (a) ULA beampattern, (b) Magnitude of beamforming coefficients, and (c) Phase of beamforming coefficients. } \label{Fig:comparision_Zhang2021}
\end{figure}
In Fig. \ref{Fig:comparision_Zhang2021}, the proposed algorithm is compared with the ADMM-based algorithm in \cite{Zhang2021_AWPL}, which considered broadened beam ULA design with CMCs but without the synthesis of isoflux radiation pattern.
A ULA with \mbox{$M=32$} is considered.
The parameters of the proposed algorithm
are set as \mbox{$(\theta_{\rm{svc}}, \theta_s^*, \theta_e) = (30^{\circ}, 35^{\circ}, 67^{\circ})$}, \mbox{$\text{SNR}_{\rm{min}} = -1.3 \ [\rm{dB}]$} and other parameters remain the same in Section \ref{sec:parameter_settings}. 
The main lobe angle set of \cite{Zhang2021_AWPL} is set as \mbox{$[-30^{\circ},30^{\circ}]$} with mainlobe variation \mbox{$\pm 1.17 \ [\rm{dB}]$}, sidelobe angle set is \mbox{$[-90^{\circ}, -35^{\circ}] \cup [35^{\circ}, 90^{\circ}]$}, and penalty parameters are \mbox{$\rho_1=1$} and \mbox{$\rho_2=0.5$}. 
In addition, we apply \cite{Fonteneau2021_EuCNC} as the initial point of the algorithm in \cite{Zhang2021_AWPL} because the algorithm does not converge if a zero vector is chosen as the initial point proposed in \cite{Zhang2021_AWPL}.
In Fig. \ref{Fig:comparision_[Zeng2021]_BP}, 
our result features the beampattern main lobe lower bound and sidelobe upper bound shaped by the curve of ${\widetilde \sigma}(\vartheta)$.
While, \cite{Zhang2021_AWPL} has a constant level of the sidelobe upper bound and main lobe variation is limited to a given range.
We do not consider the main lobe upper bound constraint since only the main lobe lower bound should be constrained for QoS guarantee in our application. 
The \mbox{$\rm{NPSL_{ULA}}$} of \cite{Zhang2021_AWPL} and the proposed algorithm are \mbox{$-13.68 \ [\rm{dB}]$} and \mbox{$-16.88 \ [\rm{dB}]$}, respectively. (Since \cite{Zhang2021_AWPL} does not consider ${\widetilde {\sigma}}(\vartheta)$, we set \mbox{${\widetilde \sigma}(\vartheta)=1$} when evaluating its $\rm{NPSL_{ULA}}$.)
Both algorithms can achieve CMCs as shown in Fig. \ref{Fig:comparision_[Zeng2021]_weight},
and our algorithm outperforms \cite{Zhang2021_AWPL} in obtaining a lower $\rm{NPSL_{ULA}}$ with a margin of \mbox{$3.2 \ [\rm{dB}]$}.

\subsubsection{Performance impact of phase quantization}
\begin{figure}
    \centering  
    \includegraphics[width=3.2in]{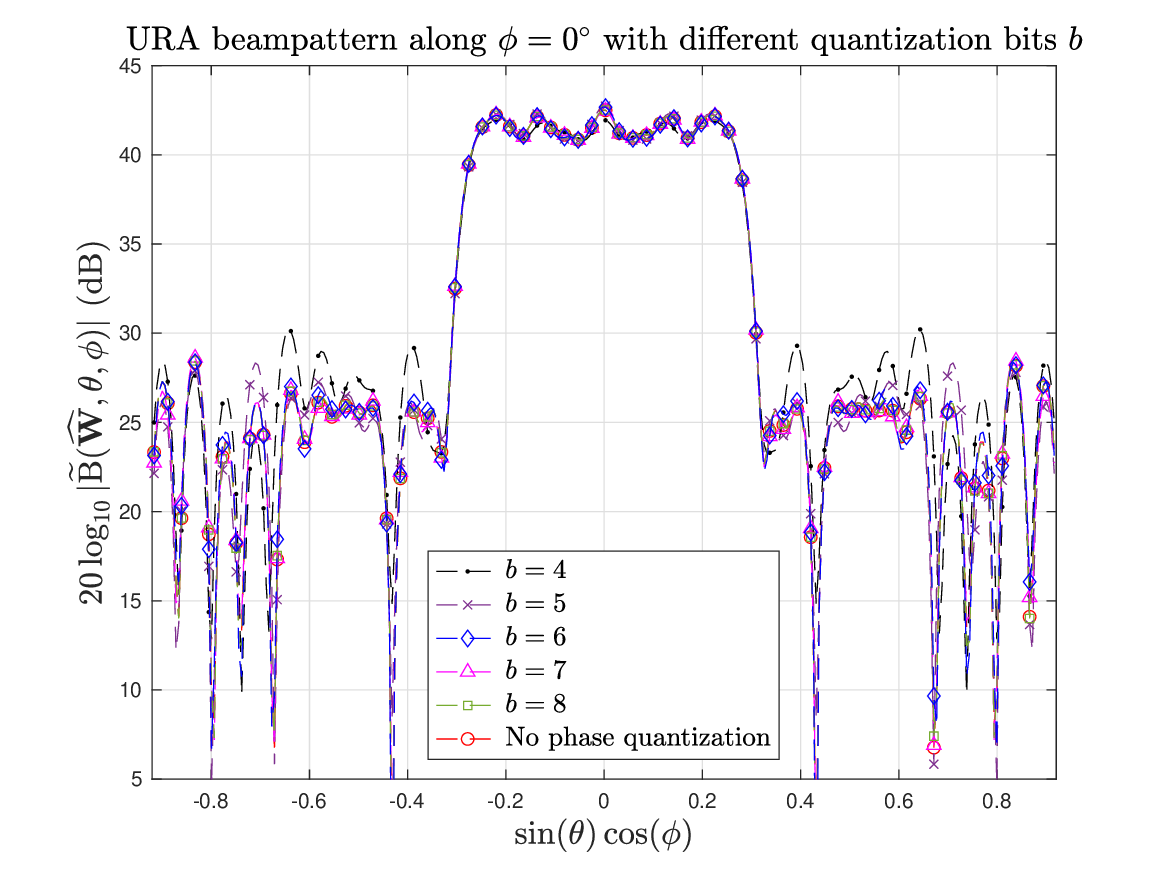}
    \caption{The URA beampattern with a beamwidth of $30^{\circ}$ along $\phi = 0^{\circ}$, considering different quantization bits $b$.} \label{fig:Phase_quantization_URA_BW30}
\end{figure}

In the subsubsection, we analyze the performance impact of using discrete phase shifters. 
In Section {\ref{sec:SAT_Tx}}, 
we assume the use of ideal phase shifters capable of continuous phase control as specified in ({\ref{eq:varphi}}).
However, phase shifters with a limited resolution are applied in practice \hbox{\cite{Chen2020,Di2020,Luo2019,Lin2017}} because phase shifters with arbitrary phases are challenging to implement \hbox{\cite{Chen2020}}.
Suppose that phase shifters with $2^b$ quantization levels are applied, where $b$ is the number of quantization bits.
Also, the discrete phase values are equally spaced over $[-\pi,\pi)$ with a quantization step: \hbox{\cite{Di2020}} 
\begin{equation} \label{def:quantization_step}
    \Delta=\frac{2\pi}{2^b}.
\end{equation}
The set of discrete phase shifts is defined as 
\begin{equation} \label{eq:set_of_discrete_phase_shifts}
\begin{aligned}
    \Xi = \{-\pi,-\pi+\Delta,-\pi+2\Delta,...,-\pi+(2^b-1)\Delta \},
\end{aligned}
\end{equation}
where $|\Xi| = 2^b$.
After obtaining the optimal URA beamforming coefficient matrix \mbox{${\bf W}_{\rm{opt}}$} from Algorithm {\ref{algorithm1}},
we have the phase of each antenna element as \mbox{$\varphi_{m, l}$}, where \mbox{$[{\bf W}_{\rm{opt}}]_{m,l}=e^{j{\varphi}_{m, l}}$}, \mbox{$\forall m \in {\mathbb Z}_{M_x}$}, \mbox{$\forall l \in {\mathbb Z}_{M_y}$}.
Then ${\varphi}_{m,l}$ is quantized to its nearest neighbor based on the closest Euclidean distance \hbox{\cite{Chen2020}}.
The quantized phases are
\begin{equation} \label{eq:phi_hat}
\begin{aligned}
    \hat{\varphi}_{m, l} = \mathop{\arg\min}_{\xi \in \Xi} \quad \big|\varphi_{m, l}-\xi \big|. 
\end{aligned}
\end{equation}
We then define the URA beamforming coefficients with phase quantization as ${\widehat{\bf W}}$, with its elements expressed as
\begin{equation} \label{eq:W_quantization}
\begin{aligned}
    \big[{\widehat{\bf W}}\big]_{m,l} = e^{j{\hat \varphi}_{m, l}}, \ \forall m \in {\mathbb Z}_{M_x}, \forall l \in {\mathbb Z}_{M_y}. 
\end{aligned}
\end{equation}
When ${\varphi}_{m, l}$ and ${\hat \varphi}_{m, l}$ are given as deterministic values, the average quantization error is evaluated as
\begin{equation} \label{def:quantization_error}
\begin{aligned}
   {\hat \varpi} = \frac{1}{M_xM_y}\sum_{m=0}^{M_x-1}\sum_{l=0}^{M_y-1}|[\widehat{{\bf W}}]_{m,l} - [{\bf W}_{\rm{opt}}]_{m,l}|^2.
\end{aligned}
\end{equation}
Moreover, quantization error can be evaluated statistically.
Let the quantization error of $\varphi_{m,l}$ be \mbox{$\epsilon_{m,l}=\hat{\varphi}_{m,l}-\varphi_{m,l}$} \hbox{\cite{Luo2019,Lin2017}} which is zero mean and uniformly distributed over the interval 
$[\frac{-\Delta}{2}, \frac{\Delta}{2}]$, 
where $\Delta$ is defined in ({\ref{def:quantization_step}}).
The mean squared quantization error (MSQE) is defined as \hbox{\cite{Luo2019,Lin2017}}
\begin{equation} \label{def:MSQE}
\begin{aligned}
    \varpi 
    =\frac{1}{M_xM_y}\sum_{m=0}^{M_x-1}\sum_{l=0}^{M_y-1}\mathbb{E}\left\{\left|[\widehat{{\bf W}}]_{m,l} - [{\bf W}_{\rm{opt}}]_{m,l}\right|^2\right\}
    \approx \frac{\Delta}{12},
\end{aligned}
\end{equation}
where $\mathbb{E}\{\cdot\}$ is the expectation operation.
The justification of ({\ref{def:MSQE}}) is shown in Appendix {\ref{appendix:justification_varpi}}.
To demonstrate the effect of phase quantization, 
we take the proposed beamformer with \mbox{$\Theta_{\rm{BW}}=30^{\circ}$}as an example and apply the same settings as in Section {\ref{sec:parameter_settings}}. 
After obtaining \mbox{${\bf W}_{\rm{opt}}$} from Algorithm {\ref{algorithm1}}, beamforming coefficients with differentquantization bits \mbox{$b=4,5,6,7,8$} are constructed through ({\ref{def:quantization_step}}) to ({\ref{eq:W_quantization}}). 
The resulting URA beampatterns along \mbox{$\phi = 0^{\circ}$} are plotted in Fig. {\ref{fig:Phase_quantization_URA_BW30}}.
The red curve with ``$\circ$" marker in Fig. {\ref{fig:Phase_quantization_URA_BW30}} corresponds to the coefficients without quantization which matches the green curve in Fig. {\ref{fig:M32_BW30_CMC_figd}}.
We can observe that in the cases of $b=6,7,8$, the beampatterns nearly overlap with the beampattern without phase quantization.
Also, as $b$ decreases, the sidelobes become more irregular. 
the average quantization errors ${\hat \varpi}$ for \mbox{$b=4,5,6,7,8$} evaluated by ({\ref{def:quantization_error}}) 
are $1.37\%$, $0.36\%$, $0.07\%$, $0.02\%$, $0.004\%$, respectively. 
Furthermore, the MSQEs $\varpi$ evaluated by ({\ref{def:MSQE}}) are $1.29\%$, $0.32\%$, $0.08\%$, $0.02\%$, $0.005\%$, respectively.
As $b$ increases, the quantization error decreases.
Given that the quantization errors evaluated by ${\hat \varpi}$ and $\varpi$ show only a small margin of difference, the use of ${\hat \varpi}$ for evaluation is thus justified.
To strike a balance between performance and hardware complexity, the number of quantization bits $b$ can be selected accordingly.
For instance, phase shifters with 6 quantization bits have been designed for the Ku-band applications \hbox{\cite{Duan2018, Wu2020}}.

\subsubsection{URA with subarray partition}
\begin{figure}[h!]
    \begin{center}
    \includegraphics[width = 3.5in]
    {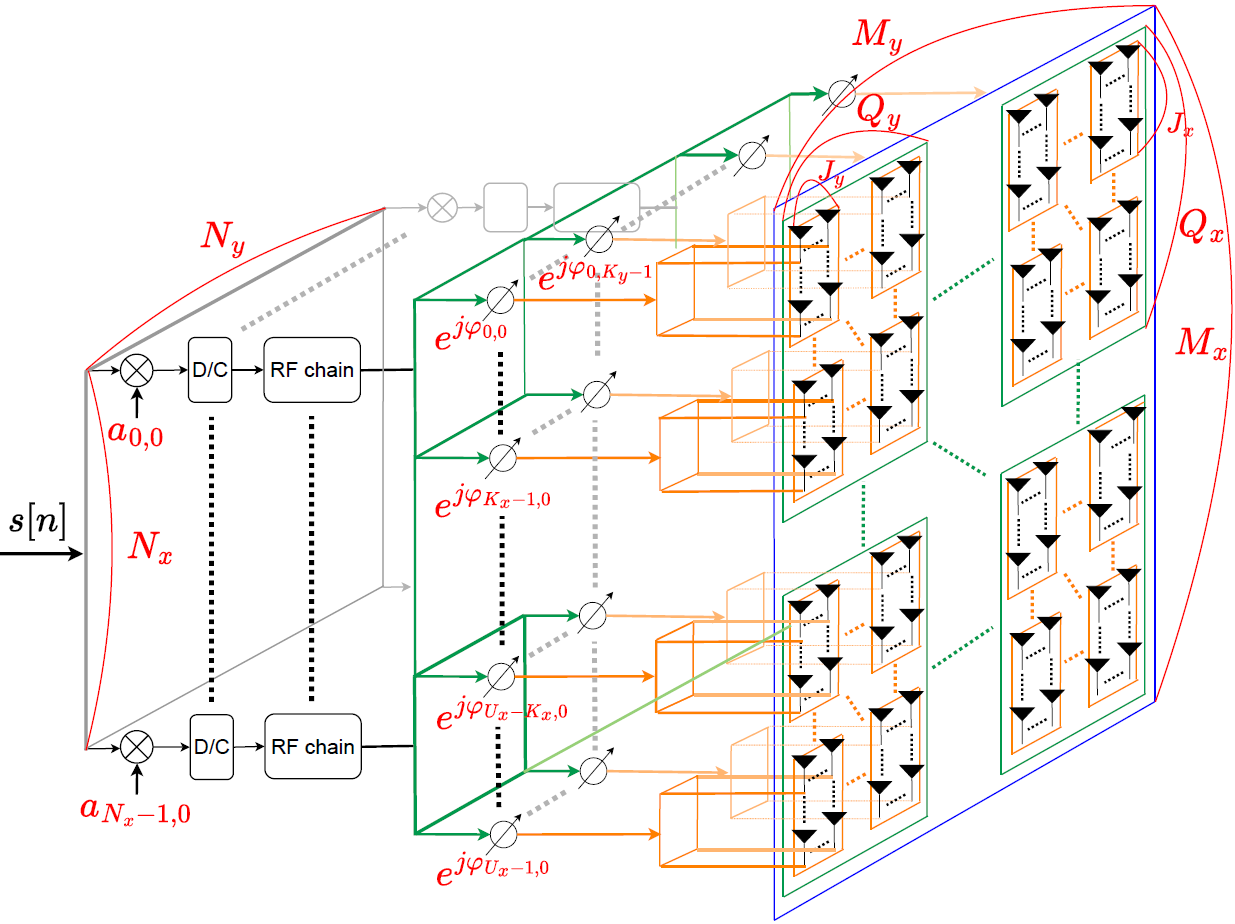}\\
    \caption{SAT transmitter system model with subarray partition.
    } \label{fig:HB_Tx_subarray}
    \end{center}
\end{figure}

\begin{figure}
    \centering  
    \includegraphics[width=3in]{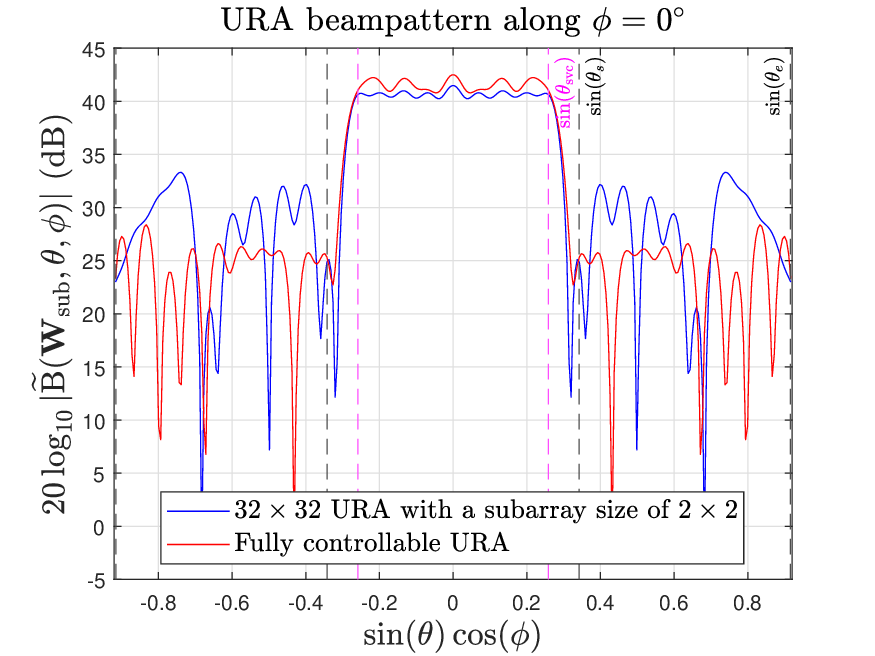} 
    \caption{The URA beampattern with a beamwidth of $30^{\circ}$ along $\phi = 0^{\circ}$,  
    considering a \mbox{$32 \times 32$} URA with a subarray size of \mbox{$2\times 2$} (i.e., $M_x=M_y=32$ and $J_x=J_y=2$).} 
    \label{fig:URABP_2by2_subarray}
\end{figure}

In the subsubsection, we discuss the possibility of reducing the feeding network complexity in terms of reducing the number of phase shifters. 
In Fig. {\ref{fig:HB_Tx_circuit_diagram}}, the number of RF chains is reduced by using hybrid beamforming; however, the required number of phase shifters is equal to the number of antenna elements.
In \hbox{\cite{Dicandia2024,Li2019,Mohammed2024,Shi2021,Abdulqader2020,Mohammed2020}},
subarray partition designs and clustered arrays were investigated to reduce the number of required phase shifters.
These approaches partition the antenna elements into smaller groups, keeping element excitations within each group identical.
To consider such a subarray parition, the transmitter system model in Fig. {\ref{fig:HB_Tx_circuit_diagram}} is generalized into that in Fig. {\ref{fig:HB_Tx_subarray}}.
The URA, with a size of \mbox{$M_x \times M_y$}, is partitioned into subarrays of size \mbox{$J_x \times J_y$}, where each subarray shares the same phase shifter.
Now, each of the $N_xN_y$ RF chains is connected to $K_xK_y$ phase shifters (where we define \mbox{$K_x=\frac{Q_x}{J_x}$} and \mbox{$K_y=\frac{Q_y}{J_y}$}).
For each RF chains, the number of phase shifters it is connected to is reduced from $Q_xQ_y$ to $K_x K_y$.
Then, the total number of required phase shifters is reduced from \mbox{$M_x M_y$} in Fig. {\ref{fig:HB_Tx_circuit_diagram}} to \mbox{$U_x U_y$} in Fig {\ref{fig:HB_Tx_subarray}} (note that  $U_xU_y=N_xN_yK_xK_y$). 
Since the element excitations are kept the same in each subarray,
we introduce ${\bf F}_x$ and ${\bf F}_y$ to map the element excitations to each subarray:
\begin{equation} 
    {\bf F}_x={\bf I}_{U_x} \otimes {\bf 1}_{J_x} \in \mathbb{R}^{M_x\times U_x}, \ {\bf F}_y={\bf I}_{U_y} \otimes {\bf 1}_{J_y} \in \mathbb{R}^{M_y\times U_y},
\end{equation}
where ${\bf I}_{U_i}$ are the identity matrices of size \mbox{$U_i\times U_i$}, and ${\bf 1}_{J_i}$ is the all one vector of size \mbox{$J_i\times 1$}, for $i=x,y$.
Additionally, we introduce the variables \mbox{${\bf g}_x \in \mathbb{C}^{U_x}$} and \mbox{${\bf g}_y \in \mathbb{C}^{U_y}$}, and incorporate the constraints, \mbox{${\bf x}={\bf F}_x{\bf g}_x$} and \mbox{${\bf y}={\bf F}_y{\bf g}_y$}, into the problem ({\ref{P1_2}}).
By applying Algorithm {\ref{algorithm1}} to the modified problem, ${\bf g}_{x,\rm{opt}}$ and ${\bf g}_{y,\rm{opt}}$ are obtained.
The URA beamforming coefficient matrix with the subarray partition is then computed as \mbox{${\bf W}_{\rm{sub}}=({\bf F}_x{\bf g}_{x,\rm{opt}})({\bf F}_y{\bf g}_{y,\rm{opt}})^T$}.
We consider a \mbox{$32 \times 32$} URA with a subarray size of \mbox{$2\times 2$}, i.e., $M_x=M_y=32$ and $J_x=J_y=2$.
The resulting beampattern is depicted by the blue curve in Fig. {\ref{fig:URABP_2by2_subarray}}.
The red curve in Fig. {\ref{fig:URABP_2by2_subarray}} matches the green curve in Fig. {\ref{fig:M32_BW30_CMC_figd}}, corresponding to the transmitter model in Fig. {\ref{fig:HB_Tx_circuit_diagram}}
(for brevity, we refer to this configuration as the fully controllable URA, where each antenna element in the URA is connected to a phase shifter).
We can observe that the PSL performance of the URA with subarray size of \mbox{$2\times 2$} is worse than that of the fully controllable URA by about 5 dB.
Although the PSL performance is degraded, subarray partitioning results in a $75\%$ reduction in the number of required phase shifters, decreasing from $1024$ to $256$.
We believe the main reason for the PSL degradation is that the degrees of freedom (DoFs) in applying a \mbox{$2\times 2$} subarray partition is less than that of the fully controllable URA.
Moreover, the URA design problem in our work is decomposed into two ULA design subproblems, resulting in the DoFs of only $M_x$ (or $M_y$) which is much less than the DoFs of \mbox{$M_x\times M_y$} if all element excitations can be designed.
In summary, the above results show that while it is possible to extend the proposed algorithm to a configuration with a reduced number of phase shifters, the resulting coefficients can suffer from great performance degradation due to the lack of DoFs.

\subsection{Channel Capacity Evaluation} \label{subsec:Capacity_evaluation}
To demonstrate that the proposed method is applicable for LEO broadcasting applications such as digital video broadcasting (DVB),
we compare the capacity of the proposed ``broadened-beam" beamformers with that of beamformers utilizing an array steering vector \cite{Tang2021, Hu2020, Lei2020}. 
For convenience, we refer to beamformers utilizing an array steering vector as ``narrow-beam" beamformers.
If beamforming is realized by adjusting the transmitted signal phase according to the array steering vector, the beamformer 3dB beamwidth can be evaluated as  \cite{VanTrees2002} 
\begin{equation} \label{def:3dB_BW}
    \Theta^{\rm 3dB}  = \sin^{-1} \left(\frac{0.891 \lambda_c}{d_a}\right) \ [\rm{deg}],
\end{equation}
where $\lambda_c$ is the signal wavelength, \mbox{$d_a = (M-1)d_e$} is the array length and $d_e$ is the antenna element spacing.
For a URA with size \mbox{$32 \times 32$} (i.e., \mbox{$M=32$}) and element spacing \mbox{$d_e=\frac{\lambda_c}{2}$}, we have \mbox{$\Theta^{\rm 3dB} = 3.3^{\circ}$}. 
Let ${\bf W}_{\rm b, \Theta_{\rm BW}}$ be the beamforming coefficient matrix of ``broadened-beam" beamformer with service beamwidth $\Theta_{\rm BW}$ obtained by Algorithm \ref{algorithm1}.  
Also, the beamforming coefficient matrix of ``narrow-beam" beamformer is denoted as ${\bf W}_{\rm n}$, 
and we set \mbox{$[{\bf W}_{\rm n}]_{m,l} = |[{\bf W}_{\rm b, \Theta_{\rm BW}}]_{m,l}| = 1, \ \forall m,l \in \mathbb{Z}_M$} (due to the CMCs) to compare channel capacity with ``broadened-beam" beamformer under the same total beamforming weight magnitude values. 
The beampattern value of ``narrow-beam" beamformer at its main lobe direction \mbox{$(\theta,\phi)=(0,0)$} is 
\begin{align} \label{eq:P_peak}
    D_{\rm n, ML} 
    = |\widetilde{\rm B}({\bf W}_{\rm n},0,0)|^2 
    = M^4,
\end{align}
where $\widetilde{\rm B}({\bf W}_{\rm n},\theta,\phi)$ is defined as (\ref{def:URA_BP}).
Let the main lobe region of ``narrow-beam" beamformer be defined within its 3dB beamwidth $\Theta^{\rm 3dB}$, we have

\vspace{-4mm}
\begin{small}
\begin{align} \label{B_n_SNRlb}
|\widetilde{\rm B}({\bf W}_{\rm n},\theta,\phi)|^2 \geq 0.5 D_{\rm n, ML}, \forall \theta \in [0, \frac{\Theta^{\rm 3dB}}{2}], \forall \phi \in [0, 2\pi].
\end{align}
\end{small}

\noindent
According to (\ref{eq:SNR_r}), 
the minimum received SNR that ``narrow-beam" beamformer can provide within its 3dB beamwidth is defined as
\begin{align} \label{eq:SNR_b_n}
    \text{SNR}_{\rm n} = \frac{\beta G_R P_s (0.5 D_{\rm n, ML} )}{\sigma^2(\theta) L_0 k T_{\rm sys} f_{\rm BW}}.  
\end{align}  
Also, $\text{SNR}_{\rm b, \Theta_{\rm BW}}$ is denoted as the minimum received SNR that ``broadened-beam" beamformer can provide within its service beamwidth $\Theta_{\rm BW}$.  
The channel capacity of ``broadened-beam" and ``narrow-beam" beamformers that can provide within their service beamwidths are evaluated based on Shannon’s capacity \cite{shannon}:
\begin{align} \label{eq:C_b}
    C_{\rm b, \Theta_{\rm BW}}= f_{\rm BW}\log_2(1+\text{SNR}_{\rm b, \Theta_{\rm BW}}),
\end{align}
\begin{align} \label{eq:C_n}
    C_{\rm n}= f_{\rm BW}\log_2(1+\text{SNR}_{\rm n}).
\end{align}  
Moreover, the beam service areas of the ``broadened-beam" beamformer with beamwidth $\Theta_{\rm BW}$ is denoted as $A_{\rm b, \Theta_{\rm BW}}$.
The beam service areas of the ``narrow-beam" beamformer with 3dB beamwidth under beam steering angle is denoted as $A_{{\rm n}, \Theta_{{\theta_T}}^{\rm 3dB}}$, 
where $\Theta_{{\theta_T}}^{\rm 3dB}$ is the 3dB beamwidth under the beam steering angle $\theta_T$ \cite{VanTrees2002} 

\vspace{-3mm}
\begin{footnotesize}
\begin{equation} \label{eq:Theta_3dB_azimuth_steering}
\begin{aligned}
    \Theta_{{\theta_T}}^{\rm 3dB} 
    = \sin^{-1}\left(\sin(\theta_T) + \frac{0.443 \lambda_c}{d_a}\right) 
    - \sin^{-1}\left(\sin(\theta_T) - \frac{0.443 \lambda_c}{d_a}\right), 
\end{aligned}
\end{equation}   
\end{footnotesize}

\noindent
where \mbox{$d_a = (M-1)d_e$}.
Note that when \mbox{$\theta_T = 0$}, \mbox{$\Theta_{{\theta_T}}^{\rm 3dB}$} can be approximated by \mbox{$\Theta^{\rm 3dB}$} (defined in (\ref*{def:3dB_BW})) using small angle approximation \cite{VanTrees2002}. 
The values of $A_{\rm b, \Theta_{\rm BW}}$ when \mbox{$\Theta_{\rm BW} = 10^{\circ}$}, $30^{\circ}$ and $60^{\circ}$ are shown in Table \ref*{table:SAT_svc_areas_broadenedbeam}, and are calculated by (\ref*{A_svc}).
Also, the values of $A_{{\rm n}, \Theta_{{\theta_T}}^{\rm 3dB}}$ when the beam steering angle is \mbox{$\theta_T = 0^{\circ}$} (no steering), $\frac{10^{\circ}}{2}$, $\frac{30^{\circ}}{2}$, and $\frac{60^{\circ}}{2}$ are shown in Table \ref*{table:SAT_svc_areas_narrowbeam} which is estimated according to Appendix \ref*{appendix:Asvc,n,T}.
Assume that UTs are uniformly distributed within $A_{\rm b, \Theta_{\rm BW}}$.
Compared with the ``broadened-beam" beamformer that can serve UTs within $A_{\rm b, \Theta_{\rm BW}}$ per single beam,
it takes ``narrow-beam" beamformer several time slots for steering and serving all UTs within $A_{\rm b, \Theta_{\rm BW}}$.
Denote $N_{\rm n, \Theta_{\rm BW}}^{(\rm{min})}$ as the minimum number of beams that the ``narrow-beam" beamformer is needed to cover the beam service areas of ``broadened-beam" beamformer $A_{\rm b, \Theta_{\rm BW}}$.
The beam transition time increases in proportion to $N_{\rm n, \Theta_{\rm BW}}^{(\rm{min})}$. 
Additionally, $N_{\rm n, \Theta_{\rm BW}}^{(\rm{min})}$ with $\Theta_{\rm BW} = 10^{\circ}$, $30^{\circ}$, and $60^{\circ}$ are evaluated as follows
\begin{equation} \label{eq:N_beam}
\begin{aligned} 
    N_{\rm n, 10}^{(\rm{min})} &= \left \lceil \frac{A_{\rm b, 10}}{A_{{\rm n}, \Theta_{5}^{\rm 3dB}}} \right \rceil, 
    N_{\rm n, 30}^{(\rm{min})} = N_{\rm n, 10}^{(\rm{min})} + \left \lceil \frac{A_{\rm b, 30} - A_{\rm b, 10}}{A_{{\rm n}, \Theta_{15}^{\rm 3dB}}} \right \rceil, \\
    N_{\rm n, 60}^{(\rm{min})} &= N_{\rm n, 30}^{(\rm{min})} + \left \lceil \frac{A_{\rm b, 60} - A_{\rm b, 30}}{A_{{\rm n}, \Theta_{30}^{\rm 3dB}}} \right \rceil.
\end{aligned}
\end{equation}
Considering beam transition time, the average channel capacity of the ``narrow-beam" beamformer for serving UTs in $A_{\rm b, \Theta_{\rm BW}}$ is evaluated as 
\begin{align} \label{eq:C_avg}
    C_{\rm n, \Theta_{\rm BW}} = \frac{C_{\rm n}}{N_{\rm n, \Theta_{\rm BW}}^{(\rm{min})}},
\end{align}
where $C_{\rm n}$ is evaluated using (\ref{eq:C_n}).
The values of $C_{\rm b, \Theta_{\rm BW}}$ and $C_{\rm n, \Theta_{\rm BW}}$ are presented in Table \ref{table:Capacity_results}.
From our analysis, the proposed ``broadened-beam" beamformers with \mbox{$\Theta_{\rm BW} = 10^{\circ}$}, $30^{\circ}$, and $60^{\circ}$ can offer capacities that are at least four times greater than ``narrow-beam" beamformers employing an array steering vector when the beam transition time is considered. 

\begin{table}[h!]
    \centering
    \begin{small}
    \caption{Beam areas of the proposed ``broadened-beam" beamformers under different beamwidths} \label{table:SAT_svc_areas_broadenedbeam}
    \vspace{-2mm}
    \begin{tabular}{ |c|c|c|c|} 
    \hline
    \textbf{Beamwidth $\Theta_{\rm BW}$}  & $10^{\circ}$ & $30^{\circ}$ & $60^{\circ}$\\ 
    \hline \hline
    $A_{\rm b, \Theta_{\rm BW}} \ [\rm{Mkm}^2]$ & $7,279$ & $68,666$ & $326,450$ \\
    \hline
    \end{tabular}
    \end{small}
\end{table} 

\begin{table}[h!]
    \centering
    \begin{footnotesize}
    \caption{Beam areas of the ``narrow-beam beamformers with 3${\rm dB}$ beamwidth under different steering angles $\theta_T$} \label{table:SAT_svc_areas_narrowbeam}   \vspace{-2mm} 
    \begin{tabular}{ |c|c|c|c|c|} 
    \hline
    \bf{\mbox{$\theta_T= \Theta_{\rm BW}/2$} } & $0^{\circ}$ (no steering) & $10^{\circ}/2$ & $30^{\circ}/2$ & $60^{\circ}/2$ \\ 
    \hline \hline
    $\Theta_{{\theta_T}}^{\rm 3dB}$ [\rm{deg}] & $3.3^{\circ}$ & $3.35^{\circ}$ & $3.41^{\circ}$ & $3.81^{\circ}$ \\ 
    \hline
    $A_{{\rm n}, \Theta_{{\theta_T}}^{\rm 3dB}} \ [\rm{M km}^2]$ & $786.23$ & $820.12$ & $1,203.9$ & $5,741.8$ \\ 
    \hline
    \end{tabular}
    \end{footnotesize}
\end{table} 

\begin{table}[h!]
    \centering
    \begin{small}
    \caption{Channel capacity evaluation of the proposed ``broadened-beam beamformers and the ``narrow-beam beamformers employing array steering vector} \label{table:Capacity_results}
    \vspace{-2mm}
    \begin{tabular}{ |c|c|c|c|c|} 
    \hline
     \textbf{Beamwidth $\Theta_{\rm BW}$}           & $10^{\circ}$  & $30^{\circ}$      & $60^{\circ}$ & Remark \\ 
    \hline \hline
    $D_{\rm n, ML}$                                 & \multicolumn{3}{|c|}{$1,048,576$} & (\ref{eq:P_peak}) \\ 
    \hline
    $\rm{SNR_n}\ [\rm{dB}]$                       & \multicolumn{3}{|c|}{$23.12$}     & (\ref{eq:SNR_b_n}) \\ 
    \hline
    $C_{\rm n} \ \rm{[Mbps]}$                       & \multicolumn{3}{|c|}{$3,844$}     & (\ref{eq:C_n})\\ 
    \hline
    $\rm{SNR_{b, \Theta_{\rm BW}}}\ [\rm{dB}]$    & $11.02$          & $5$               & $-2$          & Table \ref{tb:Angular_parameters} \\          
    \hline
    $\alpha_{\Theta_{\rm BW}}$                      & $179.35$      & $89.89$           & $40.15$       & Table \ref{tb:Angular_parameters} \\
    \hline
    $N_{\rm n, \Theta_{\rm BW}}^{(\rm{min})}$      & $9$           & $60$ & $105$      & (\ref{eq:N_beam}) \\
    \hline
    $C_{\rm n, \Theta_{\rm BW}} \ \rm{[Mbps]}$      & $427.11$      & $64.07$           & $36.61$       & (\ref{eq:C_avg}) \\ 
    \hline
    $C_{\rm b, \Theta_{\rm BW}} \ \rm{[Mbps]}$      & $1,882.2$     & $1,028.7$         & $352.86$      & (\ref{eq:C_b}) \\ 
    \hline 
    \end{tabular}
    \end{small}
\end{table} 

\section{Conclusion} \label{sec:Conclusion}
In this study, beam broadening algorithms for uniform rectangular arrays (URAs) in low Earth orbit (LEO) satellite communications (SatComs) were studied.
The proposed method is the first of its kind to jointly consider the path loss variation from satellite (SAT) to user terminal (UT) due to the Earth's curvature to guarantee quality of service (QoS), 
constant modulus constraints (CMCs) for maximizing power amplifier (PA) efficiency, and
out-of-beam radiation suppression to avoid interference. 
The URA design problem is formulated and decomposed into two uniform linear array (ULA) design subproblems by utilizing decomposition based on Kronecker product beamforming, which significantly reduces the size of the optimized variables compared to the original URA design problem.
The non-convex ULA subproblems are addressed using semidefinite relaxation (SDR) and a convex iterative algorithm.
Simulation results demonstrate that the proposed method can synthesize beampatterns with low out-of-beam radiation, achieve CMCs, and guarantee the received SNR in SAT service areas when the service beamwidths are $10^{\circ}$, $30^{\circ}$, and $60^{\circ}$.
In addition, the channel capacity evaluation shows that the proposed ``broadened-beam" beamformers can offer capacities that are at least four times greater than ``narrow-beam" beamformers employing an array steering vector when the beam transition time is considered. 
Therefore, the proposed method is a potential candidate for LEO broadcasting applications such as digital video broadcasting (DVB).
In the future, an algorithm that can directly solve the URA design problem (\ref{P1_00}) with a large antenna size, e.g., \mbox{$32 \times 32$}, is worthwhile developing to achieve optimal out-of-beam radiation suppression. 
Moreover, to save hardware costs, 
utilizing phase shifters with reduced quantization bits and applying different subarray partitions or clustered arrays for URA design is also worth further exploration.

\section{Beam areas estimation of the ``narrow-beam" beamformer with beam steering angle $\theta_T$} \label{appendix:Asvc,n,T}

\begin{figure}[h!]
    \centering
    \includegraphics[width = 5.5cm]{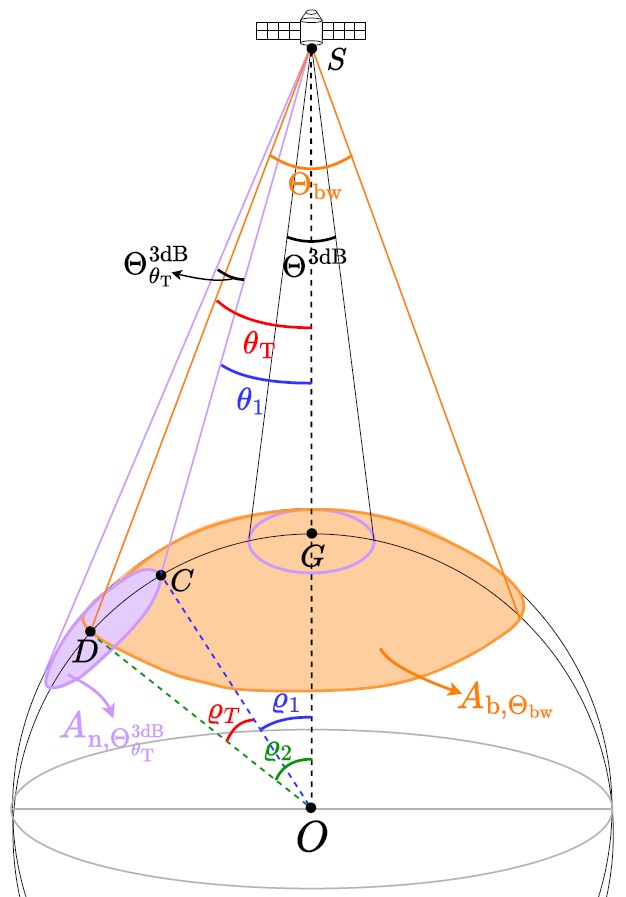}  
    \caption{Diagram for estimating the beam service areas of the ``narrow-beam" beamformer.} 
    \label{Fig:A_svc_scenario_steering} 
\end{figure}

In Fig. \ref{Fig:A_svc_scenario_steering},
$A_{{\rm n}, \Theta_{{\theta_T}}^{\rm{3dB}}}$ is the beam service areas of the ``narrow-beam" beamformer with beam steering angle $\theta_T$ that we would like to estimate. 
$\Theta_{{\theta_T}}^{\rm{3dB}}$ and $\Theta^{\rm{3dB}}$ are the ``narrow-beam" beamformer 3dB beamwidth with and without beam steering defined in (\ref{eq:Theta_3dB_azimuth_steering}) and (\ref{def:3dB_BW}), respectively. 
Also, $A_{\rm b, \Theta_{\rm{bw}}}$ is the beam service areas of the ``broadened-beam" beamformer with service beamwidth $\Theta_{\rm{bw}}$.
We estimate $A_{{\rm n}, \Theta_{{\theta_T}}^{\rm{3dB}}}$ through the following steps:
Firstly, calculate $\Theta_{{\theta_T}}^{\rm{3dB}}$ according to (\ref{eq:Theta_3dB_azimuth_steering}).
Then, we get \mbox{$\theta_1 = \theta_T - \Theta_{\theta_T}^{\rm{3dB}}/2$}.
Secondly, from $\triangle{\rm{OCS}}$, $\varrho_1$ can be obtained by (\ref{eq:beta_Spherical_cap}).
Similarly, from $\triangle{\rm{ODS}}$, $\varrho_2$ can be obtained by (\ref{eq:beta_Spherical_cap}). 
Then, we have \mbox{$\varrho_T = \varrho_2-\varrho_1$}.
Finally, $A_{{\rm n}, \Theta_{{\theta_T}}^{\rm{3dB}}}$ is obtained by (\ref{A_svc}).

\section{Justification of (\ref{def:MSQE})} \label{appendix:justification_varpi}

In the following, we generally follow the derivation approach in \hbox{\cite{Lin2017}} to justify the approximation in ({\ref{def:MSQE}}).
Assume the quantization error \mbox{$\epsilon_{m,l}=\hat{\varphi}_{m,l}-\varphi_{m,l}$} is zero mean and uniformly distributed over the interval 
$[\frac{-\Delta}{2}, \frac{\Delta}{2}]$, 
where $\Delta$ is defined in ({\ref{def:quantization_step}}).
Substitute \mbox{${\bf W}_{\rm{opt}}=e^{j\varphi_{m,l}}$} and \mbox{${\widehat{\bf W}}=e^{j\hat{\varphi}_{m,l}}$} into ({\ref{def:MSQE}}) yields 
\mbox{$\varpi = \frac{1}{M_xM_y}\sum_{m=0}^{M_x-1}\sum_{l=0}^{M_y-1}\mathbb{E}\left\{|e^{j\hat{\varphi}_{m,l}} - e^{j\varphi_{m,l}}|^2\right\}$}.
Firstly, we calculate
$e^{j\hat{\varphi}_{m,l}}-e^{j{\varphi}_{m,l}}=e^{j({\varphi}_{m,l}+{\epsilon}_{m,l})}-e^{j{\varphi}_{m,l}}=\cos({\varphi}_{m,l}+{\epsilon}_{m,l})+j\sin({\varphi}_{m,l}+{\epsilon}_{m,l})-\cos({\varphi}_{m,l})-j\sin({\varphi}_{m,l}) =\cos({\varphi}_{m,l})\cos({\epsilon}_{m,l})-\sin({\varphi}_{m,l})\sin({\epsilon}_{m,l})+j\sin({\varphi}_{m,l})\cos({\epsilon}_{m,l})+j\cos({\varphi}_{m,l})\sin({\epsilon}_{m,l})-\cos({\varphi}_{m,l})-j\sin({\varphi}_{m,l})$.
Assuming a high quantization resolution \hbox{\cite{Lin2017}}, such that \mbox{${\epsilon}_{m,l}\rightarrow 0$}, it follows \mbox{$\cos({\epsilon}_{m,l})\approx 1$} and \mbox{$\sin({\epsilon}_{m,l})\approx {\epsilon}_{m,l}$}.
We have \mbox{$e^{j\hat{\varphi}_{m,l}}-e^{j{\varphi}_{m,l}}\approx {\epsilon}_{m,l}je^{j{\varphi}_{m,l}}$}.
Then we get \mbox{$\mathbb{E}\{|e^{j\hat{\varphi}_{m,l}}-e^{j{\varphi}_{m,l}}|^2\}\approx \mathbb{E}\{{\epsilon}_{m,l}^2\}={\rm{Var}}(\epsilon_{m,l})=\frac{\Delta}{12}$}, 
where \mbox{${\rm{Var}}(\epsilon_{m,l})$} represents the variance of $\epsilon_{m,l}$.
Finally, we obtain \mbox{$\varpi\approx \frac{1}{M_xM_y}\sum_{m=0}^{M_x-1}\sum_{l=0}^{M_y-1}\rm{Var}(\epsilon_{m,l}) = \frac{\Delta}{12}$}.

\bibliographystyle{IEEEtran}
\bibliography{IEEEabrv,Bibliography}

\vfill

\end{document}